\documentclass[10pt]{article}
\PassOptionsToPackage{preprint}{tmlr}
\newcommand{\paperauthor}{\name Ryan Lim \email ryan.lim@einsteinmed.edu \\ \addr Albert Einstein College of Medicine}
\usepackage{tmlr}
\usepackage{amsmath,amssymb,amsthm,mathtools}
\usepackage{graphicx}
\usepackage{booktabs}
\usepackage{enumitem}
\usepackage{xcolor}
\usepackage{hyperref}
\usepackage{url}
\usepackage[capitalize]{cleveref}

\newtheorem{theorem}{Theorem}
\newtheorem{proposition}[theorem]{Proposition}
\newtheorem{lemma}[theorem]{Lemma}
\newtheorem{corollary}[theorem]{Corollary}
\theoremstyle{definition}
\newtheorem{definition}[theorem]{Definition}
\newtheorem{example}[theorem]{Example}
\newtheorem{conjecture}[theorem]{Conjecture}
\newtheorem{remark}[theorem]{Remark}

\newcommand{\R}{\mathbb{R}}
\newcommand{\PP}{\mathbb{P}}
\newcommand{\E}{\mathbb{E}}
\newcommand{\one}{\mathbf{1}}
\newcommand{\FDR}{\mathrm{FDR}}
\newcommand{\Bb}{\bar B}
\newcommand{\curv}{\mathfrak{D}}

\newcommand{\intuition}[1]{\par\noindent\textit{Intuition.} #1\par}

\newcommand{\gap}[1]{#1}

\newcommand{\Var}{\mathrm{Var}}
\newcommand{\Cov}{\mathrm{Cov}}

\begin{document}
\title{Almost Controlled: The FDR of Benjamini--Hochberg on Weighted Conformal p-Values}
\providecommand{\paperauthor}{Anonymous authors}
\author{\paperauthor}

\maketitle

\begin{abstract}
Weighted conformal p-values generalize conformal inference to cases with covariate shift. One way to use them is to pass
them to the Benjamini--Hochberg (BH) procedure, either to decide which test units to select or to detect outliers. But
weighted conformal p-values may not be positively dependent, so we want to know: does BH still control the false discovery
rate (FDR) in finite samples? Based on experiments, it appears to, but we prove it doesn't. We give two counterexamples,
certified in exact rational arithmetic, where the FDR exceeds its target level $q$. We also identify how large the
difference between the FDR and $q$ can be. With $n$ calibration points, $m$ test units and weight ratio $B$, the relative
excess $\FDR/q-1$ is bounded by a function of $n$, $m$, $B$ and $q$ that is of order $\sqrt{Bm/(nq)}$ when $Bm/(nq)$ is
small. The bound holds for every configuration with $B\le n$ and every $n$. This order can be reached when multiple
non-null test units share a common threshold: in a Poisson-limit model, that configuration's excess is
$c(q)\sqrt{Bm/(nq)}$ to first order as $B/n\to0$ and then $m\to\infty$. For some fixed distributions with this structure,
the excess decreases only like $n^{-1/2}$ as $n$ grows. If every test unit is null, the excess is $O(B/n)$ at fixed $m$,
and in certain configurations it grows linearly with $m$. To assist practitioners in judging how large the FDR can be, we
turn that bound, which depends only on $n$, $m$, $q$ and $B$, into a lowered BH level with proved finite-sample control,
and give a much less conservative rule of thumb that covers every worst case we were able to find but provides no
guarantee.
\end{abstract}

\section{Introduction}

Conformal p-values make a simple comparison: where does a test point's score fall among the scores of a held-out
calibration set? As long as the calibration and test data are exchangeable, these p-values are valid in finite samples
\citep{bates2023testing,jin2023selection}. But in many applications, the data aren't exchangeable. The test units come from
a covariate distribution $Q$ that differs from the calibration distribution $P$, and the plain comparison loses its
guarantee. Weighted conformal p-values \citep{tibshirani2019conformal,jin2026weighted} correct for this covariate shift by
reweighting each calibration point by the likelihood ratio $w(x)\propto dQ/dP(x)$. They're used for two tasks. In
selection, we're given $m$ test units and want the ones whose unobserved outcome clears a threshold. Outlier detection is
about flagging test units that don't come from the calibration distribution. Either way, we form a p-value for each test
unit, and we want to control the false discovery rate (FDR). The default tool is the Benjamini--Hochberg (BH) procedure
\citep{benjamini1995controlling}.

For unweighted conformal p-values, this is safe. The p-values are positively dependent, and BH controls the FDR
\citep{bates2023testing,jin2023selection}. Weights break that argument. \citet{jin2026weighted} note that weighted
conformal p-values ``cannot be passed on to classical multiple testing procedures since they may not obey a well-known
positive dependence property'', and they propose weighted conformalized selection, which has exact finite-sample
control. Does plain BH on weighted conformal p-values also control the FDR? They left that open. In their words, ``it
remains unclear whether applying existing multiple testing procedures controls the FDR in finite samples'' (\S1.2), even
though BH ``empirically controls the FDR in all of our numerical experiments'' (\S2.3). Recent work still treats the
question as open (Section~\ref{sec:related}).

\paragraph{Contributions.} This paper answers the question and measures the failure. Throughout, $n$ is the number of
calibration points, $m$ the number of test units, $q$ the target level and $B$ the ratio of the largest to the smallest
weight; $\kappa:=B/n$.
\begin{enumerate}[label=(\arabic*),leftmargin=*,itemsep=2pt]
\item \emph{BH does not control the FDR in finite samples} (Theorem~\ref{thm:counter}). We give two configurations that
satisfy \citeauthor{jin2026weighted}'s assumptions and have $\FDR>q$, certified in exact rational arithmetic: one in which
every test unit is null, and one in strict selection, where the threshold is independent of the outcome given the
covariate. In the second, with $n=80$, $m=8$ and $q=1/10$, the FDR is $1.136\,q$.
\item \emph{With non-null units, the worst excess is of order $\sqrt{Bm/(nq)}$.} In every configuration with $B\le n$ and
at every $n$, $\FDR/q-1\le\kappa+1.612\sqrt{(1+\kappa)Bm/(nq)}+O(Bm/(nq))$ (Theorem~\ref{thm:upper}). Conversely,
in a Poisson-limit model a construction with non-null ``decoy'' units that share a threshold has excess
$c(q)\sqrt{\kappa m/q}$ to first order, as $\kappa\to0$ and then $m\to\infty$, with $c(q)=(1-q)e^{-1-2q}/(2\sqrt{2\pi})$
(Theorems~\ref{thm:J2}, \ref{thm:J7} and~\ref{thm:R}). For some fixed distributions the excess decays only like $n^{-1/2}$
(Theorem~\ref{thm:F} and Example~\ref{ex:F-J3}).
\item \emph{When every test unit is null, the excess is smaller.} It is $O(B/n)$ at fixed $m$
(Theorem~\ref{thm:globalupper}), and there are configurations where $\FDR-q$ is, to first order, at least
$\tfrac12q(1-q)(m-1)(1-q+q/m)\,B/n$, linear in $m$; as $m\to\infty$ the constant per unit of $qmB/n$ is at least
$\tfrac12(1-q)$ (Theorem~\ref{thm:T2}). Whether the upper bound is also linear in $m$ is supported numerically and by a
sketch, but not proved.
\item \emph{Practical remedies} (Section~\ref{sec:practical}): a computable envelope for the FDR, a level deflation with
proved finite-sample control that uses only $n$, $m$, $q$ and $B$, and a numerical rule of thumb, about ten times less
conservative, that covered every worst case we found.
\end{enumerate}

\paragraph{How large is the excess?} The certified strict-selection construction uses a weight ratio comparable to $n$ ($B\approx1.5n$ to $1.7n$). If we
re-tune it for each $n$, it gives $\FDR=1.069\,q$, $1.103\,q$ and $1.136\,q$ at $n=20$, $40$ and $80$
(Remark~\ref{rem:F-n}). When $B\ll n$, the rates above govern the excess. Across the Poisson-limit configurations our
searches found, the relative excess stayed below $0.17\sqrt{Bm/(nq)}$ whenever $B/n\le0.1$. Among those configurations,
the largest excess at any $B/n$ was $51\%$, and it needed extreme weights ($B$ about $10n$). But in the Poisson-limit
model, the coefficient of $\sqrt{B/n}$ in the excess grows like $\sqrt m$ (Theorem~\ref{thm:J7}, as $B/n\to0$).

\paragraph{Mechanisms and techniques.} Given the calibration data $D$, the test p-values are independent, so all the dependence that BH sees passes through
$D$. Lemma~\ref{lem:F1} makes this exact. When every null is an oracle null (its threshold equals its outcome), $\FDR-q$
is a weighted sum of covariances minus a non-negative term. Each covariance is between the null mass below a BH
threshold and the probability that the other units drive BH to exactly that many rejections. The non-negative term comes
from non-null units with small oracle p-values, and in our constructions it is zero. This covariance behaves very
differently in two regimes (Section~\ref{sec:counter}). If every test unit is an oracle null, the conditional law of a
test p-value is uniform on average over $D$, so a fluctuation of the calibration data has no first-order effect. What's
left is a second-order effect of the curvature of Simes' test, of order $B/n$. Things are different when non-null units
share a threshold. Their rejection switches on and off with $D$, and the switch can be correlated with the null mass at
first order, which gives an excess of order $\sqrt{B/n}$. For the upper bound we use a cruder reduction: the FDR is at
most $q$ times the expected largest ratio, over the BH thresholds, of the conditional null mass below a threshold to the
threshold itself (Proposition~\ref{prop:E8a}). We then bound the worst relative error of a weighted empirical
distribution function on the BH grid.

Some of our lower bounds are proved in a Poisson-limit model, in which a few heavily weighted covariate cells have Poisson
calibration counts, and some statements rest on numerical evidence. We say so wherever this is the case.

\paragraph{Outline.} Section~\ref{sec:setting} sets up the notation and two reductions, and Section~\ref{sec:counter} gives the
counterexamples and the two mechanisms. Section~\ref{sec:sqrt} proves, with non-null units, an upper bound of order
$\sqrt{Bm/(nq)}$ and, in the Poisson-limit model, a lower bound of the same order. Section~\ref{sec:global} treats the
global null. Section~\ref{sec:practical} collects what a practitioner can compute, and Section~\ref{sec:open} discusses
open problems. All proofs are in the appendices; Appendix~\ref{app:counter} gives the certificates for the
counterexamples.

\subsection{Related work}\label{sec:related}
\paragraph{Multiple testing under dependence.} The BH procedure \citep{benjamini1995controlling} controls the FDR for
independent p-values and, as \citet[\S2.2]{jin2026weighted} recall, ``when applied to PRDS p-values''
\citep{benjamini2001control}. For all other forms of dependence, \citet{benjamini2001control} show that a conservative modification of BH controls the
FDR: BH at level $q/\sum_{k\le m}1/k$. \citet{hommel1983tests} gives tests of the global (intersection) null hypothesis
under arbitrary dependence, with sharp bounds on their error probability, and \citet{sarkar2002some} extends FDR control
under positive dependence from the BH step-up procedure to more general stepwise procedures with the same critical
values. \citet{fithian2022conditional}
calibrate a data-dependent rejection threshold for each hypothesis to target exact FDR control when the dependence is known;
their dBH procedure dominates BH under positive regression dependence and BY in general.

\paragraph{Conformal p-values and selection.} For unweighted conformal p-values, \citet{bates2023testing} prove that the p-values are positively dependent and that BH
controls the FDR exactly, in a marginal sense. \citet{jin2023selection} use conformal p-values with BH-type thresholds
for selection under exchangeability. Under covariate shift, weighted conformal p-values
\citep{tibshirani2019conformal,jin2026weighted} ``cannot be passed on to classical multiple testing procedures since
they may not obey a well-known positive dependence property'' \citep{jin2026weighted}. This motivates weighted
conformalized selection (WCS), which comes with exact finite-sample control. With e-values, \citet{lee2024boosting} show
that e-BH on weighted conformal e-values controls the FDR, ``but unlike before, this is no longer equivalent to applying
BH to the weighted conformal p-values'' (their \S6.2). \citet{lee2025full} develop full-conformal e-values with
finite-sample control.

\paragraph{The question remains open in recent work.} \citet{lee2025full} compare with BH on weighted conformal p-values, which ``is not necessarily FDR-controlling''.
\citet[\S5.4]{hennhofer2026between} write that applying BH directly to discrete weighted conformal p-values ``is not
theoretically justified in general'', and \citet[\S2.8]{hennhofer2026nonconform} note that ``exact finite-sample FDR
control generally requires procedures tailored to the weighted setting''. None of these works proves or disproves
finite-sample control, or bounds the excess. Neither does any other work we found.

\noindent

\section{Setting and two reductions}\label{sec:setting}

\paragraph{Data and p-values.} We observe calibration data $(X_i,Y_i)$, $i=1,\dots,n$, drawn i.i.d.\ from $P$, and $m$
test units $(X_{n+j},Y_{n+j})$ drawn from $Q$, where $dQ/dP=w(X)/\E_Pw$ for a known weight function with $1\le w\le B$.
We write $\Bb:=\sup w/\E_Pw\le B$ and $\kappa:=B/n$. Test unit $j$ has a threshold $c_{n+j}$ and is null iff
$Y_{n+j}\le c_{n+j}$. With a score $V(x,y)$ that is monotone in $y$, write $V_i:=V(X_i,Y_i)$ and
$\hat V_{n+j}:=V(X_{n+j},c_{n+j})$. The weighted conformal p-value of \citet[eq.~(4)]{jin2026weighted} is
\[
p_j=\frac{\sum_{i\le n}w(X_i)\one\{V_i<\hat V_{n+j}\}+U_j\big(w(X_{n+j})+\sum_{i\le n}w(X_i)\one\{V_i=\hat V_{n+j}\}\big)}
{\sum_{i\le n}w(X_i)+w(X_{n+j})},
\]
with $U_j\sim\mathrm{Unif}(0,1)$ for tie-breaking. The oracle p-value $p^*_j$ is the same expression with $\hat V_{n+j}$
replaced by the true score $V(X_{n+j},Y_{n+j})$. A null unit with $c_{n+j}=Y_{n+j}$ (an \emph{oracle null}) has
$p_j=p^*_j$, and every null unit has $p_j\ge p^*_j$. BH at level $q$ uses the grid $t_k:=qk/m$: it finds the largest $k$ such that at least $k$
p-values are at most $t_k$, and rejects the units whose p-values are at most that $t_k$. We write $D$ for the
calibration data. The assumptions of \citet{jin2026weighted} (i.i.d.\ test tuples and $U_j$, independent of $D$) hold
throughout, and some results need only the weaker assumptions (A1)--(A3) of Assumption~\ref{ass:A}
(Appendix~\ref{app:envelope}). \emph{Strict selection} means that $c_{n+j}$ is independent of $Y_{n+j}$ given $X_{n+j}$; in
the \emph{global null} every test unit is null. In \emph{outlier detection} the inliers are drawn from $Q$ and are the
nulls, and the outliers come from another distribution.

Two exact facts organise the analysis. Oracle p-values are exactly uniform, ties included; and, given $D$, the test units
are i.i.d., which leads to an exact identity for the excess and to two reductions.

\begin{lemma}[Exact uniformity]\label{lem:U}
$\PP(p^*\le s) = s$ for all $s\in[0,1]$.
\end{lemma}

\intuition{Given the multiset of the $n+1$ points, the test point is each of them with probability proportional to its
weight. The randomised weighted CDF maps each tie group to an interval whose length is exactly that group's probability.
So the intervals tile $[0,1]$ and $p^*$ is uniform, ties included.}

\begin{lemma}[Exact covariance identity]\label{lem:F1}
Let $A_k(D) := \PP(\text{unit }1\text{ null},\ p_1\le t_k\mid D)$ and $r_k(D) := \PP(R_1 = k\mid D)$, where $R_1$ is
the number of BH rejections among $(0,p_2,\dots,p_m)$. If every null unit is an oracle null and
$\varepsilon_k := \PP(\text{unit 1 non-null},\ p^*_1\le t_k)$, then for every $n$,
\[
\FDR - q \;=\; m\sum_{k=1}^m \frac1k\,\mathrm{Cov}\big(A_k(D),\,r_k(D)\big) \;-\; m\sum_{k=1}^m\frac1k\,\varepsilon_k\,\E\, r_k(D).
\]
\end{lemma}

\intuition{Given the calibration data, BH sees $m$ i.i.d.\ p-values, and the leave-one-out identity turns the FDR into an
average of the null mass below the threshold the other units set. If that null mass did not move with what the other
units do, the average would be $q$ (minus the $\varepsilon$-term). So the whole excess is a covariance between the null
mass below $t_k$ and the chance that the others drive BH to exactly $k$ rejections.}

Lemma~\ref{lem:F1} is the basis of the fixed-distribution analysis in Section~\ref{sec:sqrt}. For upper bounds we use a
cruder reduction that needs no oracle assumption.

\begin{proposition}[Calibration-conditional reduction]\label{prop:E8a}
Let $F^*_D(t) := \PP(p^*\le t\mid D)$ for a test point from $Q$ and $M_D := \max_{k\le m}F^*_D(t_k)/t_k$. Then
$\FDR\le q\,\E[M_D]$, for every configuration under JC's assumptions (non-null units, shared thresholds and score atoms
allowed).
\end{proposition}

\intuition{Conditionally on $D$ the test p-values are independent, so BH behaves as under independence, except that a
null's conditional law may put more than $t$ of its mass below $t$. BH can exploit that excess at whichever grid point it
is largest. So the FDR is at most $q$ times the expected worst ratio $F^*_D(t_k)/t_k$.}

In the global null, BH makes a false discovery as soon as it rejects anything, so the FDR is the probability that Simes'
test rejects when applied to $m$ i.i.d.\ draws from the conditional law of a test p-value.

\begin{lemma}[Global-null reduction]\label{lem:L1}
If every test unit is null, $\FDR = \E\,\psi_m(F_D)$, where $\psi_m$ is the rejection probability of Simes' test for $m$
i.i.d.\ draws with CDF $F_D(t):=\PP(p_1\le t\mid D)$. If moreover every null is an oracle null ($c=Y$), then $\E F_D(t)=t$;
in general $\E F_D(t)\le t$.
\end{lemma}

\paragraph{Heavy cells and the Poisson limit.} Our constructions combine many \emph{light} covariate values of weight $1$
with a few \emph{heavy} cells of weight $\kappa n=B$ and $P$-mass $\lambda/n$, where $\lambda=\omega/\kappa$. If $N$ is
the number of calibration points in a heavy cell, its normalised calibration weight $K:=\kappa N$ has mean $\omega$ and,
in the Poisson-limit model, standard deviation $\sqrt{\kappa\omega}$; this fluctuation drives all our lower-bound
constructions. In the
Poisson-limit model the heavy calibration counts are independent Poisson($\lambda$) variables and the light calibration
scores have exactly the uniform CDF; this is the limit $n\to\infty$ at fixed $\kappa$. Lemma~\ref{lem:L2} proves that the
global-null FDR converges to its value in this model; for the non-null structures of Section~\ref{sec:sqrt} that
convergence is not proved, and the results there are stated in the model itself.

\section{Counterexamples and the two mechanisms}\label{sec:counter}

We first settle the question.

\begin{theorem}[No finite-sample control]\label{thm:counter}
There are configurations satisfying JC's assumptions with $\FDR>q$, certified by computer in exact rational arithmetic:
(i) global null, $n=5$, $m=2$, $q=4/5$: $\FDR = 393728107849/492032000000\approx0.80021$; (ii) strict selection
($c\perp Y\mid X$; $Y\mid X$ is non-degenerate on the light covariate and fixed on the two heavy ones), $n=80$, $m=8$,
$q=1/10$: $\FDR = 1.13585\,q$ (an exact rational; Appendix~\ref{app:counter}).
\end{theorem}

Instance (i) uses a large level ($q=4/5$) and tiny sizes, so that the excess $\FDR-q$, about $2\times10^{-4}$, can be
certified by full enumeration. Instance (ii) is a genuine strict-selection problem at $q=1/10$: the threshold depends only on
the covariate, and the outcome is random on the light covariate; its weight ratio, $B=137$, exceeds $n=80$. Both instances
use a few covariate cells with large weights, and their excesses come from the two mechanisms below, respectively.

\paragraph{Mechanism 1 (global null, second order).} When every test unit is an oracle null, the conditional law of a test
p-value given $D$ is uniform on average over $D$ (Lemma~\ref{lem:L1}), so a fluctuation of the heavy calibration weight has
no first-order effect. What remains is second order: Simes' rejection probability is a curved function of the conditional
distribution function on the grid, and averaging a curved function over the fluctuation produces an excess of order
$\Var(K)=\kappa\omega$, that is, of order $B/n$.

\begin{figure}[t]
\centering
\includegraphics[width=0.48\linewidth]{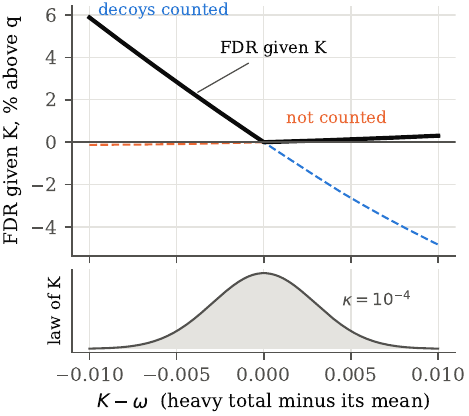}\hfill
\includegraphics[width=0.48\linewidth]{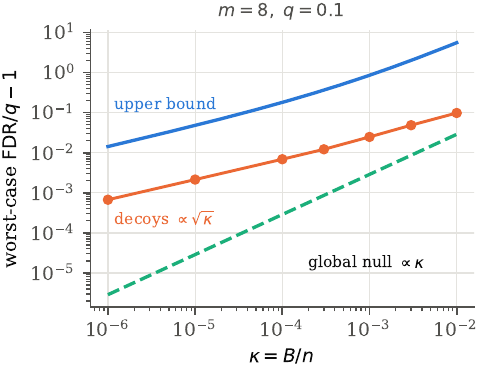}
\caption{\textbf{Left:} the FDR given the heavy calibration total $K$ (the configuration of Example~\ref{ex:J3}, $m=8$, $q=0.1$,
Poisson-limit model; exact evaluation). Each branch is linear through $q$ at $K=\omega$; the switch keeps the counted branch to the left and
the other to the right, so the conditional FDR has a kink at the mean and its average exceeds $q$ by order
$\sqrt{\kappa}$. \textbf{Right:} the worst-case excess against $\kappa=B/n$ at $m=8$: exact Poisson-limit FDR of the decoy
construction ($\propto\sqrt\kappa$), the global-null linear lower bound (Theorem~\ref{thm:T2}, $\propto\kappa$), and the upper bound of
Theorem~\ref{thm:upper} with $\Bb=\kappa n/(1+\omega)$.}
\label{fig:mechanism}
\end{figure}

\paragraph{Mechanism 2 (non-null selection, first order).} Suppose a block of non-null units (\emph{decoys}) shares one
threshold, so that their p-values coincide and sit near a BH threshold. Whether BH counts them then depends on the
calibration draw. At the mean draw the null law is exactly uniform below $q$ (saturation, Lemma~\ref{lem:J1}), so the FDR is
exactly $q$ on both sides of the switch. Away from the mean, each side's FDR moves linearly in the heavy weight, with
different slopes, and the switch pairs a lighter heavy total with the side on which the decoys are counted, which is where
the null units gain (Figure~\ref{fig:mechanism}). The average of such a kinked function exceeds its value at the mean by an
amount proportional to the standard deviation $\sqrt{\kappa\omega}$: a first-order effect.

\section{Non-null selection: a worst-case excess of order $\sqrt{Bm/(nq)}$}\label{sec:sqrt}

This section shows that, when non-null units can share a threshold, the worst excess is of order $\sqrt{Bm/(nq)}$. We
give an upper bound valid in every configuration with $B\le n$ at every $n$ (Theorem~\ref{thm:upper}), a lower bound of the
same order in the Poisson-limit model (Theorems~\ref{thm:J2} and~\ref{thm:J7}), and a fixed-distribution result as
$n\to\infty$ at fixed $B$ (Theorem~\ref{thm:F}). Theorem~\ref{thm:R} collects them.

\paragraph{Upper bound.}

\begin{theorem}[Upper bound, every configuration]\label{thm:upper}
Let $\kappa=B/n\le1$, $x := \Bb m/(nq)$ and $c := 2+(1+\kappa)q/3$. Then
\[
\frac{\FDR}{q}-1 \;\le\; \E[M_D]-1 \;\le\; \kappa + 1.612\sqrt{(1+\kappa)\,x} + 2.7726\,c\,x,
\]
and $\E[M_D]-1\le\kappa+J$ with $J$ an explicit one-dimensional integral (computable, slightly tighter).
\end{theorem}

\intuition{The reduction turns the FDR into the worst relative error of a weighted empirical CDF on the grid. At $t_k$
that relative error is about $\sqrt{\Bb m/(nqk)}$, so the first grid points dominate. One-sided Bernstein tails summed over
the grid and integrated exactly give the constant $\sqrt2\int_0^\infty\min(1,1/(e^{\nu^2}-1))\,d\nu = 1.612$.}

The constant $1.612$ is far above the lower-bound constant $c(q)$ below. Bounds proved through the reduction cannot go
below about $0.399$:

\begin{proposition}[No better constant through the reduction]\label{prop:E9}
In the Poisson-limit model, a heavy atom placed below every light score with $\omega=t_1/(1-t_1)$ gives, as $\kappa\to0$,
$\E[M_D]-1 = \frac{1-1/m}{\sqrt{2\pi}}\sqrt{\Bb m/(nq)}\,(1+o(1))$. So no bound proved through Proposition~\ref{prop:E8a} has a
leading constant below $(1-1/m)/\sqrt{2\pi}$, which tends to $1/\sqrt{2\pi}\approx0.399$.
\end{proposition}

\intuition{With all heavy mass at the very bottom, filling $[0,t_1]$ on average, every ratio $F^*_D(t_k)/t_k$ equals $1$ at
the mean draw and falls linearly in the heavy total, steepest at $k=1$ (slope about $m/q$). A light draw pushes $M_D$ up
through $k=1$. The Gaussian mean of that positive part, $\sigma/\sqrt{2\pi}$, is the whole effect.}

\paragraph{Lower bounds: decoys in the Poisson limit.} The lower bounds use Mechanism~2 in the Poisson-limit model. There
is one heavy cell, at score quantile $g$, with normalised calibration weight $K$ of mean $\omega$. Given $K$, a light unit
with score quantile $u$ has p-value $f_K(u):=(u+K\one\{g<u\})/(1+K)$, and heavy test units have p-values spread over the
\emph{heavy window} $[g/(1+K),(g+K)/(1+K)]$. A fraction $\zeta$ of the light units, those with $u>1-\zeta$, are decoys:
they share the threshold quantile $v_0$, so all of them have the p-value $f_K(v_0)$. We write $A_k(K)$ for the null mass
below $t_k$ given $K$. The \emph{deterministic-$K$ FDR} is the FDR of the limiting model at a fixed value of $K$, with the
decoy atom counted from $t_{k_0}$ on (side $-$) or from $t_{k_0+1}$ on (side $+$), where $t_{k_0}$ is the grid point it
crosses (Appendix~\ref{app:sqrt} gives the details).

\begin{lemma}[Saturation at the mean]\label{lem:J1}
In the decoy structure (one heavy cell of mean weight $\omega$; light nulls; a non-null decoy block with a common
threshold; $f_\omega(1-\zeta)\ge q$), the deterministic-$K$ FDR equals $q$ at $K=\omega$ on both sides of the switch.
\end{lemma}

\begin{theorem}[The $\sqrt\kappa$ expansion]\label{thm:J2}
In the Poisson-limit model, let the decoy atom cross the grid point $t_{k_0}$ ($1\le k_0\le m$) at the switch point
$k^* = \omega + z_0\sqrt{\kappa\omega}$, and assume strict saturation ($f_\omega(1-\zeta)>q$), $v_0>g$, and genericity at
$K=\omega$ (no grid point at an edge of the heavy window or at a jump of the light map). Then, uniformly in $z_0$ in compacts,
as $\kappa\to0$,
$\FDR = q + \sqrt{\kappa\omega}\,(h_+-h_-)\,\varphi(z_0) + O(\kappa)$, with $h_s = m\sum_k\frac1k A_k'(\omega)P_s(R_1=k)$, where
under $P_-$ ($P_+$) the decoy atom is counted from $t_{k_0}$ (from $t_{k_0+1}$) on.
\end{theorem}

\intuition{At the mean heavy total the FDR is exactly $q$ on both sides of the switch. Each side moves linearly in the
heavy total, with different slopes. The switch selects the steeper side exactly when the total is low, so the average
picks up the slope difference times $\E(K-\omega)^+\approx\sqrt{\kappa\omega}\,\varphi(z_0)$.}

A concrete instance at $m=8$:

\begin{example}\label{ex:J3}
$q=0.1$, $m=8$, one heavy cell at $g=0.018418$ with $\omega=0.086271$, decoy fraction $\zeta=0.91847$:
$\FDR/q-1 = 0.6707\sqrt\kappa + 1.68\kappa$ (the coefficient is the formula of Theorem~\ref{thm:J2}, evaluated numerically;
the $\kappa$ term is fitted to exact Poisson-limit evaluations down to $\kappa=10^{-6}$).
\end{example}

The coefficient is large: at $\kappa=10^{-3}$ this gives $\FDR/q-1\approx0.023$. It also grows with $m$. Placing the
decoys at the second BH threshold gives a family whose coefficient has an explicit limit as $m\to\infty$.

\begin{theorem}[$\sqrt m$ growth with an explicit constant]\label{thm:J7}
In the Poisson-limit model, for a decoy switching at $t_2$ with $\omega = q/m$ and decoy fraction $1/m$ (the structure of
Theorem~\ref{thm:C-J2prime}), the coefficient
$C_m := \lim_{\kappa\to0}(\FDR/q-1)/\sqrt\kappa$ satisfies
$\lim_{m\to\infty}C_m/\sqrt{m/q} = c(q) := (1-q)e^{-1-2q}/(2\sqrt{2\pi})$ ($=0.054$ at $q=0.1$), an iterated limit
($\kappa\to0$ first).
\end{theorem}

\intuition{The decoy p-value sits on $t_2$, so the calibration draw decides whether BH counts it at the second threshold. A
light heavy draw lowers every null p-value and pulls the decoy below $t_2$, so BH stops at $R=2$ instead of $R=1$ exactly
when the nulls are favoured. The fluctuation is $\sqrt{\kappa\omega}$ with $\omega\approx q/m$, but its effect is summed
over $m$ units, which leaves $\sqrt{\kappa m/q}$.}

\paragraph{Fixed distributions.} The Poisson-limit model lets the distribution change with $n$. For a single fixed
distribution the same mechanism can still give an excess of order $n^{-1/2}$, but only if the decoy threshold sits exactly at
the population switching point (condition (F2) in Appendix~\ref{app:finite}); otherwise the switch freezes for large $n$, and
Remark~\ref{rem:A-tuning} outlines why the excess is then $o(n^{-1/2})$. Under (F1)--(F4) the limit $c_0$ below can still
be zero (for example when the decoys sit at $t_1$); $c_0>0$ is certified for the distributions of Example~\ref{ex:F-J3}.

\begin{theorem}[Fixed distribution, finite $n$]\label{thm:F}
For a fixed $P$ with one heavy atom of weight $B$ and $P$-mass $\pi_H$, uniform light scores, and the decoy threshold at
the population switching point (conditions (F1)--(F4)), $\sqrt n\,(\FDR_n-q)\to c_0$ as $n\to\infty$, with $c_0$ in
closed form (Appendix~\ref{app:finite}).
\end{theorem}
\begin{example}[The parameters of Example~\ref{ex:J3}]\label{ex:F-J3}
For the parameters of Example~\ref{ex:J3} with a fixed weight $B$, the sign $c_0>0$ is certified in exact rational arithmetic at
$B=2,5,10,100$. Numerically, $c_0/(q\sqrt B) = 0.316,\,0.522,\,0.595,\,0.663$ at these $B$, approaching that example's Poisson-limit
coefficient $0.6707$ as $B$ grows (numerically; the limit in $B$ is not proved).
\end{example}

\intuition{For a fixed distribution each conditional null mass fluctuates by about $1/\sqrt n$, so smooth dependence on the
calibration data contributes only $O(1/n)$ to the covariance identity (Lemma~\ref{lem:F1}). The decoys' rejection is an on/off
switch of the same fluctuation, and $\E[X\one\{X>0\}]$ is of order $\mathrm{sd}(X)$. So the excess is of order $1/\sqrt n$,
with a constant that, numerically, tends to the Poisson-limit coefficient as $B$ grows.}

\begin{figure}[t]
\centering
\includegraphics[width=0.48\linewidth]{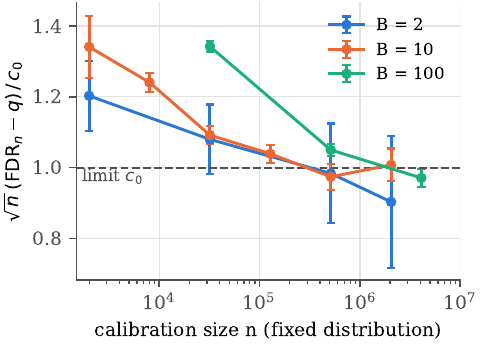}
\caption{Theorem~\ref{thm:F} at finite $n$: $\sqrt n(\FDR_n-q)$ divided by the predicted limit $c_0$, for one fixed distribution
per $B$ (the parameters of Example~\ref{ex:J3}). Rao--Blackwellised simulation of the calibration data with
$\E[\mathrm{FDP}\mid D]$ computed exactly; bars are one standard error.}
\label{fig:F}
\end{figure}

\begin{remark}[Outlier detection]
In outlier detection, in the Poisson-limit model, the first-order $\sqrt\kappa$ term cannot produce an excess once
$\kappa$ is below a small threshold (Proposition~\ref{prop:C-J5}). At moderate $B/n$ a switching outlier block does exceed
the largest global-null FDR our searches found (numerically; Appendix~\ref{app:sqrt}).
\end{remark}

\paragraph{The rate.} The next theorem combines the upper and lower bounds and states exactly what is proved where.

\begin{theorem}[The rate]\label{thm:R}
Write $x := \Bb m/(nq)$.
\begin{enumerate}[label=(\roman*),leftmargin=*,itemsep=1pt]
\item \emph{Upper bound, every $n$.} For every configuration under JC's assumptions with weight ratio at most $B$ and
$\kappa = B/n\le1$, $\FDR/q-1\le\kappa+1.612\sqrt{(1+\kappa)x}+2.7726\,c\,x$ (Theorem~\ref{thm:upper}).
\item \emph{Lower bound, Poisson-limit model, iterated limit.} For the structure of Theorem~\ref{thm:J7} in the
Poisson-limit model, take $\kappa\to0$ first and then $m\to\infty$:
\[
\lim_{m\to\infty}\ \lim_{\kappa\to0}\ \frac{\FDR/q-1}{\sqrt{\kappa m/q}} \;=\; c(q)=\frac{(1-q)e^{-1-2q}}{2\sqrt{2\pi}} .
\]
The limits are not shown to commute, and no rate is given for the inner one. That the finite-$n$ FDR converges to its
Poisson-limit value as $n\to\infty$ at fixed $\kappa=B/n$ is not proved for this structure.
\item \emph{Finite-$n$ support.} For one fixed distribution (fixed $B$, $n\to\infty$), Theorem~\ref{thm:F} gives
$\sqrt n(\FDR_n-q)\to c_0$, and $c_0>0$ for the distributions certified in Example~\ref{ex:F-J3}. So for those
distributions the excess is of order $n^{-1/2}$ for all large $n$, with no explicit $n_0$. Evaluating $c_0$ for the
structures of Theorem~\ref{thm:J7} at fixed $B$ shows the same $\sqrt m$ growth (numerically).
\end{enumerate}
So the worst excess is of order $\sqrt{Bm/(nq)}$: the upper side holds at every $n$, and the lower side holds in the
Poisson-limit model in the iterated limit (ii), with (iii) as finite-$n$ support.
\end{theorem}

\paragraph{Search for a larger lower constant (numerical).} With a saturation margin ($f_\omega(1-\zeta)\ge q+0.005$), structures with bottom-placed or stacked heavy cells and up to three decoy blocks do not beat \cref{thm:J7}'s family: the best coefficients found are $0.0553$, $0.0547$, $0.0544$ and $0.0542$ (each $\pm0.0005$) in units of $\sqrt{m/q}$ at $m=32,64,128,256$, against $0.0552$, $0.0546$, $0.0544$ and $0.0542$ for the structure of \cref{thm:J7}. Without the margin the optimiser finds a spurious gain at a degenerate boundary, where the non-null-oracle term of \cref{lem:F1} becomes first order; exact Poisson-limit sums rule it out. So we have no
evidence that the lower constant $c(q)$ can be improved within these families; the gap to the upper constant $1.612$ is
discussed in Section~\ref{sec:open}.

\section{Global null: an $O(B/n)$ excess, with a lower bound linear in $m$}\label{sec:global}

When every test unit is null, all rejections are false, and the FDR is the probability that Simes' test rejects
(Lemma~\ref{lem:L1}). Mechanism~1 is then the only effect, and it is second order: the excess is of order $B/n$ rather than
$\sqrt{B/n}$. We give a lower bound that is linear in $m$, an upper bound that is $O(B/n)$ at every fixed $m$, and a
conditional result on the constant.

\begin{theorem}[Lower bound]\label{thm:T2}
For every $m\ge2$, $q\in(0,1)$ and $\epsilon>0$, and all small $\kappa$ with $n$ large, there is a global-null configuration
with $\FDR\ge q+\big(L^*_m(q)-\epsilon\big)B/n$, where $L^*_m(q)$, the best first-order coefficient over the Poisson window
configurations of Appendix~\ref{app:global}, satisfies $L^*_m(q)\ge\tfrac12q(1-q)(m-1)(1-q+q/m)$ and
$\liminf_{m\to\infty}L^*_m(q)/(qm)\ge\tfrac12(1-q)$. So the excess is linear in $m$, and as $m\to\infty$ its constant per unit
of $qmB/n$ is at least $\tfrac12(1-q)$.
\end{theorem}

\intuition{Heavy cells in different bands fluctuate independently, so their contributions add. The band curvatures of
Simes' test sum to exactly $m(m-1)(1-q)$, because Simes stays exact when a p-value is duplicated
($\sum_{k=1}^{m-1}\curv_k = m(m-1)(1-q)$). Spreading cells over all bands collects that sum.}

So some global-null configurations reach an excess of order $mB/n$, against $\sqrt{Bm/(nq)}$ with non-null units, which
is much smaller when $Bm/n$ is small. Theorem~\ref{thm:globalupper} gives $O(B/n)$ at fixed $m$; an upper bound linear in
$m$ is not proved. The upper bound comes from the envelope of Theorem~\ref{thm:envelope}, combined with a high-probability bound on the
conditional distribution function of a test p-value; it holds for every score law, ties included.

\begin{theorem}[Upper bound, every score law]\label{thm:globalupper}
In the global null, for $\rho>1+\kappa$,
$\FDR\le q\big[1+2(m-1)\beta_m(\rho)\tfrac{B}{n+1}+(m-1)\tfrac{e^{-2a'}}{1-e^{-a'}}\big]$, with
$a' = nq(\rho-1-\kappa)^2/(2\Bb m(3\rho-2-2\kappa))$ and $\beta_m(\rho)$ explicit. Also
$\FDR\le q\big[1+m(m-1)^2B/(n+1)\big]$ for all $n$ and $B$; in particular, at fixed $m$ the excess is $O(B/n)$. Both bounds
hold for non-oracle nulls too, since then $p_j\ge p^*_j$ and BH is monotone.
\end{theorem}
\begin{remark}[Linearity in $m$ is not proved]
The bound is linear in $m$ wherever $\beta_m(\rho)$ stays bounded in $m$. Numerically it does ($\beta_m(1)\approx1.78$ at $m=1000$,
$q=0.1$, $\tau_B=0$), and an $m$-free majorant via a Chernoff bound is sketched but not written out; it also needs
$\rho(q+mB/(n+1))<1$. So linearity of the upper bound in $m$ is supported numerically and by a sketch, but not proved.
\end{remark}

Is $\tfrac12(1-q)$ the right constant? Within the family of configurations used for the lower bound it is, under a
condition that we have verified in exact arithmetic for many $(m,q)$ but have not proved.

\begin{proposition}[Conditional sharpness]\label{prop:H}
Simes' Hessian at the uniform point has non-positive off-diagonal entries and, for $0<q\le1/e$, is positive
semidefinite on grid points $2,\dots,m$. \emph{If} all band curvatures are positive (the positive-curvature condition of
Appendix~\ref{app:global}, which is open), then within the Poisson window family of Theorem~\ref{thm:T2} (windows strictly
inside one band each) the first-order constant per unit of $qmB/n$ is at most $\tfrac12(1-q)(m-1)/m$; so the constant
$\tfrac12(1-q)$ is attained as $m\to\infty$ and is not exceeded in that family. For $0<q\le1/e$ the Hessian result extends this bound on
the quadratic form to windows above $t_1$ that contain grid points, but the first-order expansion is not proved for such
windows.
\end{proposition}

We conjecture that the answer is the same over all global-null configurations.

\begin{conjecture}\label{conj:C1}
In the global null the linear-regime constant equals $\tfrac12(1-q)$.
\end{conjecture}

\section{What a practitioner can compute}\label{sec:practical}

The counterexamples show that BH on weighted conformal p-values can exceed its level, and Theorem~\ref{thm:upper} shows
that the excess is small when $Bm/(nq)$ is small. We give three tools: an envelope that bounds the FDR by the expectation of a quantity computable from
data, a deflated level with proved finite-sample control, and a much less conservative numerical rule. When the thresholds
are shared across units, weighted conformal selection controls the FDR exactly and is the safe choice (end of this
section).

\begin{theorem}[Envelope]\label{thm:envelope}
$\FDR\le \frac qm\sum_j\E\,\Gamma^*_j\le\frac qm\sum_j\E\,\Gamma^{\sup}_j\le\frac qm\sum_j\E\,\tilde\Gamma_j$, where
$\tilde\Gamma_j = R^0(p_{-j}-\bar\tau_j)/R^0(p_{-j}+\bar\tau_j)$ ($R^0$: BH count with an appended zero) and
$\bar\tau_j\le B/(n+1)$; $\tilde\Gamma_j$ and $\Gamma^{\sup}_j$ are computable from data.
\end{theorem}

Each $\tilde\Gamma_j$ is at least $1$, and it equals $1$ unless shifting the other p-values by $\pm\bar\tau_j$ changes the
BH count. So the envelope is close to $q$ when few p-values lie within $B/(n+1)$ of a BH threshold, and it flags the data
sets on which the weights could matter.

\begin{figure}[t]
\centering
\includegraphics[width=0.48\linewidth]{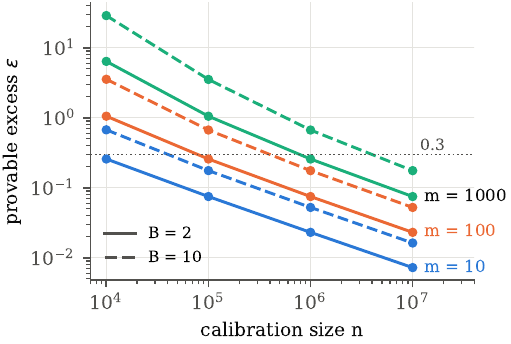}
\caption{Provable excess $\varepsilon$ (Theorem~\ref{thm:upper}, integral form $J$) against $n$, $q=0.1$, for $m=10,100,1000$ and
weight ratios $B=2$ (solid) and $B=10$ (dashed). BH at the level $q'$ solving $q'(1+\varepsilon(q'))=q$ controls the
FDR at $q$.}
\label{fig:deflation}
\end{figure}

\paragraph{Provable level deflation (conservative).} By Theorem~\ref{thm:upper}, BH at level $q'$ with $q'(1+\varepsilon(q'))\le q$
controls the FDR at $q$ in finite samples, where $\varepsilon$ uses only $n$, $m$, $q$ and a weight bound $B$
(Figure~\ref{fig:deflation}). At $q=0.1$, $\varepsilon$ is at most $0.3$ once $n/B\gtrsim3.8\cdot10^4$ ($m=100$) or
$3.8\cdot10^5$ ($m=1000$). The level has to be solved from this inequality rather than set to $q/(1+\varepsilon(q))$,
because $\varepsilon$ grows as the level falls: at $n=10^5$, $m=100$, $B=2$ and $q=0.1$ the solution is $q'=0.0770$,
while $q/(1+\varepsilon(q))=0.0795$ would only guarantee $\FDR\le0.1028$.

\paragraph{Rule of thumb (numerical, not a guarantee).} Across every configuration our searches found (909 optimised
Poisson-limit configurations over selection with decoys, strict selection, outlier detection, non-null signal families and
global-null families, $q\in\{0.05,0.1,0.2\}$, $m$ from $2$ to $1024$; plus the bottom-placed and stacked structures of
Section~\ref{sec:sqrt}), with $\kappa=B/n\le0.1$,
\[
\frac{\FDR}{q}-1\;\le\;0.17\sqrt{\frac{Bm}{nq}} ,
\]
and at any $\kappa$ the FDR never exceeded $1.52\,q$. The scaled worst case falls with $m$ and then levels off (0.167 at
$m=2$, 0.115 at $m=8$, 0.057 at $m=32$, and close to $0.054$ for $m$ from $64$ to $256$, from the structures of
Section~\ref{sec:sqrt}). Used as a deflation, it suggests running BH at the $q'$ that solves
$q'(1+0.17\sqrt{Bm/(nq')})=q$. This is a heuristic: we did not rerun the worst cases at lowered levels. Its constant is
about ten times smaller than the proved one (Table~\ref{tab:rules}).

\begin{table}[h]
\caption{Bounds on the relative excess $\FDR/q-1$ when BH is run at level $q$. The proved values are for $q=0.1$ (they depend on $(m,B,n)$ almost only through $x$). The rule of thumb covers every worst case found, but carries
no guarantee.}
\label{tab:rules}
\centering
\begin{tabular}{@{}lcccc@{}}\toprule
$x = Bm/(nq)$ & $0.002$ & $0.02$ & $0.1$ & $1$ \\\midrule
Proved bound $\varepsilon$ (Theorem~\ref{thm:upper}, form $J$) & 0.075 & 0.258 & 0.669 & 3.55 \\
Rule of thumb $0.17\sqrt{x}$ (numerical) & 0.0076 & 0.024 & 0.054 & 0.17 \\\bottomrule
\end{tabular}
\end{table}

\paragraph{Shared thresholds in selection.} Use weighted conformal selection with pruning, which has exact finite-sample
control \citep[Thm.~3.1]{jin2026weighted}.

\section{Discussion and open problems}\label{sec:open}

BH on weighted conformal p-values does not control the FDR in finite samples. When non-null units can share a threshold
the worst excess is of order $\sqrt{Bm/(nq)}$ (the upper bound at every $n$ with $B\le n$, the lower bound in the
Poisson-limit model); in the global null it is $O(B/n)$ at fixed $m$, with a lower bound that grows linearly in $m$. In practice the excess is small when $Bm/(nq)$ is small, and a deflated level restores provable
control. Several questions remain.

\paragraph{The constants.} The upper and lower constants of Theorem~\ref{thm:R} are $1.612$ and $c(q)$, which is $0.054$
at $q=0.1$. Of the factor of about $30$ between them, at most about $3.1$ is lost in evaluating the reduction $\FDR\le q\,\E[M_D]$
(about $1.64$ from the Chernoff tails and at most about $1.9$ from the union over grid points). The rest separates the
reduction from the best constructions we found: numerically, the reduction's own constant is at least about $9.6$ times
$c(q)$. So either the true constant is well above $c(q)$, or closing the gap needs an upper bound that does not go through
$\E[M_D]$ (Proposition~\ref{prop:E9}).

\paragraph{Finite $n$.} Theorem~\ref{thm:F} gives no explicit $n_0$, and the convergence of the finite-$n$ FDR to its
Poisson-limit value is proved only for the global null (Lemma~\ref{lem:L2}), not for the non-null structures of
Section~\ref{sec:sqrt}. An explicit $n_0$ would make Theorem~\ref{thm:R}(iii) hold at explicit $n$; that convergence,
ideally with a rate, would carry Theorem~\ref{thm:R}(ii) to large finite $n$ at fixed $\kappa$, as Lemma~\ref{lem:L2} does
for the global null.

\paragraph{The global null.} The positive-curvature condition (Appendix~\ref{app:global}) and
Conjecture~\ref{conj:C1} are open, and so is an $m$-free bound on $\beta_m(\rho)$, which would make the upper bound of
Theorem~\ref{thm:globalupper} linear in $m$.

\providecommand{\paperacknowledgments}{}
\paperacknowledgments

\bibliography{refs}
\bibliographystyle{tmlr}

\appendix
\providecommand{\clip}{\operatorname{clip}}
\providecommand{\sd}{\operatorname{sd}}
\providecommand{\id}{\mathrm{id}}

\section{Proofs for the exact identities and the fixed-distribution theorem}\label{app:finite}

This appendix proves Lemma~\ref{lem:U}, Lemma~\ref{lem:F1} and Theorem~\ref{thm:F}. It uses the weighted
exchangeability of Lemma~\ref{lem:A} and the leave-one-out identity of Lemma~\ref{lem:LOO}.

\subsection{Exact uniformity of the oracle p-value}\label{app:A-U}

Under (A1), the calibration pairs $(X_i,Y_i)_{i\le n}$ are i.i.d.\ $P$ and the test pair $(X_{n+1},Y_{n+1})$ has joint
law $Q$, with $dQ/dP=w(X)/\E_Pw$. Write $V_i:=V(X_i,Y_i)$, so that $V_{n+1}$ is the true score of the test point. The
randomised oracle p-value of \citet[eq.~(4)]{jin2026weighted} is
\begin{equation}\label{eq:A-pstar}
p^* \;=\; \frac{\sum_{i\le n}w(X_i)\one\{V_i<V_{n+1}\}+U\big(w(X_{n+1})+\sum_{i\le n}w(X_i)\one\{V_i=V_{n+1}\}\big)}
{\sum_{i\le n}w(X_i)+w(X_{n+1})},
\end{equation}
where $U\sim U(0,1)$ is independent. Lemma~\ref{lem:U} states that $\PP(p^*\le s)=s$ for every $s\in[0,1]$.

\begin{proof}[Proof of Lemma~\ref{lem:U}]
Condition on the multiset of the $n+1$ pairs $(X_i,Y_i)$, $i\le n+1$; Lemma~\ref{lem:A} covers exactly this
conditioning. By Lemma~\ref{lem:A}, given the multiset, the test point is pair $i$ with probability
$w(X_i)/\sum_{l\le n+1}w(X_l)$, and $U$ is uniform and independent of that choice. Lemma~\ref{lem:A} is self-contained and
handles ties; it is the covariate-shift case of the weighted exchangeability of
\citet[\S3.2, Lemma~2]{tibshirani2019conformal}.

Group the $n+1$ pairs by distinct score value $a$. Let $W_a$ be the total weight of group $a$, $W_{<a}$ the total weight
of the pairs with smaller scores, and $\Sigma_w:=\sum_{l\le n+1}w(X_l)$.
\begin{itemize}[leftmargin=*,itemsep=1pt]
\item The test point lands in group $a$ with probability $W_a/\Sigma_w$.
\item Given that, \eqref{eq:A-pstar} reads $p^*=(W_{<a}+UW_a)/\Sigma_w$, so $p^*$ is uniform on
$\big[W_{<a}/\Sigma_w,\,(W_{<a}+W_a)/\Sigma_w\big]$.
\end{itemize}
These intervals tile $[0,1]$, and each has length equal to its probability. So, given the multiset, $p^*$ is exactly
uniform on $[0,1]$, ties included, and hence $\PP(p^*\le s)=s$ for every $s\in[0,1]$.
\end{proof}

\subsection{The exact covariance identity}\label{app:A-F1}

Assume JC's independence: the $m$ test units are i.i.d.\ and independent of the calibration data $D$; and, given $D$,
the test p-values are i.i.d., each a function of its own unit, its own $U_j$ and $D$. Let $A_k(D)$, $r_k(D)$, $R_1$ and
$\varepsilon_k$ be as in Lemma~\ref{lem:F1}. We prove two statements; (b) is the identity of Lemma~\ref{lem:F1}.
\begin{enumerate}[label=(\alph*),leftmargin=*,itemsep=1pt]
\item At every $n$,
\begin{equation}\label{eq:A-F1a}
\FDR \;=\; m\sum_{k=1}^m\frac1k\,\E\big[A_k(D)\,r_k(D)\big].
\end{equation}
\item Suppose every null unit is an oracle null ($p=p^*$), and put
$\varepsilon_k:=\PP(\text{unit 1 non-null},\ p^*_1\le t_k)$. Then $\E A_k(D)=t_k-\varepsilon_k$ and
\begin{equation}\label{eq:A-F1b}
\FDR-q \;=\; m\sum_{k=1}^m\frac1k\Cov\big(A_k(D),r_k(D)\big)\;-\;m\sum_{k=1}^m\frac1k\,\varepsilon_k\,\E\,r_k(D).
\end{equation}
\end{enumerate}

\begin{proof}[Proof of Lemma~\ref{lem:F1}]
\emph{(a)} By the leave-one-out identity (Lemma~\ref{lem:LOO}), unit $j$ is rejected iff $p_j\le t_{R_j}$, and then
$R=R_j$; here $R_j$ is the number of BH rejections when $p_j$ is replaced by $0$. So
\[
\FDR=\sum_{j=1}^m\E\Big[\frac{\one\{j\text{ null}\}\,\one\{p_j\le t_{R_j}\}}{R_j}\Big].
\]
Given $D$, unit $j$ is independent of $p_{-j}$, and $R_j$ depends only on $p_{-j}$. So the conditional expectation of
$\one\{j\text{ null},\ p_j\le t_k\}$ given $(D,p_{-j})$ is $A_k(D)$. Exchangeability of the units then gives
\[
\FDR=m\,\E\big[A_{R_1}(D)/R_1\big]=m\sum_{k=1}^m\frac1k\,\E\big[A_k(D)\,\one\{R_1=k\}\big].
\]
Conditioning on $D$ gives \eqref{eq:A-F1a}.

\emph{(b)} Since every null unit is an oracle null, Lemma~\ref{lem:U} gives
\[
\E A_k(D)=\PP(p^*_1\le t_k)-\PP(\text{unit 1 non-null},\ p^*_1\le t_k)=t_k-\varepsilon_k .
\]
Write $\E[A_kr_k]=\E A_k\,\E r_k+\Cov(A_k,r_k)$ in \eqref{eq:A-F1a}. Since $t_k=qk/m$ and $R_1\in\{1,\dots,m\}$,
\[
m\sum_{k=1}^m\frac1k\,t_k\,\E r_k(D)=q\sum_{k=1}^m\E r_k(D)=q,
\]
which gives \eqref{eq:A-F1b}.
\end{proof}

\begin{example}[A two-state illustration of Lemma~\ref{lem:F1}]\label{ex:A-F1}
Take $m=2$ and $q=0.2$, so $t_1=0.1$ and $t_2=0.2$. Let $D$ take two states, $a$ and $b$, with probability $\tfrac12$ each.
Units are null with probability $\tfrac12$, with oracle p-values; the other half are decoys whose oracle p-value exceeds
$q$. The null mass below the thresholds is $A_1=0.10$, $A_2=0.25$ in state $a$ and $A_1=0.10$, $A_2=0.15$ in state $b$;
the means $0.10$ and $0.20$ are as Lemma~\ref{lem:U} requires. The decoys' reported p-value is $0.18$ in state $a$ (below
$t_2$) and $0.5$ in state $b$. With $m=2$, $R_1=2$ iff the other unit's p-value is at most $t_2$, so $r_2(a)=0.25+0.5=0.75$
and $r_2(b)=0.15$.
\begin{itemize}[leftmargin=*,itemsep=1pt]
\item Directly: the FDR given the state is $2[0.10\cdot0.25+0.25\cdot0.75/2]=0.2375$ in state $a$ and
$2[0.10\cdot0.85+0.15\cdot0.15/2]=0.1925$ in state $b$; the mean is $0.2150$.
\item By the identity: $\Cov(A_1,r_1)=0$ and $\Cov(A_2,r_2)=\tfrac12(0.25\cdot0.75+0.15\cdot0.15)-0.20\cdot0.45=0.015$, so
$\FDR-q=2\cdot\tfrac12\cdot0.015=0.015$.
\end{itemize}
In state $a$, more null mass sits below $q$ \emph{and} the decoys sit below $q$, so BH reaches $2$ rejections exactly when
the threshold is generous to the nulls. If the decoys' p-value did not depend on the state ($0.5$ in both), then
$r_2=A_2$, so $\Cov(A_2,r_2)=\Var(A_2)=0.0025$, while $\Cov(A_1,r_1)=0$ because $A_1$ is constant. This gives
$\FDR=q+0.0025$ (directly, $\tfrac12(0.2125+0.1925)=0.2025$): smaller, and of second order in the fluctuation.
\end{example}

\subsection{The fixed-distribution theorem}\label{app:A-F}

\paragraph{Setting $\mathcal S(B,\pi_H,g,\zeta,v_0;q,m,k_0)$.} Selection with thresholds $c=c(X)$; a unit is null iff
$Y\le c$. The covariate $X$ is either \emph{light}, with a latent $u\sim U(0,1)$ ($P$-probability $1-\pi_H$, weight
$w=1$), or the \emph{heavy atom} ($P$-probability $\pi_H$, weight $w=B$). $Y=u$ for light units and $Y=g$ for the heavy
atom, and the score is $V=Y$. The thresholds are:
\begin{itemize}[leftmargin=*,itemsep=1pt]
\item heavy: $c=g=Y$ (null);
\item light with $u\le1-\zeta$: $c=u=Y$ (null);
\item light with $u>1-\zeta$: $c=v_0<Y$ (non-null; the \emph{decoys}).
\end{itemize}
Write $\bar w:=\E_Pw=1-\pi_H+B\pi_H$; $\gamma:=(1-\pi_H)/\bar w$, the $Q$-density of light units in $u$;
$Q_H:=B\pi_H/\bar w$; and $Q_D:=\gamma\zeta$. The law $P$ does not depend on $n$.

\paragraph{Conditions.} Throughout, $0<\pi_H<1$, $B\ge1$, $m\ge2$ and $0<q<1$. (The formulas below divide by $\pi_H^2$,
and the bound $W\ge n$ used in the proof needs $B\ge1$.)
\begin{itemize}[leftmargin=3em,itemsep=1pt]
\item[(F1)] $0<g<v_0<1-\zeta<1$.
\item[(F2)] \emph{Zero offset:} $\bar p_D:=\big((1-\pi_H)v_0+B\pi_H\big)/\bar w=t_{k_0}$ for some $k_0\in\{1,\dots,m\}$.
\item[(F3)] \emph{Decoys above $q$:} $\big((1-\pi_H)(1-\zeta)+B\pi_H\big)/\bar w>q$. This also keeps every light level
$u_k$ (defined below) below $1-\zeta$.
\item[(F4)] \emph{Genericity:} no $t_k$ equals $b_H:=(1-\pi_H)g/\bar w$ or $T_H:=\big((1-\pi_H)g+B\pi_H\big)/\bar w$.
\end{itemize}
By (F1), $\bar p_D>T_H$.

\paragraph{Population quantities.} For $G=(G_1,\dots,G_m)\in[0,1]^m$, let $\Pi_k(G):=\PP(R_1=k)$ when the $m-1$ other
p-values are i.i.d.\ with CDF values $G$ on the grid $(t_1,\dots,t_m)$. The others' CDF on the grid when the decoys are
counted from $t_{k_0+1}$ is $\bar G^0_j:=t_j+Q_D\one\{j>k_0\}$; when they are counted from $t_{k_0}$ it is
$\bar G^1:=\bar G^0+Q_D\mathbf e_{k_0}$, where $\mathbf e_{k_0}$ is the $k_0$-th unit vector. The jump of $r_k$ at the
switch is
\[
\Delta r_k:=\Pi_k(\bar G^1)-\Pi_k(\bar G^0).
\]

\paragraph{Gaussian limit.} Let $e\sim\mathcal N\big(0,\pi_H(1-\pi_H)\big)$, and let $b$, independent of $e$, be a
Brownian bridge divided by $\sqrt{1-\pi_H}$:
\[
\Cov\big(b(x),b(y)\big)=\frac{\min(x,y)-xy}{1-\pi_H}.
\]
Define
\begin{equation}\label{eq:A-etaD}
\eta_D:=\frac{\big(B-v_0-t_{k_0}(B-1)\big)\,e+(1-\pi_H)\,b(v_0)}{\bar w},
\end{equation}
and define $\xi_k$ by the region of $t_k$:
\begin{itemize}[leftmargin=*,itemsep=2pt]
\item $t_k<b_H$ (light nulls below $g$): with $u_k:=t_k\bar w/(1-\pi_H)$,
\[
\xi_k:=\gamma\Big[\frac{\big(t_k(B-1)+u_k\big)\,e}{1-\pi_H}-b(u_k)\Big];
\]
\item $b_H<t_k<T_H$ (heavy window): with $N_k:=t_k\bar w-(1-\pi_H)g$,
\[
\xi_k:=\frac{Q_H\big[\big(t_k(B-1)+g\big)\,e-(1-\pi_H)\,b(g)\big]}{B\pi_H}-\frac{Q_H\,N_k\,e}{B\pi_H^2};
\]
\item $t_k>T_H$ (light nulls above $g$): with $u_k:=(t_k\bar w-B\pi_H)/(1-\pi_H)$,
\[
\xi_k:=\gamma\Big[\frac{\big(t_k(B-1)+u_k-B\big)\,e}{1-\pi_H}-b(u_k)\Big].
\]
\end{itemize}

\begin{theorem}[Precise form of Theorem~\ref{thm:F}]\label{thm:A-F}
In the setting $\mathcal S(B,\pi_H,g,\zeta,v_0;q,m,k_0)$, under (F1)--(F4), as $n\to\infty$,
\begin{equation}\label{eq:A-c0}
\sqrt n\,(\FDR_n-q)\;\to\;c_0:=-m\sum_{k=1}^m\frac1k\cdot\frac{\Delta r_k\,\Cov(\xi_k,\eta_D)}{\sqrt{2\pi}\,\sd(\eta_D)}.
\end{equation}
In particular, if $c_0>0$, then $\FDR_n\ge q+c_0/(2\sqrt n)$ for every $n\ge n_0$, for some $n_0$.
\end{theorem}

The threshold $n_0$ and the $o(n^{-1/2})$ term are not explicit: the proof uses the central limit theorem and a
uniform-integrability step. An explicit-rate version would need Berry--Esseen-type bounds for the multinomial counts and
for the empirical quantile; these are not given here.

\begin{proof}[Proof of Theorem~\ref{thm:F} (Theorem~\ref{thm:A-F})]

Let $n\to\infty$ with $P$ fixed. $C$, $c$, $c'$ denote positive constants that depend on $P$, $q$, $m$ but not on $n$.

\emph{Step 1 (identity).} Every null unit has $c=Y$, so it is an oracle null, and Lemma~\ref{lem:F1} gives
\begin{equation}\label{eq:A-step1}
\FDR_n-q=m\sum_{k=1}^m\frac1k\Cov(A_k,r_k)-m\sum_{k=1}^m\frac1k\,\varepsilon_k\,\E r_k.
\end{equation}
A decoy's oracle p-value is at least $S(1-\zeta)/(W+1)$, where $S(x)$ is the calibration weight below $x$ and $W$ the total
calibration weight. By (F3) and Hoeffding's inequality \citep{hoeffding1963probability}, $\varepsilon_k\le e^{-cn}$ for some $c>0$.

\emph{Step 2 (exact formulas given $D$).} Let $N$ be the number of heavy calibration points, $L=n-N$, $\hat G$ the
empirical CDF of the $L$ light scores, and $W=L+BN$. Eq.~(4) of \citet{jin2026weighted} gives:
\begin{itemize}[leftmargin=*,itemsep=1pt]
\item light null with latent $u$: $p=\big(L\hat G(u)+BN\one\{g<u\}+U\big)/(W+1)$;
\item heavy test unit: $p=\big(L\hat G(g)+U(BN+B)\big)/(W+B)$;
\item decoy: $p_D=\big(L\hat G(v_0)+BN+U\big)/(W+1)$. There are almost surely no ties at $v_0$.
\end{itemize}
So, with $\clip(x,0,1):=\min\{1,\max\{0,x\}\}$,
\[
A_k=\gamma\int_0^{1-\zeta}\clip\big(t_k(W+1)-L\hat G(u)-BN\one\{g<u\},0,1\big)\,du
+Q_H\,\clip\Big(\frac{t_k(W+B)-L\hat G(g)}{BN+B},0,1\Big).
\]
The decoys add $Q_D\,d_k$ to the others' CDF at $t_k$, where $d_k:=\clip\big(t_k(W+1)-L\hat G(v_0)-BN,0,1\big)$. So
$r_k=\Pi_k(A+Q_Dd)$, with $A=(A_1,\dots,A_m)$ and $d=(d_1,\dots,d_m)$, and $\Pi_k$ is a polynomial with bounded derivatives
on $[0,1]^m$.

\emph{Step 3 (good event).} Let $x_{(1)}<\dots<x_{(L)}$ be the light calibration scores in increasing order. Let $E_n$ be
the intersection of the following events:
\begin{itemize}[leftmargin=*,itemsep=1pt]
\item $|N/n-\pi_H|\le n^{-1/2}\log n$;
\item $L\ge n(1-\pi_H)/2$;
\item $\sup_x|\hat G(x)-x|\le n^{-1/2}\log n$;
\item the largest spacing between consecutive light scores, including the gaps at $0$ and $1$, is at most $\log^2n/n$.
\end{itemize}
By Hoeffding's inequality \citep{hoeffding1963probability}, and by the Dvoretzky--Kiefer--Wolfowitz inequality
\citep{dvoretzky1956asymptotic} with Massart's constant \citep{massart1990tight} applied conditionally on $N$, the first
three events fail with probability at most $Ce^{-c\log^2n}$.
For the spacing: a fixed interval of length $\log^2n/(2n)$ is empty with probability
$\big(1-\log^2n/(2n)\big)^L\le e^{-c'\log^2n}$. If some spacing exceeds $\log^2n/n$, then one of about $2n/\log^2n$ fixed such
intervals is empty. So $\PP(E_n^c)\le Cn\,e^{-c\log^2n}$, which is still super-polynomially small. On $E_n$, for
$n\ge n_1$:
\begin{enumerate}[label=(\roman*),leftmargin=*,itemsep=1pt]
\item $d_j=0$ for $j<k_0$ and $d_j=1$ for $j>k_0$, because $\bar p_D=t_{k_0}$ sits strictly between $t_{k_0-1}$ and
$t_{k_0+1}$;
\item each $t_k$ stays in its population region, by (F4);
\item $A_k$ is within $C\log^2n/n$ of the following expression. Below $g$ it is $\gamma x_{(j)}$. In the heavy region it is
$\gamma g+Q_H\big(t_kW-L\hat G(g)\big)/(BN)$. Above $g$ it is $\gamma x_{(j')}+Q_H$; here $\gamma x_{(j')}$ already includes
the light mass below $g$. The indices $j$ and $j'$ are the order-statistic indices fixed by $t_k$, $W$ and $N$. The exact
light part is $\gamma\big[x_{(j)}+f\,(x_{(j+1)}-x_{(j)})\big]$ for some $f\in[0,1]$, so the error includes one spacing:
the bound $C\log^2n/n$ absorbs one spacing, the $+1$ and $+B$ terms in the denominators, the $U$-smoothing, and the
integer parts.
\end{enumerate}
Step~4 does not use (iii). Step~5 needs only the sandwich $\gamma x_{(j)}\le(\text{light part})\le\gamma x_{(j+1)}$, and
that $|x_{(j)}-u_k|\le\sup|\hat G-\id|+|j/L-u_k|+1/L$ is sub-Gaussian at scale $n^{-1/2}$. Step~6 needs the error to be
$o(n^{-1/2})$, which $\log^2n/n$ is. Off $E_n$, every quantity lies in $[0,1]$.

\emph{Step 4 (smooth part plus switch).} Put
\[
\delta_D:=\frac{L\hat G(v_0)+BN}{W+1}-t_{k_0}.
\]
Then $d_{k_0}=\clip\big(-(W+1)\delta_D,0,1\big)$. Let $G^0:=A+Q_D\one\{\cdot>k_0\}$ and $G^1:=G^0+Q_D\mathbf e_{k_0}$; by
Step~3(i), $A+Q_Dd=G^0+d_{k_0}Q_D\mathbf e_{k_0}$ on $E_n$. On $E_n$:
\begin{itemize}[leftmargin=*,itemsep=1pt]
\item $\Pi_k$ is a polynomial in its $k_0$-th coordinate. So for $d\in[0,1]$,
\[
\Pi_k(G^0+d\,Q_D\mathbf e_{k_0})=(1-d)\,\Pi_k(G^0)+d\,\Pi_k(G^1)+O\big(d(1-d)\big).
\]
\item Since $|d_{k_0}-\one\{\delta_D\le0\}|\le\one\{|\delta_D|\le1/(W+1)\}$, this gives
\begin{equation}\label{eq:A-step4}
r_k=\Pi_k\big(G^0(A)\big)+\one\{\delta_D\le0\}\,\Delta_k(A)+O\big(\one\{|\delta_D|\le2/n\}\big),
\end{equation}
where $\Delta_k(A):=\Pi_k(G^1(A))-\Pi_k(G^0(A))$ is a polynomial in $A$ with $\Delta_k(\bar A)=\Delta r_k$, and
$\bar A:=(t_1,\dots,t_m)$.
\end{itemize}

\emph{Step 5 (covariances).} Write $\Delta A_k:=A_k-\E A_k$. Then $\E|\Delta A_k|^p\le C_pn^{-p/2}$ for $p\le4$: on $E_n$,
$A_k$ is Lipschitz in $N/n$, $\hat G(g)$ and one light order statistic, each with sub-Gaussian deviations at scale
$n^{-1/2}$; off $E_n$, $A_k$ is bounded. Also $|\E A-\bar A|\le e^{-cn}$ by Step~1. Hence:
\begin{itemize}[leftmargin=*,itemsep=1pt]
\item $\Cov\big(A_k,\Pi_k(G^0(A))\big)=\sum_i\partial_i\Pi_k\,\Cov(A_k,A_i)+O\big(\E\big[|\Delta A_k|\,|A-\bar A|^2\big]\big)=O(1/n)$,
with $\partial_i\Pi_k$ taken at $G^0(\bar A)$;
\item $\big|\Cov\big(A_k,\one\{\delta_D\le0\}(\Delta_k(A)-\Delta r_k)\big)\big|\le C\,\E\big[|\Delta A_k|\,|A-\bar A|\big]=O(1/n)$;
\item $\big|\Cov\big(A_k,O(\one\{|\delta_D|\le2/n\})\big)\big|\le C\,(\E\Delta A_k^2)^{1/2}\,\PP(|\delta_D|\le2/n)^{1/2}
=O(n^{-1/2})\cdot o(1)$. Here $\PP(|\sqrt n\,\delta_D|\le2/\sqrt n)\to0$, because $\sqrt n\,\delta_D$ converges to the
continuous law of $\eta_D$ (Step~6);
\item the contribution of $E_n^c$ is $O(n\,e^{-c\log^2n})$, still super-polynomially small.
\end{itemize}
So
\begin{equation}\label{eq:A-step5}
\Cov(A_k,r_k)=\Delta r_k\,\E\big[\Delta A_k\,\one\{\delta_D\le0\}\big]+o(n^{-1/2}).
\end{equation}

\emph{Step 6 (joint CLT).} We show
\begin{equation}\label{eq:A-jointCLT}
\big(\sqrt n\,\Delta A_1,\dots,\sqrt n\,\Delta A_m,\ \sqrt n\,\delta_D\big)\ \Rightarrow\ (\xi_1,\dots,\xi_m,\eta_D).
\end{equation}
Up to $O(\log^2n/n)$ on $E_n$, each coordinate is a smooth function of: $N/n$; $\hat G$ at the fixed points $g$ and $v_0$;
and at most one light order statistic $x_{(j)}$ per $k$, with $j/L\to u_k$. The vector formed by $N$ and the light counts in a
fixed finite partition is multinomial. Given $N$, the light process satisfies $\sqrt L(\hat G-\id)\to$ Brownian bridge at fixed
points, whatever the value of $N$. This gives the limits $e$ and $b$, and they are independent. Both statements are the
multivariate central limit theorem \citep[Ch.~2]{vandervaart1998asymptotic}: for the multinomial counts, and, at
finitely many fixed points, for the i.i.d.\ indicator vectors of the light scores, whose limit is the Brownian bridge at
those points \citep[Ch.~19]{vandervaart1998asymptotic}.

\emph{Order statistics.} The claim is joint, and $j/L$ is random through $N$.
\begin{itemize}[leftmargin=*,itemsep=1pt]
\item \emph{The index.} Below $g$, $j:=\lfloor t_k(W+1)\rfloor$; above $g$, $j:=\lfloor t_k(W+1)-BN\rfloor$. In both cases
$j/L=h_k(N/n)+O(1/n)$ for a smooth function $h_k$ with $h_k(\pi_H)=u_k$.
\item Call a value of $N$ \emph{admissible} if it satisfies $E_n$'s conditions on $N$, and condition on an admissible $N$.
Then $L$ and $j$ are fixed, and $\vartheta_n:=j/L$ is within $Cn^{-1/2}\log n$ of $u_k$.
\item Put $\mathcal X_n:=\sqrt n\,(x_{(j)}-\vartheta_n)$ and $\mathcal Y_n:=\sqrt n\,(\hat G(\vartheta_n)-\vartheta_n)$. For
each fixed $z$,
\[
\{\mathcal X_n\le z\}=\{L\hat G(\vartheta_n+z/\sqrt n)\ge j\}=\{\sqrt n\,(\hat G(\vartheta_n+z/\sqrt n)-\vartheta_n)\ge0\}.
\]
This holds for any sample, since $\hat G$ is right-continuous.
\item Given $N$, $\sqrt n\,[\hat G(\vartheta_n+z/\sqrt n)-\hat G(\vartheta_n)]$ has mean exactly $z$ and variance at most
$\sqrt n|z|/L=O(n^{-1/2})$. So, by Chebyshev's inequality, for every $\epsilon>0$ both
$\PP(\mathcal X_n\le z,\ \mathcal Y_n<-z-\epsilon\mid N)$ and $\PP(\mathcal X_n>z,\ \mathcal Y_n>-z+\epsilon\mid N)$
tend to $0$, uniformly over admissible $N$ and over $z$ in a compact range.
\item $\mathcal Y_n$ is tight uniformly in $N$: $\Var(\mathcal Y_n\mid N)\le n/(4L)\le1/(2(1-\pi_H))$. So for any $\alpha>0$
there is $\bar z$ with $\PP(|\mathcal Y_n|>\bar z\mid N)\le\alpha$. Take an $\epsilon$-mesh grid of $z$ in
$[-\bar z-2,\bar z+2]$, with $\epsilon\in(0,1]$.
  \begin{itemize}[leftmargin=*,itemsep=0pt]
  \item If $\mathcal X_n+\mathcal Y_n>2\epsilon$ and $|\mathcal Y_n|\le\bar z$, pick a grid point
  $z\in(-\mathcal Y_n+\epsilon,\,-\mathcal Y_n+2\epsilon]$. Then $\mathcal X_n>z$ and $\mathcal Y_n>-z+\epsilon$.
  \item If $\mathcal X_n+\mathcal Y_n<-2\epsilon$ (and $|\mathcal Y_n|\le\bar z$), pick
  $z\in[-\mathcal Y_n-2\epsilon,\,-\mathcal Y_n-\epsilon)$. Then $\mathcal X_n\le z$ and $\mathcal Y_n<-z-\epsilon$.
  \item A union bound over the finitely many grid points gives
  $\PP(|\mathcal X_n+\mathcal Y_n|>2\epsilon\mid N)\le\alpha+o(1)$, uniformly in admissible $N$.
  \end{itemize}
This is the lemma in \citet{ghosh1971new}'s proof of the Bahadur representation, written out.
\item Also, given $N$, $\sqrt n\,[(\hat G(\vartheta_n)-\vartheta_n)-(\hat G(u_k)-u_k)]$ has variance at most
$n|\vartheta_n-u_k|/L\to0$. Hence
\[
\sqrt n\,(x_{(j)}-\vartheta_n)=-\sqrt n\,(\hat G(u_k)-u_k)+o_P(1),
\]
where $u_k$ is a fixed point.
\item A remainder that is small uniformly given $N$, together with $\PP(N\text{ not admissible})\to0$, gives an
unconditional $o_P(1)$.
\item Finally, $\sqrt n\,(\vartheta_n-u_k)=h_k'(\pi_H)\sqrt n\,(N/n-\pi_H)+o(1)$, uniformly on $E_n$; at this scale it is
linear. Here $h_k'(\pi_H)=t_kB/(1-\pi_H)^2$ below $g$ and $(t_k-1)B/(1-\pi_H)^2$ above $g$. These match the coefficients
of $e$ in $\xi_k$.
\item So every coordinate of \eqref{eq:A-jointCLT} is, up to $o_P(1)$, a smooth function of $\sqrt n\,(N/n-\pi_H)$ and of
$\sqrt n\,(\hat G-\id)$ at the fixed points $g$, $v_0$ and $u_k$. The joint CLT for the multinomial counts of (heavy,
light in the fixed partition cells) gives joint convergence.
\end{itemize}
The delta method gives the linear forms $\xi_k$ and $\eta_D$ defined above. As a numerical check, the covariances
they imply agree with simulated covariances at $n=10^7$ on eight random parameter sets, within $1.5$ standard errors. For the delta method see \citet[Ch.~3]{vandervaart1998asymptotic}.

\emph{Step 7 (the kink).} $\sup_n\E[n\,\Delta A_k^2]<\infty$ gives uniform integrability. Also $\PP(\eta_D=0)=0$, since
$\sd(\eta_D)>0$: the $b(v_0)$ term alone has positive variance. So, by continuous mapping and convergence of expectations
under uniform integrability \citep[Ch.~2]{vandervaart1998asymptotic},
\[
\sqrt n\,\E\big[\Delta A_k\,\one\{\delta_D\le0\}\big]\ \to\ \E\big[\xi_k\,\one\{\eta_D\le0\}\big]
=-\frac{\Cov(\xi_k,\eta_D)\,\varphi(0)}{\sd(\eta_D)},\qquad\varphi(0)=\frac1{\sqrt{2\pi}}.
\]
The last equality holds for a centred Gaussian pair: write $\xi_k=\big(\Cov(\xi_k,\eta_D)/\Var(\eta_D)\big)\,\eta_D$ plus a
part independent of $\eta_D$.

Steps~1, 5 and 7 together give $\sqrt n\,(\FDR_n-q)\to c_0$; the bound for $n\ge n_0$ follows from the limit.
\end{proof}

\begin{remark}[Two points of care]\label{rem:A-care}
Step~4's $O(d(1-d))$ bound needs $\Pi_k$ to be a polynomial of bounded degree in one coordinate. That holds, with degree at
most $m-1$. Step~6 is the only place where the joint law matters. The covariance formulas for $\xi_k$ and $\eta_D$ come from
it, and the exact-arithmetic evaluations of $c_0$ use exactly those formulas.
\end{remark}

\begin{remark}[Two checks on $c_0$]\label{rem:A-checks}
\begin{enumerate}[label=(\roman*),leftmargin=*,itemsep=1pt]
\item Under (F2), $u_{k_0}=v_0$ ($t_{k_0}=\bar p_D>T_H$, so the third region applies), and therefore $\eta_D=-\xi_{k_0}$
coefficient by coefficient (also confirmed numerically, to $10^{-14}$). The $k_0$-term of $c_0$ is therefore
$m\,\Delta r_{k_0}\,\sd(\eta_D)/(k_0\sqrt{2\pi})$, which is $\ge0$, and $>0$ when $k_0\ge2$. Indeed, couple the $m-1$
draws under $\bar G^0$ and $\bar G^1$ so that each decoy draw sits in band $k_0+1$, respectively band $k_0$, and all other
draws agree. The counts at the grid points then agree except at index $k_0$, where they are larger under $\bar G^1$. So
the two BH counts with a forced zero, $R^0$ and $R^1$, agree unless $R^0<k_0$ and $R^1=k_0$, and
$\Delta r_{k_0}=\PP(R^0<k_0,\ R^1=k_0)\ge0$. For $k_0\ge2$ this is positive: if $k_0-1$ of the draws are decoys and the
other $m-k_0$ lie above $t_m$ (which has positive probability: $Q_D>0$ by (F1), and the mass above $t_m$ is positive by
(F3)), then $R^0=1$ and $R^1=k_0$. For $k_0=1$, $R^0=R^1$ always, so every $\Delta r_k$ is $0$ and $c_0=0$.
\item At $B=1$ (no weighting), $c_0=0$ in every configuration checked (60 configurations; numerical). This is consistent
with exact FDR control in the unweighted case \citep[Theorem~2.3]{jin2023selection}.
\end{enumerate}
\end{remark}

\begin{remark}[Without the tuning (F2)]\label{rem:A-tuning}
If $\bar p_D$ is not a grid point, the switch is frozen on $E_n$, and $\sqrt n\,(\FDR_n-q)\to0$. This is recorded with
only that one-line reason, as an outline rather than a separate proof. So a fixed distribution shows the $\sqrt n$ excess
only when its decoy threshold sits exactly at a switching point. The worst case over distributions with weight ratio at
most $B$ still has the $\sqrt{B/n}$ order, because the tuning is available to the adversary.
\end{remark}

\paragraph{The values for the parameters of Example~\ref{ex:J3}.} Example~\ref{ex:F-J3} evaluates \eqref{eq:A-c0} at
$q=0.1$, $m=8$, $k_0=8$, $g=0.018418$, $\pi_H=0.086271/B$ and $\zeta=0.91847$, with $v_0$ from (F2) ($v_0=0.021680$ at
$B=10$); this is the structure of Example~\ref{ex:J3}. For $B\in\{2,5,10,100\}$, (F1)--(F4) hold and
$-m\sum_k\Delta r_k\Cov(\xi_k,\eta_D)/k>0$ in exact rational arithmetic, with an independent enumeration of the BH count;
the values of $c_0$ themselves involve one square root in floating point (for example $c_0=0.18806$ at $B=10$, so
$\FDR_n\approx0.1+0.188/\sqrt n$).
\gap{We have not proved that $c_0/(q\sqrt B)$ tends to $0.6707$, the Poisson-limit coefficient of Example~\ref{ex:J3},
as $B\to\infty$. The evidence is numerical only: in floating point, $c_0/(q\sqrt B)$ is $0.6699$ at $B=10^3$ and $0.6707$
at $B=10^5$.}
Figure~\ref{fig:F} compares $\sqrt n\,(\FDR_n-q)$ at finite $n$ with $c_0$.

\section{Proofs for the calibration-conditional upper bound}\label{app:upper}

This appendix proves Proposition~\ref{prop:E8a}, Theorem~\ref{thm:upper} and Proposition~\ref{prop:E9}. It does not use
the envelope of Section~\ref{sec:practical}.

\paragraph{Setting.} We assume (A1) and (A3) of Section~\ref{sec:setting}, together with the independence assumption of
\citet{jin2026weighted}: the $m$ test tuples $(X_{n+j},Y_{n+j},c_{n+j},U_j)$ are mutually independent and independent of
the calibration data $D:=(Z_1,\dots,Z_n)$, $Z_i=(X_i,Y_i)$. In selection, $(X_{n+j},Y_{n+j})\sim Q$, and $c_{n+j}$ may be
any function of $X_{n+j}$ and of noise independent of everything else. In outlier detection, the null units (inliers)
have $(X,Y)\sim Q$ and the outliers have any law. The set of nulls may be fixed or random; if it is random, it must be
independent of $D$, and the tuples must be independent given the null set.

\paragraph{Notation.} $V_i := V(X_i,Y_i)$, $\hat V_{n+j} := V(X_{n+j},c_{n+j})$ and $w_i := w(X_i)$. For $v\in\R$ let
\[
W := \sum_{i\le n} w_i,\qquad S_<(v) := \sum_{i\le n} w_i\one\{V_i<v\},\qquad S_=(v) := \sum_{i\le n} w_i\one\{V_i=v\}.
\]
Eq.~(4) of \citet{jin2026weighted} reads $p_j = \big(S_<(\hat V_{n+j}) + U_j\,(w_{n+j}+S_=(\hat V_{n+j}))\big)/(W+w_{n+j})$.
The oracle p-value $p^*_j$ is the same expression with the true score $V_{n+j}$ in place of $\hat V_{n+j}$ and the same
$U_j$. For a test point $(X,Y)\sim Q$ let $F^*_D(t) := \PP(p^*_j\le t\mid D)$; it does not depend on $j$ or on the
thresholds $c$. Let $M_D := \max_{1\le k\le m}F^*_D(t_k)/t_k$, with $t_k = qk/m$, and
$\Bb := (\operatorname{ess\,sup} w)/\E_P[w]$; with $1\le w\le B$ we have $1\le\Bb\le B$. Write $\kappa := B/n$. We assume
$q\in(0,1)$ throughout.

\begin{remark}[The assumption $q<1$]\label{rem:B-q}
In the proofs of Proposition~\ref{prop:E8a}, Lemma~\ref{lem:B-explicit} and Theorem~\ref{thm:upper} no step needs $q<1$.
When $s_k = t_k(1+\kappa)\ge1$, the tail event $\{F^*_D(t_k)>s_k+d\}$ is empty because $F^*_D\le 1$. The assumption
$q\in(0,1)$ is kept only as a harmless standing assumption.
\end{remark}

\subsection{A one-sided Bernstein inequality}

\begin{lemma}[One-sided Bernstein]\label{lem:B1}
Let $\xi_1,\dots,\xi_n$ be independent with $\E\xi_i=0$ and $\xi_i\le M$ almost surely, for some $M>0$, and let
$\sum_i\Var\xi_i\le\sigma^2$. Then for every $\varrho>0$,
\[
\PP\Big(\sum_{i=1}^n\xi_i\ge\varrho\Big)\;\le\;\exp\Big(-\frac{\varrho^2}{2(\sigma^2+M\varrho/3)}\Big).
\]
\end{lemma}

\begin{proof}
If $\sigma^2=0$, every $\xi_i$ is $0$ almost surely and the claim is trivial, so assume $\sigma^2>0$. The condition $M>0$
is no restriction: with mean zero, $M<0$ is impossible and $M=0$ forces $\xi_i\equiv0$.
\begin{enumerate}[label=\textit{Step \arabic*.},leftmargin=*,itemsep=2pt]
\item For $\tau>0$ put $g_\tau(\vartheta) := (e^{\tau\vartheta}-1-\tau\vartheta)/\vartheta^2$, with
$g_\tau(0):=\tau^2/2$. Then
\[
g_\tau(\vartheta) = \tau^2\int_0^1(1-r)\,e^{\tau\vartheta r}\,dr
\]
(integrate by parts to check the identity), which is non-decreasing in $\vartheta$ because each integrand is. So for
$\vartheta\le M$, $e^{\tau\vartheta}\le 1+\tau\vartheta+\vartheta^2 g_\tau(M)$.
\item Let $\Psi(r) := e^r-1-r$, so that $g_\tau(M) = \Psi(\tau M)/M^2$. Taking expectations,
\[
\E e^{\tau\xi_i}\;\le\;1+\Var(\xi_i)\,\frac{\Psi(\tau M)}{M^2}\;\le\;\exp\Big(\Var(\xi_i)\,\frac{\Psi(\tau M)}{M^2}\Big).
\]
\item For $0\le r<3$,
$\Psi(r) = \sum_{j\ge2}r^j/j! \le (r^2/2)\sum_{j\ge0}(r/3)^j = r^2/(2(1-r/3))$, because $j!\ge 2\cdot3^{j-2}$.
\item Chernoff's bound (Markov's inequality applied to $e^{\tau\sum_i\xi_i}$, and independence) gives, for
$0<\tau<3/M$,
\[
\PP\Big(\sum_i\xi_i\ge\varrho\Big)\;\le\;\exp\Big(-\tau\varrho+\frac{\sigma^2\tau^2}{2(1-\tau M/3)}\Big).
\]
\item Take $\tau := \varrho/(\sigma^2+M\varrho/3)$. Then $\tau M<3$ and $1-\tau M/3 = \sigma^2/(\sigma^2+M\varrho/3)$. So
$\sigma^2\tau^2/(2(1-\tau M/3)) = \tau\varrho/2$, and the exponent is
$-\tau\varrho/2 = -\varrho^2/(2(\sigma^2+M\varrho/3))$.\qedhere
\end{enumerate}
\end{proof}

\subsection{Chernoff bounds for Poisson and binomial counts}

The proofs in Appendices~\ref{app:upper}--\ref{app:envelope} use the following standard tail bounds.

\begin{lemma}[Chernoff bounds]\label{lem:B-chernoff}
Let $X$ be Poisson with mean $\mu>0$, or a sum of independent Bernoulli variables with total mean $\mu>0$ (for example a
binomial variable). Let $h(u):=u\log u-u+1$ for $u>0$ and $h(0):=1$. Then
\[
\PP(X\ge x)\le e^{-\mu h(x/\mu)}\quad(x\ge\mu),\qquad \PP(X\le x)\le e^{-\mu h(x/\mu)}\quad(0\le x\le\mu).
\]
In particular, for every $c>0$, $\PP(|X-\mu|\ge c\mu)\le2e^{-\mu h_c}$ with
$h_c:=\min\{h(1+c),h(\max\{1-c,0\})\}>0$.
\end{lemma}

\begin{proof}
For real $\theta$, $\E e^{\theta X}\le\exp\big(\mu(e^\theta-1)\big)$: with equality for the Poisson law, and, for a sum of
Bernoulli($p_i$) variables, from $1+p_i(e^\theta-1)\le\exp(p_i(e^\theta-1))$. By Markov's inequality applied to
$e^{\theta X}$, $\PP(X\ge x)\le\exp\big(-\theta x+\mu(e^\theta-1)\big)$ for $\theta\ge0$, and
$\PP(X\le x)\le\exp\big(-\theta x+\mu(e^\theta-1)\big)$ for $\theta\le0$. For $x>0$ take $\theta=\log(x/\mu)$, which has
the required sign in each case; the exponent is then $-x\log(x/\mu)+x-\mu=-\mu h(x/\mu)$. For $x=0$, let
$\theta\to-\infty$: $\PP(X\le0)\le e^{-\mu}=e^{-\mu h(0)}$. For the last claim, $\{|X-\mu|\ge c\mu\}$ is contained in
$\{X\ge(1+c)\mu\}\cup\{X\le(1-c)\mu\}$, where the second event is empty if $c>1$, and $h>0$ on
$[0,\infty)\setminus\{1\}$ because $h$ is strictly convex with $h(1)=h'(1)=0$.
\end{proof}

\subsection{The reduction}

\begin{proof}[Proof of Proposition~\ref{prop:E8a}]
Under the setting above we show $\FDR\le q\,\E[M_D]$, and in outlier detection with $m_0$ nulls
$\FDR\le(m_0/m)\,q\,\E[M_D]$; with a random null set, $\FDR\le(\E[m_0]/m)\,q\,\E[M_D]$.
\begin{enumerate}[label=\textit{Step \arabic*.},leftmargin=*,itemsep=2pt]
\item By the leave-one-out identity (Lemma~\ref{lem:LOO}), unit $j$ is rejected iff $p_j\le qR_j/m$, where
$R_j := R^0(p_{-j})$ is the number of BH rejections when $p_j$ is replaced by $0$; on that event the number of
rejections is $R_j$. So
\[
\FDR = \sum_{j=1}^m\E\Big[\frac{\one\{p_j\le qR_j/m,\ j\text{ null}\}}{R_j}\Big].
\]
\item Given $D$, the p-value $p_j$ and the null indicator of unit $j$ are functions of unit $j$'s tuple, while $R_j$ is a
function of the other units. So they are conditionally independent given $D$, and
\[
\E\Big[\frac{\one\{p_j\le qR_j/m,\ j\text{ null}\}}{R_j}\,\Big|\,D\Big] = \E\Big[\frac{G_{D,j}(qR_j/m)}{R_j}\,\Big|\,D\Big],
\qquad G_{D,j}(t) := \PP(p_j\le t,\ j\text{ null}\mid D).
\]
\item On $\{j\text{ null}\}$ in selection, $Y_{n+j}\le c_{n+j}$, so $V:=V_{n+j}\le\hat V:=\hat V_{n+j}$ by (A3). Then
$p_j\ge p^*_j$, in two cases.
\begin{itemize}[leftmargin=*,itemsep=1pt]
\item If $V=\hat V$, then $p_j=p^*_j$.
\item If $V<\hat V$, the calibration points with $V_i=V$ are counted in $S_<(\hat V)$, so
\[
(W+w_{n+j})(p_j-p^*_j)\;\ge\;(1-U_j)\,S_=(V)+U_j\,S_=(\hat V)\;\ge\;0;
\]
the $U_j w_{n+j}$ terms cancel.
\end{itemize}
In outlier detection $p_j=p^*_j$ for nulls. So $G_{D,j}(t)\le F^*_D(t)$.
\item $R_j$ takes values $r\in\{1,\dots,m\}$, and
$F^*_D(t_r)/r = (q/m)\,F^*_D(t_r)/t_r\le(q/m)\,M_D$. Summing over $j$ gives $\FDR\le q\,\E[M_D]$. In outlier detection
only the $m_0$ nulls contribute; with a random null set, condition on it first.\qedhere
\end{enumerate}
\end{proof}

\subsection{Explicit forms of the bound}

\begin{lemma}[Explicit forms]\label{lem:B-explicit}
In the setting above:
\begin{enumerate}[label=(\roman*),leftmargin=*,itemsep=2pt]
\item For all $n,m,q,B$ and every score law (atoms allowed),
\[
\E[M_D]\;\le\;1+\kappa+2m\sqrt{\frac{\pi(1+\kappa)\Bb}{nq}}+12(1+\log m)\,\frac{m\Bb}{nq}.
\]
\item Let $\theta := \sqrt{5\Bb mL/(nq)}$ with $L := \log(0.8\,nm/(\Bb q))$. If $L\ge1$, $\theta\le1$ and $\kappa\le1$,
then $\E[M_D]\le1+\kappa+2\theta$.
\item If $\kappa\le1$, then
\[
\E[M_D]\;\le\;1+\kappa+7.4\sqrt{\frac{\Bb m}{nq}}+33\,\frac{\Bb m}{nq}.
\]
\end{enumerate}
\end{lemma}

With Proposition~\ref{prop:E8a}, (i) gives $\FDR\le q[1+O(m\sqrt{B/n})]$ for every configuration covered by the
assumptions of \citet{jin2026weighted}: non-null units, decoys, clustered thresholds and score atoms included. No
anti-concentration of the test p-values is needed. Bound (iii) is smaller than (i) for every $m\ge6$ (at $m=5$ only where
both bounds exceed $5$), and it has no logarithmic factor, unlike (ii).

\begin{proof}[Proof of Lemma~\ref{lem:B-explicit}(i)]
Fix $t=t_k$ and write $s:=t(1+\kappa)$.
\begin{enumerate}[label=\textit{Step \arabic*.},leftmargin=*,itemsep=2pt]
\item \emph{From $p^*$ to a weighted empirical CDF.} For $u\in[0,1]$ let
$\hat\Lambda(v,u) := (S_<(v)+u\,S_=(v))/W$. For a test point $(V,w,U)$,
\[
p^*\;\ge\;\frac{S_<(V)+U\,S_=(V)}{W+w} = \hat\Lambda(V,U)\,\frac{W}{W+w}.
\]
Since $w/W\le B/n=\kappa$, $p^*\le t$ implies $\hat\Lambda(V,U)\le s$.
\item \emph{Population version.} Let $\Lambda(v,u) := Q(V<v)+u\,Q(V=v)$. With $U$ uniform and independent,
$\Lambda(V,U)\sim\mathrm{Unif}(0,1)$ under $Q$ (the randomised probability integral transform). For completeness: fix
$s\in(0,1)$ and let $v_s:=\inf\{v:Q(V\le v)\ge s\}$, so that $Q(V<v_s)\le s\le Q(V\le v_s)$. Then $\Lambda(v,u)\le
Q(V\le v)<s$ for $v<v_s$, and, if $Q(V=v_s)>0$, $\Lambda(v_s,u)<s$ for $u<u_s:=(s-Q(V<v_s))/Q(V=v_s)$. So
$\PP(\Lambda(V,U)<s)\ge Q(V<v_s)+u_s\,Q(V=v_s)=s$ (and $\ge Q(V<v_s)=s$ if $Q(V=v_s)=0$). Hence $\Lambda(V,U)$ is
stochastically smaller than $\mathrm{Unif}(0,1)$; it also has mean $\PP(V'<V)+\tfrac12\PP(V'=V)=\tfrac12$ for an
independent copy $V'$, by symmetry, so it is $\mathrm{Unif}(0,1)$. Also
$\PP\big((V,U)<_{\mathrm{lex}}(v,u)\big) = \Lambda(v,u)$ in the lexicographic order. Both $\hat\Lambda$ and $\Lambda$ are
non-decreasing in that order, and as $(v,u)$ increases, $\Lambda$ takes every value in $[0,1)$.
\item \emph{Localising.} Fix $d>0$ with $\Lambda$-level $\ell := s+d<1$, and pick $(v^+,u^+)$ with
$\Lambda(v^+,u^+)=\ell$. If $\hat\Lambda(v^+,u^+)>s$, then every $(v,u)\ge_{\mathrm{lex}}(v^+,u^+)$ has $\hat\Lambda>s$, so
$\{\hat\Lambda(V,U)\le s\}\subset\{(V,U)<_{\mathrm{lex}}(v^+,u^+)\}$ and $F^*_D(t)\le\ell=s+d$. Hence
\[
\PP\big(F^*_D(t)>s+d\big)\;\le\;\PP\big(\hat\Lambda(v^+,u^+)\le s\big).
\]
If $s+d\ge1$, the probability on the left is $0$, since $F^*_D\le1$.
\item \emph{Bernstein.} Let $w'_i := w_i/\E_P[w]$ (so $\E_P w'=1$ and $w'\le\Bb$) and
$\chi_i := \one\{V_i<v^+\}+u^+\one\{V_i=v^+\}\in[0,1]$. Then $\E_P[w'\chi] = \Lambda(v^+,u^+) = \ell$, because
$dQ/dP = w/\E_P w$. The bad event $\{\hat\Lambda(v^+,u^+)\le s\}$ is $\{\sum_i w'_i(\chi_i-s)\le0\}$. Put
$\eta_i := w'_i(\chi_i-s)$, which has mean $d$, and $\tilde\eta_i := \eta_i-d$.
\begin{itemize}[leftmargin=*,itemsep=1pt]
\item $|\tilde\eta_i|\le\Bb$: $\eta_i\in[-\Bb s,\Bb(1-s)]$ and $d\le\Bb d$, so $\tilde\eta_i\ge-\Bb(s+d)\ge-\Bb$ and
$\tilde\eta_i\le\Bb(1-s)\le\Bb$.
\item $\E\tilde\eta_i^2\le\E[w'^2(\chi-s)^2]\le\Bb\,\E_P[w'(\chi-s)^2] = \Bb\,\E_Q[(\chi-s)^2]
= \Bb(\Var_Q\chi+d^2)\le\Bb(\ell(1-\ell)+d^2)$. (Under $Q$, $\chi\in[0,1]$ has mean $\ell$.)
\item Lemma~\ref{lem:B1}, applied to $\xi_i := -\tilde\eta_i$ with $M=\Bb$ (only the bound $\tilde\eta_i\ge-\Bb$ is used
here), $\sigma^2 := n\Bb(\ell+d^2)\ge\sum_i\Var\xi_i$ and $\varrho = nd$, gives
\[
\PP\Big(\sum_i\tilde\eta_i\le-nd\Big)\;\le\;\exp\Big(-\frac{nd^2}{2[\Bb(\ell+d^2)+\Bb d/3]}\Big).
\]
\item With $s\le1$ and $d\le1$: $\ell+d^2+d/3\le s+d+d+d/3\le s+3d$. So
\[
\PP\big(F^*_D(t)>s+d\big)\;\le\;\exp\Big(-\frac{nd^2}{2\Bb(s+3d)}\Big).
\]
\end{itemize}
\item \emph{Integrating.} $\E(F^*_D(t)-s)^+ = \int_0^1\PP(F^*_D(t)>s+d)\,\mathrm{d}d$. For $d\le s/3$ the exponent is at
least $nd^2/(4\Bb s)$, which contributes at most $\sqrt{\pi\Bb s/n}$ (a half-Gaussian integral). For $d\ge s/3$ it is at
least $nd/(12\Bb)$, which contributes at most $12\Bb/n$. So
\[
\E(F^*_D(t)-s)^+\;\le\;\sqrt{\pi\Bb s/n}+12\Bb/n.
\]
\item \emph{Summing over the grid.} With $s_k := t_k(1+\kappa)$,
$M_D\le(1+\kappa)+\sum_k(F^*_D(t_k)-s_k)^+/t_k$. Taking expectations, and using
$\sum_{k\le m}t_k^{-1/2}\le 2m/\sqrt q$ and $\sum_{k\le m}1/t_k\le(m/q)(1+\log m)$, gives (i).\qedhere
\end{enumerate}
\end{proof}

\begin{proof}[Proof of Lemma~\ref{lem:B-explicit}(ii)]
Take $d_k := \theta t_k$ in Step~4 of the proof of (i). Under the stated conditions $\ell = t_k(1+\kappa+\theta)\le3t_k$,
$d_k^2\le t_k$ and $d_k/3\le t_k/3$, so $\ell+d_k^2+d_k/3\le5t_k$ and
\[
\PP\big(F^*_D(t_k)>s_k+\theta t_k\big)\;\le\;\exp\big(-n\theta^2t_k/(10\Bb)\big) = e^{-c_\theta k},
\qquad c_\theta := \frac{n\theta^2q}{10\Bb m}.
\]
Now $M_D\le1+\kappa+\theta+\sum_k\one\{F^*_D(t_k)>s_k+\theta t_k\}/t_k$, and
\[
\E\sum_k\frac{\one\{F^*_D(t_k)>s_k+\theta t_k\}}{t_k}\;\le\;\frac mq\sum_k e^{-c_\theta k}
\;\le\;\frac mq\,\frac{e^{-c_\theta}}{1-e^{-c_\theta}}.
\]
If $e^{-c_\theta}\le q\theta/(2m)$ (which is $\le\frac12$), this is at most $\theta$. With $\theta^2 = 5\Bb mL/(nq)$ we get
$c_\theta = L/2$, and $e^{-L/2}\le q\theta/(2m)$ is equivalent to $L\ge\log(0.8\,nm/(\Bb q))-\log L$. So the choice
$L := \log(0.8\,nm/(\Bb q))$ works whenever $L\ge1$. Hence $\E[M_D]\le1+\kappa+2\theta$.
\end{proof}

\begin{proof}[Proof of Lemma~\ref{lem:B-explicit}(iii)]
With $\kappa\le1$ and $s_k := t_k(1+\kappa)$, let $M'_D := \max_k(F^*_D(t_k)-s_k)^+/t_k$, so that
$M_D\le1+\kappa+M'_D$.
\begin{enumerate}[label=\textit{Step \arabic*.},leftmargin=*,itemsep=2pt]
\item \emph{Tail at one grid point.} Step~4 of the proof of (i), with $d:=u\,t_k$, gives for every $u>0$ (the probability
is $0$ when $s_k+d\ge1$, so $d\le1$ and $s_k<1$ there)
\[
\PP\big(F^*_D(t_k)>s_k+u\,t_k\big)\;\le\;\exp\Big(-\frac{nu^2t_k^2}{2\Bb(s_k+3ut_k)}\Big).
\]
Since $s_k\le2t_k$, the exponent is at least $k\,a\,g(u)$, with $a := nq/(2\Bb m)$ and $g(u) := u^2/(2+3u)$.
\item \emph{Union over the grid.}
$\PP(M'_D>u)\le\sum_{k\ge1}e^{-k a g(u)} = e^{-ag(u)}/(1-e^{-ag(u)})$. The tails decay geometrically in $k$, because
relative fluctuations shrink like $k^{-1/2}$; this is why no $\log m$ appears.
\item \emph{Integrating.} $g$ is increasing. Let $u_0$ solve $a\,g(u_0)=\log2$. For $u\ge u_0$ the tail is at most
$2e^{-ag(u)}$. Also $g(u)\ge\min(u^2/4,u/6)$ (split at $u=2/3$). So
\[
\E[M'_D]\;\le\;u_0+2\int_0^\infty e^{-a\min(u^2/4,\,u/6)}\,du\;\le\;u_0+2\sqrt{\pi/a}+12/a,
\]
and $u_0\le\sqrt{4\log2/a}+6\log2/a\le1.666/\sqrt a+4.16/a$. Hence
\[
\E[M'_D]\;\le\;\frac{5.211}{\sqrt a}+\frac{16.16}{a}\;\le\;7.37\sqrt{\frac{\Bb m}{nq}}+32.4\,\frac{\Bb m}{nq}.\qedhere
\]
\end{enumerate}
\end{proof}

\begin{remark}[The bound does not vanish as $B\to1$]\label{rem:B-limit}
$\E[M_D]-1$ does not vanish as $B\to1$. Even with constant weights it is about $\sqrt{m/(nq)}$: in simulation,
$\E[M_D]=1.18$ ($m=10$) and $1.44$ ($m=50$) at $n=1000$, $q=0.1$, where BH controls the FDR at exactly $q$. So the bounds of
this appendix are informative only when $m\ll nq/\Bb$, and ``$O(m\sqrt{B/n})$'' hides a factor $1/\sqrt q$. For weights
near $1$ the envelope bounds (Theorem~\ref{thm:envelope}), which vanish at $B=1$, are better.
\end{remark}

\subsection{Tighter constants: proof of Theorem~\ref{thm:upper}}

We prove Theorem~\ref{thm:upper} in the following form. In the setting above let $\kappa=B/n\le1$, $x := \Bb m/(nq)$ and
$c := 2+(1+\kappa)q/3$ (so $c\le2+2q/3$). With $s_k := t_k(1+\kappa)$, define the per-point tail bounds
\[
T_k(u) := \exp\Big(-\frac{nu^2t_k^2}{2\Bb\,[\,s_k(1+ut_k/3)+ut_k(1+ut_k)\,]}\Big)\quad\text{if }ut_k<1-s_k,
\qquad T_k(u) := 0\ \text{ otherwise}.
\]
Then
\[
\E[M_D]\;\le\;1+\kappa+J,\qquad J := \int_0^\infty\min\Big(1,\sum_{k=1}^m T_k(u)\Big)\,du,
\]
and, in closed form,
\[
\E[M_D]\;\le\;1+\kappa+1.6120\sqrt{(1+\kappa)\,x}+2.7726\,c\,x.
\]
Together with Proposition~\ref{prop:E8a}, $\FDR/q-1\le\E[M_D]-1$, this is Theorem~\ref{thm:upper}.

\begin{proof}[Proof of Theorem~\ref{thm:upper}]
The proof changes Step~4 of the proof of Lemma~\ref{lem:B-explicit}(i) and the proof of Lemma~\ref{lem:B-explicit}(iii).
\begin{enumerate}[label=\textit{Step \arabic*.},leftmargin=*,itemsep=2pt]
\item \emph{One-sided Bernstein.} In Step~4 of the proof of Lemma~\ref{lem:B-explicit}(i), the bad event is
$\sum_i\eta_i\le0$, with $\eta_i = w'_i(\chi_i-s)$ and $\E\eta_i = d$.
\begin{itemize}[leftmargin=*,itemsep=1pt]
\item Put $\tilde\eta_i := \eta_i-d$, which has mean zero. Since $\eta_i\ge-w'_is\ge-\Bb s$, we have
$-\tilde\eta_i\le M := d+\Bb s$. Only this upper bound is needed; the two-sided $|\tilde\eta_i|\le\Bb$ used in
Lemma~\ref{lem:B-explicit}(i) is wasteful.
\item $\Var(\tilde\eta_i)\le\E\eta_i^2-d^2\le\Bb(\ell(1-\ell)+d^2)-d^2 =: \sigma_d^2$, as in
Lemma~\ref{lem:B-explicit}(i).
\item Lemma~\ref{lem:B1}, applied to $\xi_i := -\tilde\eta_i$ with $\sigma^2 = n\sigma_d^2$ and $\varrho = nd$
($\sigma_d^2>0$ whenever $0<\ell<1$), gives
\[
\PP\big(F^*_D(t)>s+d\big)\;\le\;\exp\Big(-\frac{nd^2}{2(\sigma_d^2+Md/3)}\Big).
\]
\item Also $\sigma_d^2+Md/3\le\Bb\,[\,s(1+d/3)+d(1+d)\,]$. The slack is exactly $\Bb\ell^2+\tfrac23d^2\ge0$, so no
assumption on $\Bb$ is needed.
\end{itemize}
\item \emph{Exact $s_k$.} Write $d=u\,t_k$ and $s_k = t_k(1+\kappa)$. The probability is $0$ once $s_k+d\ge1$, so
otherwise $ut_k<1$. Then $s_k(1+d/3)+d(1+d)\le t_k[(1+\kappa)+c\,u]$. The exponent is therefore at least $k\,a\,h(u)$, with
$a := nq/(2\Bb m)$ and $h(u) := u^2/((1+\kappa)+c\,u)$.
\item \emph{Union and integral.} With $M'_D := \max_k(F^*_D(t_k)-s_k)^+/t_k$ (so $M_D\le1+\kappa+M'_D$),
\[
\PP(M'_D>u)\;\le\;\min\Big(1,\sum_{k\ge1}e^{-kah(u)}\Big) = \varphi_0(ah(u)),\qquad
\varphi_0(r) := \min\big(1,1/(e^r-1)\big).
\]
So $\E[M'_D]\le\int_0^\infty\varphi_0(ah(u))\,du$. The same union bound with the sharper per-point tails $T_k(u)$ of
Step~1 (with $d=ut_k$) gives the computable form $\E[M'_D]\le J$.
\item \emph{Closed form.} Substitute $\nu = \sqrt{a\,h(u)}$. Since $h$ is increasing from $0$ to $\infty$, $u=u(\nu)$ is a
bijection, and solving the quadratic gives $u(\nu)\le\bar u(\nu) := \nu\sqrt{(1+\kappa)/a}+c\nu^2/a$.
\begin{itemize}[leftmargin=*,itemsep=1pt]
\item Write $\varphi_0(\nu^2) = \int_\nu^\infty\big(-\tfrac{d}{dr}\varphi_0(r^2)\big)\,dr$, with a non-negative
integrand. Then by Tonelli,
\begin{align*}
\int_0^\infty\varphi_0(\nu^2)\,u'(\nu)\,d\nu &= \int_0^\infty u(r)\Big(-\frac{d}{dr}\varphi_0(r^2)\Big)dr\\
&\le\int_0^\infty\bar u(r)\Big(-\frac{d}{dr}\varphi_0(r^2)\Big)dr = \int_0^\infty\varphi_0(\nu^2)\,\bar u'(\nu)\,d\nu,
\end{align*}
using $u(0)=\bar u(0)=0$. The left side is $\int_0^\infty\varphi_0(ah(u))\,du$.
\item The last integral is $\sqrt{(1+\kappa)/a}\cdot I+(c/a)\int_0^\infty\varphi_0(r)\,dr$, with
$I := \int_0^\infty\min\big(1,1/(e^{\nu^2}-1)\big)\,d\nu$.
\item $\int_0^\infty\varphi_0 = \ln2+\big[\ln(1-e^{-r})\big]_{\ln2}^\infty = 2\ln2$.
\item With $1/a = 2\Bb m/(nq) = 2x$, this is $\sqrt2\,I\sqrt{(1+\kappa)x}+4\ln2\cdot c\,x$, the closed form.\qedhere
\end{itemize}
\end{enumerate}
\end{proof}

\begin{remark}[Constants and variants]\label{rem:B-constants}
The constants are $\sqrt2\,I$ and $4\ln2$, with $I = 1.139773$ (numerical quadrature). The exact values are
$\sqrt2\,I = 1.611883$ and $4\ln2 = 2.772589$, so $1.6120$ and $2.7726$ are upward roundings. For $q=0.1$ and small $\kappa$
the closed form reads $1+\kappa+1.612\sqrt x+5.64x$, against $7.4\sqrt x+33x$ in Lemma~\ref{lem:B-explicit}(iii). For all
$\kappa\le1$ and $q\le1$ it is at most $1+\kappa+2.28\sqrt x+7.4x$. By Step~2, $T_k(u)\le e^{-kah(u)}$, so $J$ never exceeds
the closed form. Using $\Bb=B$ (or any upper bound on $\Bb$) is valid: every $T_k(u)$, and hence $J$, is increasing in
$\Bb$.
\end{remark}

\subsection{The reduction cannot do better: proof of Proposition~\ref{prop:E9}}

Proposition~\ref{prop:E9} is a lower bound for the reduction quantity $\E[M_D]$ of Proposition~\ref{prop:E8a}, not for the
FDR. It says how good any bound obtained through Proposition~\ref{prop:E8a} can possibly be.

\paragraph{Model.} We use the Poisson-limit calibration model (Section~\ref{sec:setting}, Lemma~\ref{lem:L2}):
\begin{itemize}[leftmargin=*,itemsep=1pt]
\item light units have weight $1$ and continuous scores, with latent $u\sim U(0,1)$;
\item one heavy atom has weight $\kappa n$ and $P$-mass $\lambda/n$, with $\lambda=\omega/\kappa$;
\item the heavy atom's score lies below every light score;
\item $\omega := t_1/(1-t_1)$. Then at $K=\omega-\kappa$ the heavy window, which includes the test point's own weight
$\kappa$, is exactly $[0,t_1]$; at $K=\omega$ it is $[0,t_1]$ up to $O(\kappa)$. (Numerically, among bottom-placed atoms
this $\omega$ is optimal to leading order.)
\end{itemize}

\paragraph{Precise statement.} Fix $q\in(0,1)$ and $m\ge2$. Then, as $\kappa\to0$,
\[
\E[M_D]-1 = C^{\mathrm{red}}_m\sqrt\kappa+O_m(\kappa),\qquad
C^{\mathrm{red}}_m := \sqrt{\frac{m}{2\pi q}}\,\Big[\big(1-\tfrac qm\big)^{3/2}-\frac{(1-q)\big(1-\tfrac qm\big)^{1/2}}{m}\Big].
\]
In the units $x := \Bb m/(nq)$ of Theorem~\ref{thm:upper}, with $\Bb = \kappa n/\E_P w$ and $\E_P w\to1+\omega$, this reads
\[
\E[M_D]-1 = \frac{1-1/m}{\sqrt{2\pi}}\,\sqrt x\,(1+o(1)).
\]
So for every $m$, the supremum of $\E[M_D]-1$ over laws with weight ratio $\Bb$ is at least
$((1-1/m)/\sqrt{2\pi})\sqrt{\Bb m/(nq)}$ in the Poisson limit. Any bound proved through Proposition~\ref{prop:E8a} therefore
has leading constant at least $0.399(1-1/m)$.

\paragraph{From finite $n$ to the Poisson limit.} For fixed $\kappa$, the finite-$n$ laws converge to the Poisson-limit law
(the coupling of Lemma~\ref{lem:L2}). $M_D\le m/q$ is bounded and is a continuous function of $(F^*_D(t_k))_k$, so
$\E_n[M_D]$ converges to its Poisson-limit value. This uses the convergence $F^*_D(t_k)\to F_N(t_k)$ in probability
established in Steps 1--5 of the proof of Lemma~\ref{lem:L2}; that lemma is stated for the FWER and for $g_h\in(0,1)$, but
those steps apply unchanged to an atom at $g=0$. The rest is the continuous mapping theorem and bounded convergence
\citep[Ch.~2]{vandervaart1998asymptotic}. A bound valid at every
finite $n$ therefore applies to the limit. This uses only the convergence in distribution that Lemma~\ref{lem:L2}
provides. 

\begin{proof}[Proof of Proposition~\ref{prop:E9}]
Let $K := \kappa N$, with $N\sim\mathrm{Poisson}(\lambda)$. In the Poisson limit, write $Q_H = \omega/(1+\omega)$ for the
heavy $Q$-mass and $Q_L = 1/(1+\omega)$ for the light $Q$-density, and write $F_K$ for $F^*_D$, which depends on the
calibration data through $K$.
\begin{enumerate}[label=\textit{Step \arabic*.},leftmargin=*,itemsep=2pt]
\item \emph{Conditional law.}
\begin{itemize}[leftmargin=*,itemsep=1pt]
\item A heavy test point carries weight $K+\kappa$, including its own, and nothing lies below it. So
$p^*\sim\mathrm{Unif}[0,h_K]$, with $h_K := (K+\kappa)/(1+K+\kappa)$.
\item A light test point with latent $u$ has all heavy weight below it, so $p^* = (K+u)/(1+K)$.
\item Hence, for $t\le q$,
\[
F_K(t) = Q_H\min(1,t/h_K)+Q_L\big(t(1+K)-K\big)^+.
\]
\item As a consistency check, $\E_N F_K(t) = t$ (the limit form of Lemma~\ref{lem:U}); this was checked numerically to
$10^{-15}$.
\end{itemize}
\item \emph{Near $K=\omega$.} Fix $\delta_0>0$ small, so that $K+\kappa\ge\omega/2$ and $h_K<t_2$ when $|K-\omega|\le\delta_0$
and $\kappa$ is small. Put $\rho_k(K) := F_K(t_k)/t_k-1$.
\begin{itemize}[leftmargin=*,itemsep=1pt]
\item For $k\ge2$: $t_k>h_K>K/(1+K)$, so $F_K(t_k) = Q_H+Q_L(t_k(1+K)-K)$. This gives $\rho_k(K) = -a_k(K-\omega)$ exactly,
with $a_k := Q_L(1-t_k)/t_k = (1-t_1)(1-t_k)/t_k$. (Note $1/(1+\omega) = 1-t_1$, and $\rho_k(\omega)=0$.)
\item For $k=1$ there are three cases.
\begin{itemize}[leftmargin=*,itemsep=1pt]
\item $K\le\omega-\kappa$: then $h_K\le t_1$, and $\rho_1 = -(1-t_1)^2(K-\omega)/t_1$ exactly.
\item $K\ge\omega$: then $\rho_1 = Q_H/h_K-1$. This equals $-\kappa/((1+\omega)(\omega+\kappa)) = O(\kappa)$ at
$K=\omega$, has derivative $-Q_H/(\omega+\kappa)^2 = -(1-t_1)^2/t_1+O(\kappa)$ there, and has a bounded second derivative
on the window.
\item On $[\omega-\kappa,\omega]$, $\rho_1$ interpolates between the two cases within $O(\kappa)$.
\end{itemize}
\item So with $a_1 := (1-t_1)^2/t_1$, $\rho_k(K) = -a_k(K-\omega)+e_k(K)$ with $|e_k(K)|\le C'(\kappa+(K-\omega)^2)$,
uniformly in $k$.
\end{itemize}
\item \emph{The maximum.} $a_1/a_k = k(1-t_1)/(1-t_k)\ge1$, and $a_k$ decreases in $k$. So
$\max_k\big(-a_k(K-\omega)\big) = a_1(\omega-K)^+-a_m(K-\omega)^+$. The maximum is $1$-Lipschitz in the sup norm, so on
$|K-\omega|\le\delta_0$,
\[
M_D-1 = a_1(\omega-K)^+-a_m(K-\omega)^++O\big(\kappa+(K-\omega)^2\big).
\]
\item \emph{Expectation.}
\begin{itemize}[leftmargin=*,itemsep=1pt]
\item Off the window, $M_D-1\in[-1,m/q]$, and $\PP(|K-\omega|>\delta_0) = \PP(|N-\lambda|>\delta_0/\kappa)\le2e^{-c_1/\kappa}$
for some $c_1>0$, by Lemma~\ref{lem:B-chernoff} with $c=\delta_0/\omega$.
\item $\E(K-\omega)=0$ and $\E(K-\omega)^2 = \kappa\omega$, so
\[
\E[M_D]-1 = (a_1-a_m)\,\E(K-\omega)^++O(\kappa).
\]
\item The exact Poisson identity $\E[(N-\lambda)\one\{N\ge j\}] = \lambda\,\PP(N=j-1)$, with $j = \lfloor\lambda\rfloor+1$
(as in the proof of Theorem~\ref{thm:J2}), gives $\E(K-\omega)^+ = \kappa\lambda\,\PP(N=\lfloor\lambda\rfloor) =
\omega\,\PP(N=\lfloor\lambda\rfloor)$.
\item By Stirling's formula, $j!=\sqrt{2\pi j}\,(j/e)^j(1+O(1/j))$. With $j=\lfloor\lambda\rfloor=\lambda-f$, $f\in[0,1)$,
this gives $\PP(N=j)=(2\pi j)^{-1/2}\exp\{j-\lambda+j\log(\lambda/j)\}(1+O(1/\lambda))$, and
$j\log(\lambda/j)=-j\log(1-f/\lambda)=f+O(1/\lambda)$. So $\PP(N=\lfloor\lambda\rfloor)=(2\pi\lambda)^{-1/2}(1+O(1/\lambda))$, and
$\E(K-\omega)^+ = \sqrt{\kappa\omega}/\sqrt{2\pi}+O(\kappa^{3/2})$.
\end{itemize}
\item \emph{Algebra.} With $\omega = t_1/(1-t_1)$ and $t_1 = q/m$,
\[
(a_1-a_m)\sqrt\omega = (1-t_1)^{3/2}t_1^{-1/2}-(1-q)(1-t_1)^{1/2}t_1^{1/2}/q,
\]
which, divided by $\sqrt{2\pi}$, gives $C^{\mathrm{red}}_m$.
\begin{itemize}[leftmargin=*,itemsep=1pt]
\item Units: $\kappa = (1+\omega)\,\Bb/n\cdot(1+o(1))$ and $1+\omega = (1-t_1)^{-1}$.
\item So $C^{\mathrm{red}}_m\sqrt\kappa = \sqrt{x/(2\pi)}\,\big[(1-t_1)-(1-q)/m\big] = \sqrt{x/(2\pi)}\,(1-1/m)$.\qedhere
\end{itemize}
\end{enumerate}
\end{proof}

\begin{remark}[Scope of Proposition~\ref{prop:E9}]\label{rem:B-E9scope}
This is a Poisson-limit statement. Heuristically, the $O_m(\kappa)$ remainder is small once $\kappa m/q\ll1$, since the
spread of $K$ is $\sqrt{\kappa\omega}$ (numerically, at $m=512$, $q=0.05$, $\kappa=10^{-6}$ the ratio to the asymptotic value is
already $0.997$). It is a lower bound only: the single bottom atom is \emph{not} the worst case for $\E[M_D]$. Several heavy
cells just above the bottom, stacked so that each fills one grid step, do better numerically: at $m=8$, two cells give
$0.412$--$0.422$ against $0.349$ here; at $m=32$, four cells give $0.518\pm0.0005$ (Poisson Monte Carlo) against $0.387$, and the
value still rises with more cells. Each cell puts its kink at a different grid point, so $M_D$ becomes the maximum of a
random walk rather than one positive part. So the true supremum of the reduction's coefficient is unknown: it is at least
$(1-1/m)/\sqrt{2\pi}$ (proved here) and, numerically, at least $0.518$ at $m=32$.
\end{remark}

\providecommand{\PhiDet}{\Phi_{\mathrm{det}}}
\providecommand{\Pois}{\mathrm{Pois}}
\providecommand{\kcut}{k_{\mathrm{cut}}}

\section{Proofs for the $\sqrt{\kappa}$ regime in the Poisson limit}\label{app:sqrt}

All results in this appendix are statements about the Poisson-limit model: the heavy calibration count is exactly Poisson,
and the light calibration scores have exactly the uniform CDF. The expansions are as $\kappa\to0$ within that model. That
the finite-$n$ FDR converges to its Poisson-limit value as $n\to\infty$ at fixed $\kappa=B/n$ is proved for the global
null (Lemma~\ref{lem:L2}), not for the decoy structures of this appendix; Theorem~\ref{thm:F} gives finite-$n$ support. Numerical evaluations of the formulas below (the
configuration of Example~\ref{ex:J3}, finite-$n$ simulations, and optimised structures switching at $t_2$) are not
reproduced here.

\subsection{Setting}\label{sec:C-setting}

We work in the Poisson-limit selection model with one heavy cell.
\begin{itemize}[leftmargin=*,itemsep=1pt]
\item Light covariates have weight $1$ and latent score quantile $u\sim U(0,1)$, with $Y=y(X)$ deterministic and score
$V=Y$.
\item Light units with $u\le1-\zeta$ get oracle thresholds $c=Y$. They are null, and their p-value is $f_K(u)$.
\item Light units with $u>1-\zeta$ are \emph{decoys}: their threshold quantile is $v_0<u$, so they are non-null, and all of
them get the same p-value $p_D=f_K(v_0)$.
\item One heavy cell at score quantile $g<v_0$: weight $\kappa n$, $P$-mass $\lambda/n$ with $\lambda=\omega/\kappa$, $Y=g$
and $c=Y$ (null). Its calibration total is $K=\kappa N$ with $N\sim\mathrm{Poisson}(\omega/\kappa)$, so
$K=\omega+\sqrt{\kappa\omega}\,Z+O(\kappa)$ in law, $Z$ standard normal.
\item $f_K(u):=(u+K\one\{g<u\})/(1+K)$.
\end{itemize}
Under $Q$ the light units have total mass $1/\Xi$ and the heavy cell has mass $\omega/\Xi$, where $\Xi:=1+\omega$. In the
deterministic-$K$ quantities below, a heavy test unit's p-value is spread uniformly over the \emph{heavy window}
$[g/(1+K),(g+K)/(1+K)]$, the gap of the light map $u\mapsto f_K(u)$ at $u=g$; the exact window at $K$ is this window at
$K+\kappa$ (Step~1 of the proof of Theorem~\ref{thm:J2}).

Given $K$, the test units are i.i.d.\ (of three types: light null, decoy, heavy), and the decoys' p-value is the
deterministic atom $p_D=f_K(v_0)$. Let $k_0\in\{1,\dots,m\}$ be the grid index that the decoy atom crosses at the switch
point $k^*$ defined below. For $s\in\{-,+\}$ define the deterministic-$K$ FDR on each side of the switch,
\[
\PhiDet(K,s) \;:=\; m\sum_{k=1}^m\frac1k\,A_k(K)\,\PP_{K,s}(R_1=k),
\]
where
\begin{itemize}[leftmargin=*,itemsep=1pt]
\item $A_k(K):=\PP(\text{null},\ p\le t_k\mid K)$ (the Poisson-limit counterpart of $A_k(D)$);
\item $R_1$ is the BH count with a forced zero among $m-1$ i.i.d.\ units, as in Lemma~\ref{lem:F1};
\item under $\PP_{K,s}$ the decoy atom is counted in the BH counts at $t_k$ for $k\ge k_0$ ($s=-$), or for $k\ge k_0+1$
($s=+$; for $k_0=m$ it is then not counted at all).
\end{itemize}
We write $\PP_s:=\PP_{\omega,s}$ and $\E_s$ for the corresponding expectation, and $\varphi$ for the standard normal
density.

\subsection{Saturation at the mean}\label{sec:C-J1}

\begin{proof}[Proof of Lemma~\ref{lem:J1}]
Assume $f_\omega(1-\zeta)\ge q$, so all decoys' oracle p-values lie above $q$. We show first that $A_k(\omega)=t_k$ for
every $k\le m$.
\begin{itemize}[leftmargin=*,itemsep=1pt]
\item At $K=\omega$ the full oracle law (all light $u$, plus the heavy cell) is exactly uniform on $[0,1]$. This is the
deterministic-$K$ version of exact uniformity (Lemma~\ref{lem:U}), as in the global-null multi-cell expansion of the
Poisson-limit model at $\Delta K=0$. For the structure of Theorem~\ref{thm:C-J2prime} the computation is written out in Step~1
of its proof.
\item Below $q$ only null units contribute.
\end{itemize}
So $A_k(\omega)=\PP(p^*\le t_k\mid K=\omega)=t_k$ for $k\le m$. Given $K$ the units are i.i.d., so the leave-one-out
identity (Lemma~\ref{lem:LOO}) applies conditionally on $K$ and gives, for both $s$,
\[
\PhiDet(\omega,s)\;=\;m\sum_{k=1}^m\frac1k\cdot\frac{qk}{m}\cdot\PP_s(R_1=k)\;=\;q\sum_{k=1}^m\PP_s(R_1=k)\;=\;q.
\qedhere
\]
\end{proof}

\subsection{The $\sqrt\kappa$ expansion}\label{sec:C-J2}

\paragraph{Full statement of Theorem~\ref{thm:J2}.} Write $\FDR(\kappa)$ for the FDR of the Poisson-limit model at
$\kappa$. Assume:
\begin{enumerate}[label=(\alph*),leftmargin=*,itemsep=1pt]
\item\label{C-J2-k0} $1\le k_0\le m$ (Theorem~\ref{thm:C-J2prime} writes out the case $k_0=2$, $z_0=0$, with explicit
derivatives);
\item\label{C-J2-sat} Lemma~\ref{lem:J1}'s condition, strengthened to $f_\omega(1-\zeta)>q$; in particular
$f_\omega(1-\zeta)\ne t_k$ for every $k$. (With equality, $A_m$ has a kink at $K=\omega$, its one-sided slopes differ by
$(1-q)/(1+\omega)$, and $h_s$ below is undefined.)
\item\label{C-J2-gen} genericity at $K=\omega$ for the other breakpoints: no grid point is an edge of the heavy window or a
jump of the light map;
\item\label{C-J2-v0} $v_0>g$.
\end{enumerate}
Let the decoy threshold depend on $\kappa$, $v_0=v_0(\kappa)$, so that the switch point $k^*$ (the solution of
$f_{k^*}(v_0)=t_{k_0}$) satisfies $k^*-\omega=z_0\sqrt{\kappa\omega}$, with $z_0$ in a compact set. In particular $z_0=0$
is allowed: then $v_0=t_{k_0}(1+\omega)-\omega$, which does not depend on $\kappa$. No assumption on the position of $k^*$
relative to $K_w$ (the value of $K$ at which the heavy window's top edge crosses $t_{k_0}$) is needed: the proof uses only
$K_w\ne\omega$, which is part of \ref{C-J2-gen}. Then, as $\kappa\to0$, uniformly in $z_0$,
\begin{equation}\label{eq:C-J2}
\FDR(\kappa) \;=\; q+\sqrt{\kappa\omega}\,(h_+-h_-)\,\varphi(z_0)+O(\kappa),
\qquad h_s:=\partial_K\PhiDet(K,s)\big|_{K=\omega}.
\end{equation}
By Lemma~\ref{lem:J1} the terms containing $\partial_K\PP_{K,s}(R_1=k)$ cancel at $K=\omega$: since $A_k(\omega)=t_k$,
they add up to $m\sum_k\frac1k t_k\,\partial_K\PP_{K,s}(R_1=k)=q\,\partial_K\sum_k\PP_{K,s}(R_1=k)=0$. Hence
\[
h_s \;=\; m\sum_{k=1}^m\frac1k\,A_k'(\omega)\,\PP_s(R_1=k).
\]

\begin{proof}[Proof of Theorem~\ref{thm:J2}]
Since $v_0>g$, $p_D=f_K(v_0)=(v_0+K)/(1+K)$, which is increasing in $K$. Let $k^{**}<k^*$ be the value of $K$ at which
$p_D=t_{k_0-1}$ (if $k_0\ge2$ and this value is positive), and $K^\uparrow>k^*$ the value at which $p_D=t_{k_0+1}$ (if
$k_0<m$). Then the decoy atom is counted from $t_{k_0}$ on when $k^{**}<K\le k^*$, and from $t_{k_0+1}$ on (for
$k_0=m$: not at all) when $k^*<K\le K^\uparrow$. Both $\omega-k^{**}$ and $K^\uparrow-\omega$ are bounded below by positive
constants for small $\kappa$, because $k^*-\omega\to0$ while $k^*-k^{**}$ and $K^\uparrow-k^*$ stay bounded away from $0$.

\emph{Step 1 (conditioning on $K$).} Let $\FDR(K)$ denote the FDR given $K$. For $|K-\omega|\le\omega/2$ with
$k^{**}<K\le K^\uparrow$ (drop the first condition if $k^{**}$ is not defined or not positive, and the second if $k_0=m$),
\[
\FDR(K)=\PhiDet\big(K,\mathrm{side}(K)\big)+O(\kappa)\quad\text{uniformly},\qquad \mathrm{side}(K)=-\iff K\le k^*.
\]
(At $K=k^*$, $p_D=t_{k_0}$ is counted, because BH rejects $p\le t$.) The exact heavy window at $K$ is the limiting window at
$K+\kappa$, and the test masses agree exactly. The restriction $|K-\omega|\le\omega/2$ is needed: as $K\to0$ the heavy
window shrinks and the Lipschitz constant grows like $1/K$. The event $\{|K-\omega|>\omega/2\}$ has exponentially small
Poisson mass, by Lemma~\ref{lem:B-chernoff}. (A numerical check in the configuration of Example~\ref{ex:J3} gives
$\sup|\FDR(K)-\PhiDet(K,\mathrm{side}(K))|/\kappa=0.054$ over $\pm6$ standard deviations around $\omega$.)

The relation fails for $K<k^{**}$, where $k^{**}<k^*$ is the point at which the decoy atom drops into bin $k_0-1$; there
further switches follow. On that region $|\FDR(K)-\PhiDet(K,-)|\le C|K-\omega|$, and the event $\{K<k^{**}\}$ sits at
distance $\omega-k^{**}>0$ from $\omega$, so Step~4 absorbs it. For $k_0<m$ the relation also fails for $K>K^\uparrow$, where
the atom is counted only from $t_{k_0+2}$ on, or later. There $\FDR(K)$ and $\PhiDet(K,+)$ both lie in $[0,1]$ (each is an
FDR), and $\{K>K^\uparrow\}$ sits at distance $K^\uparrow-\omega>0$ from $\omega$, so Step~4 absorbs it in the same way.

\emph{Step 2 (smoothness).} Each $\PhiDet(\cdot,s)$ is smooth near $\omega$ under the remaining genericity
\ref{C-J2-sat}--\ref{C-J2-gen}. Since $\PhiDet$ is bounded and $\PhiDet(\omega,s)=q$ (Lemma~\ref{lem:J1}),
\[
\big|\PhiDet(K,s)-q-h_s(K-\omega)\big|\le C(K-\omega)^2\qquad\text{for all }K,
\]
with $C$ of order $1/\delta^2$, where $\delta$ is the distance from $\omega$ to the nearest breakpoint.

\emph{Step 3 (averaging; an exact identity).} $\E(K-\omega)=0$, and for $N\sim\mathrm{Poisson}(\lambda)$,
$\E[(N-\lambda)\one\{N\ge j\}]=\lambda\,\PP(N=j-1)$ (from $n\,\PP(N=n)=\lambda\,\PP(N=n-1)$). Taking
$j=\lfloor k^*/\kappa\rfloor+1$, so that $\{K>k^*\}=\{N\ge j\}$,
\[
\E\big[(K-\omega)\one\{K>k^*\}\big]=\omega\,\PP\big(N=\lfloor k^*/\kappa\rfloor\big)=\sqrt{\kappa\omega}\,\varphi(z_0)+O(\kappa),
\]
uniformly in $z_0$, by the local central limit theorem for the Poisson law:
$\sqrt\lambda\,\PP(N=j)=\varphi(z)+O(\lambda^{-1/2})$ uniformly for $z:=(j-\lambda)/\sqrt\lambda$ in a compact set. (By Stirling's
formula $\log j!=j\log j-j+\tfrac12\log(2\pi j)+O(1/j)$, so $\log\PP(N=j)=j-\lambda-j\log(j/\lambda)-\tfrac12\log(2\pi j)
+O(1/\lambda)$; with $j/\lambda=1+z/\sqrt\lambda$, $j\log(j/\lambda)=z\sqrt\lambda+\tfrac12z^2+O(\lambda^{-1/2})$ and
$j-\lambda=z\sqrt\lambda$, which gives $\log\PP(N=j)=-\tfrac12\log(2\pi\lambda)-\tfrac12z^2+O(\lambda^{-1/2})$.) Here
$(j-\lambda)/\sqrt\lambda=z_0+O(\lambda^{-1/2})$ and $\varphi$ is Lipschitz.

\emph{Step 4 (remainders).} $\E(K-\omega)^2=\kappa\omega$, and the far region $\{|K-\omega|>\delta'\}$ has exponentially
small mass once $\kappa\ll\delta'^2/\omega$ (the Poisson tail bound of Step~1). Combining Steps 1--4, and using
$\E(K-\omega)=0$ in the form $\E[(K-\omega)\one\{K\le k^*\}]=-\E[(K-\omega)\one\{K>k^*\}]$,
\begin{align*}
\FDR(\kappa)&=\E\,\PhiDet\big(K,\mathrm{side}(K)\big)+O(\kappa)
=q+h_-\E\big[(K-\omega)\one\{K\le k^*\}\big]+h_+\E\big[(K-\omega)\one\{K>k^*\}\big]+O(\kappa)\\
&=q+(h_+-h_-)\,\E\big[(K-\omega)\one\{K>k^*\}\big]+O(\kappa),
\end{align*}
and Step~3 gives \eqref{eq:C-J2}.
\end{proof}

\begin{remark}[Scope]\label{rem:C-J2-scope}
For $k_0<m$, side $+$ means that the decoy atom is counted from $t_{k_0+1}$ on, not that it is not counted; the upper switch
at $K^\uparrow$ ($p_D$ crossing $t_{k_0+1}$) is handled in Step~1 like $k^{**}$. Theorem~\ref{thm:C-J2prime} is the case
$k_0=2$, $z_0=0$, with the derivatives $A_k'(\omega)$ computed explicitly. For $k_0=1$ the theorem holds but is empty:
the two sides differ only in the count at $t_1$, which never changes $R_1$ ($k=1$ is always feasible), so $h_+=h_-$.
\end{remark}

\subsection{A switch at $t_2$}\label{sec:C-J2prime}

\begin{theorem}[Switch at $t_2$]\label{thm:C-J2prime}
Consider the Poisson-limit selection model with one heavy cell (score $g$, mean weight $\omega$) and one decoy block of
light units (fraction $\zeta$, threshold quantile $v_0$), and let $q<1$. Assume:
\begin{enumerate}[label=(\roman*),leftmargin=*,itemsep=1pt]
\item\label{C-J2p-window} $t_1(1+\omega)<g<t_2(1+\omega)-\omega$, so that the heavy window
$[g/(1+\omega),(g+\omega)/(1+\omega)]$ lies strictly inside $(t_1,t_2)$ at $K=\omega$. Writing $a:=\omega m/q$, this needs
$a<m/(m-q)$. The second inequality is $g<v_0$;
\item\label{C-J2p-v0} $v_0:=t_2(1+\omega)-\omega$ (so $v_0>g$, and the decoy p-value equals $t_2$ at $K=\omega$);
\item\label{C-J2p-sat} $1-\zeta>v_0$, and $f_\omega(1-\zeta)>q$.
\end{enumerate}
Then, as $\kappa\to0$,
\[
\frac{\FDR}{q}-1=\sqrt\kappa\,C_m+O(\kappa),\qquad C_m=\sqrt{\frac{\omega}{2\pi}}\,\frac{h_+-h_-}{q},\qquad
h_s=m\sum_{k=1}^m\frac1k\,A_k'(\omega)\,\PP_s(R_1=k),
\]
with $A_1'(\omega)=t_1/\Xi$ and $A_k'(\omega)=(t_k-1)/\Xi$ for $k\ge2$ ($\Xi=1+\omega$), and $\PP_s(R_1=k)$ the law of BH's
leave-one-out count when each other unit's p-value has CDF $t_k$ at grid point $k$, plus a decoy atom of mass $\zeta/\Xi$
counted from $k=2$ ($s=-$) or from $k=3$ ($s=+$).
\end{theorem}

\begin{proof}
We list only the differences from the proof of Theorem~\ref{thm:J2}.

\emph{Step 1 (saturation, as in Lemma~\ref{lem:J1}).} At $K=\omega$ the null p-values are exactly uniform on
$[0,1-\zeta/\Xi]$: light nulls below $g$ map to $[0,g/(1+\omega))$, the heavy test units fill
$[g/(1+\omega),(g+\omega)/(1+\omega)]$ with mass $\omega/\Xi$, and light nulls above $g$ map on with density $1$. Since
$1-\zeta/\Xi=f_\omega(1-\zeta)>q$, $A_k(\omega)=t_k$ for all $k$, so
$\PhiDet(\omega,s)=m\sum_k\frac1k\,t_k\,\PP_s(R_1=k)=q$ for both sides, and the $\partial\PP_s$ terms cancel in $h_s$
because $\sum_k\partial\PP_s(R_1=k)=0$.

\emph{Step 2 (derivatives).} Under assumption~\ref{C-J2p-window} no grid point is in the heavy window or the light-map
jump near $K=\omega$, so $A_k$ is affine in $K$ there. For $k=1$ only light nulls below $g$ count ($u\le t_1(1+K)$), which
gives $A_1'=t_1/\Xi$. For $k\ge2$ the heavy cell is fully counted (constant mass $\omega/\Xi$), and light nulls above $g$
satisfy $u\le t_k(1+K)-K$, which gives $A_k'=(t_k-1)/\Xi$.

\emph{Step 3 (switches).} $p_D(K)=(v_0+K)/(1+K)$ is increasing. Side $-$ (decoy counted from $k=2$) holds iff $K\le k^*$,
with $k^*=\omega$; side $+$ (counted from $k=3$) holds for $k^*<K<K_3$, where $p_D(K_3)=t_3$, i.e.\
$K_3-\omega=(t_3-t_2)(1+\omega)/(1-t_3)>0$. This upper switch is at a fixed distance from $\omega$, so it is handled
exactly like $k^{**}$ in the proof of Theorem~\ref{thm:J2} (a bounded difference times an exponentially small Poisson
tail). Below $k^*$, $p_D$ would cross $t_1$ only at $K-\omega=-(t_2-t_1)(1+\omega)/(1-t_1)<-\omega$, i.e.\ at $K<0$, so
this is never reached ($p_D(0)=v_0>t_1$).

\emph{Step 4.} Steps 1--4 of the proof of Theorem~\ref{thm:J2} then apply verbatim with $z_0=0$; since $\varphi(0)=1/\sqrt{2\pi}$, \eqref{eq:C-J2} divided by $q$ is the display.
\end{proof}

\begin{remark}\label{rem:C-J2p}
$C_m$ does not depend on $g$: neither $A_k'$ nor $\PP_s$ involves it. The constant in the $O(\kappa)$ remainder is of order
$1/\delta^2$, where $\delta$ is the distance in $K$ from $\omega$ to the nearest breakpoint (Step~2 of the proof of
Theorem~\ref{thm:J2}). So for each generic structure the expansion holds, and $C_m$ is the coefficient.
\end{remark}

\subsection{$\sqrt m$ growth with an explicit constant}\label{sec:C-J7}

\begin{theorem}[General form of Theorem~\ref{thm:J7}]\label{thm:C-J7gen}
Fix $q\in(0,1)$, $b>0$ and $a\in(0,1]$. For each $m$, take the structure of Theorem~\ref{thm:C-J2prime} with
$\omega=aq/m$ and $\zeta=b/m$ ($g$ generic, $v_0=t_2(1+\omega)-\omega$), and let $C_m$ be its coefficient:
$\FDR/q-1=C_m\sqrt\kappa+O(\kappa)$. Then
\[
C_m=\sqrt{\frac{a\,m}{2\pi q}}\;\Xi^{-1}\,\Delta_m(q,b),\qquad
\Delta_m:=\E_-\Big[\frac{\one\{R_1\ge2\}}{R_1}\Big]-\E_+\Big[\frac{\one\{R_1\ge2\}}{R_1}\Big],
\]
and $\Delta_m\to\Delta(q,b):=\E\big[\one\{R^-\ge2\}/R^-\big]-\E\big[\one\{R^+\ge2\}/R^+\big]$ as $m\to\infty$. Here
\[
R^s:=\max\{k\ge1:\ 1+\Pi(k)+D\one\{k\ge k_s\}\ge k\},
\]
with $\Pi$ a Poisson process of rate $q$ on the integers' scale ($\Pi(k)\sim\mathrm{Poisson}(qk)$), $D\sim\mathrm{Poisson}(b)$
independent of $\Pi$, $k_-=2$ and $k_+=3$. Hence $C_m/\sqrt m\to\sqrt{a/(2\pi q)}\,\Delta(q,b)$.
\end{theorem}

\begin{remark}[Scope; an iterated limit]\label{rem:C-J7-scope}
\begin{itemize}[leftmargin=*,itemsep=1pt]
\item The conditions of Theorem~\ref{thm:C-J2prime} ($1-\zeta>v_0$, $f_\omega(1-\zeta)>q$, $a<m/(m-q)$, genericity of
$g$) hold only for $m\ge m_0(q,b)$, with $m$ large relative to $b$.
\item $C_m$ is the $\kappa\to0$ coefficient at \emph{fixed} $m$. The constant of the $O(\kappa)$ remainder (of order
$1/\delta^2$, Remark~\ref{rem:C-J2p}) depends on $m$, so ``$C_m/\sqrt m\to\dots$'' is an iterated limit: first
$\kappa\to0$, then $m\to\infty$. It is not a joint limit.
\end{itemize}
\end{remark}

\begin{proof}[Proof of Theorem~\ref{thm:C-J7gen}]
\emph{Step 1 (exact identity).} By Theorem~\ref{thm:C-J2prime}, $h_s=m\sum_k\frac1k A_k'\PP_s(R_1=k)$ with
$A_1'=t_1/\Xi$ and $A_k'=(t_k-1)/\Xi$ for $k\ge2$. Writing $(t_k-1)/k=t_k/k-1/k$ and $t_k/k=q/m$,
\[
\frac{h_s}{m}=-\,\Xi^{-1}\,\E_s\Big[\frac{\one\{R_1\ge2\}}{R_1}\Big]+\frac{q}{m\,\Xi}.
\]
The last term is the same for both sides, so $(h_+-h_-)/m=\Xi^{-1}\Delta_m$, and $C_m=\sqrt{\omega/(2\pi)}\,(h_+-h_-)/q$
with $\omega=aq/m$ gives the display.

\emph{Step 2 (law of $R_1$).} The other $m-1$ p-values are i.i.d.\ with grid CDF $t_k+(\zeta/\Xi)\one\{k\ge k_s\}$, and
$R_1=\max\{k:1+N_m(k)\ge k\}$, where $N_m(k)$ counts those at or below $t_k$. Fix an integer cutoff $\kcut$. The increments
of $(N_m(1),\dots,N_m(\kcut))$ are multinomial with cell probabilities $q/m$ (and the decoy cell $b/(m\Xi)$), so they
converge to independent Poisson variables. (For fixed counts $c_1,\dots,c_r$ with total $c$ and cell probabilities
$p_{i,m}$ with $mp_{i,m}\to\mu_i$, the multinomial probability
$\frac{(m-1)!}{(m-1-c)!\,\prod_ic_i!}\prod_ip_{i,m}^{c_i}\big(1-\sum_ip_{i,m}\big)^{m-1-c}$ tends to
$\prod_ie^{-\mu_i}\mu_i^{c_i}/c_i!$, since $(m-1)!/(m-1-c)!\sim m^c$ and $(1-\sum_ip_{i,m})^{m-1-c}\to e^{-\sum_i\mu_i}$.)
\[
(N_m(k))_{k\le\kcut}\ \Rightarrow\ \big(\Pi(k)+D\one\{k\ge k_s\}\big)_{k\le\kcut}.
\]

\emph{Step 3 (uniform tail).} $N_m(k)$ is binomial with mean $\mu_k\le qk+b$. By the Chernoff bound
$\PP(X\ge x)\le\exp(-\mu\,h(x/\mu))$ for $x\ge\mu$, $h(u)=u\log u-u+1$ (Lemma~\ref{lem:B-chernoff}), applied with $x=k-1$,
\[
\PP\big(1+N_m(k)\ge k\big)\le\exp(-c_{q,b}\,k)\qquad\text{for all }k>k_{\min}(q,b)\text{ and all }m,
\]
with $c_{q,b}>0$ because $q<1$. Two conditions are used here: $\mu\,h(x/\mu)$ decreases in $\mu$ for $\mu<x$, so
$\mu_k$ may be replaced by $qk+b$; and this needs $k>(1+b)/(1-q)$, strictly (at $k=(1+b)/(1-q)$ the exponent is $0$). The
constant $c_{q,b}$ depends on $b$ as well as on $q$, and the exponent per $k$ tends to $\log(1/q)-1+q>0$. So
$\PP(R_1\ge\kcut)\le\sum_{k\ge\kcut}e^{-c_{q,b}k}$ uniformly in $m$, once $\kcut>k_{\min}$.

\emph{Step 4 (conclusion).} Steps 2 and 3 give $\PP_s(R_1=k)\to\PP(R^s=k)$ for every $k$ (on $\{R_1<\kcut\}$ the event
$\{R_1=k\}$ is a function of $(N_m(j))_{j\le\kcut}$), and tightness, so $R_1\Rightarrow R^s$. Since $\one\{R\ge2\}/R$ is
bounded, $\Delta_m\to\Delta$. Finally $\Xi=1+aq/m\to1$.
\end{proof}

\begin{lemma}[A ballot probability]\label{lem:C-ballot}
Let $\Pi$ be a Poisson process of rate $q\in(0,1)$. Then $\PP\big(\Pi(j)\le j-1\text{ for every integer }j\ge1\big)=1-q$.
\end{lemma}

\begin{proof}
Let $X_i:=\Pi(i)-\Pi(i-1)$, i.i.d.\ $\Pois(q)$, and $Y_0:=0$, $Y_j:=\sum_{i\le j}(X_i-1)$. The event is
$E:=\{Y_j<0\text{ for all }j\ge1\}$; write $p:=\PP(E)$. The steps $X_i-1$ are integers $\ge-1$.
\begin{enumerate}[label=\arabic*.,leftmargin=*,itemsep=1pt]
\item \emph{Weak records.} Call $j\ge0$ a weak record if $Y_j\ge Y_i$ for all $i<j$; $j=0$ is one. Each weak record is
  followed by a next one, namely the first later time at which $Y$ is at least its current value, with probability
  $\PP(\exists j\ge1:Y_j\ge0)=1-p$, independently of the past (strong Markov property). So the number $L$ of weak records
  has $\E L=\sum_{\ell\ge1}(1-p)^{\ell-1}=1/p$ (with $1/0:=\infty$).
\item \emph{Duality.} Let $\sigma:=\inf\{j\ge1:Y_j<0\}$. For $n\ge1$, reversing the order of the first $n$ steps preserves
  their joint law and turns $Y_j$ into $Y_n-Y_{n-j}$, so it maps $\{Y_1\ge0,\dots,Y_n\ge0\}$ onto $\{n\text{ is a weak
  record}\}$. Hence $\PP(\sigma>n)=\PP(n\text{ is a weak record})$ for all $n\ge0$, and
  $\E\sigma=\sum_{n\ge0}\PP(\sigma>n)=\E L=1/p$.
\item \emph{Wald.} The steps are at least $-1$, so $Y_\sigma=-1$ on $\{\sigma<\infty\}$, and $Y_j\ge0$ for $j<\sigma$. Since
  $Y_j+(1-q)j$ is a martingale, $(1-q)\E[\sigma\wedge n]=-\E Y_{\sigma\wedge n}\le1$, so $\E\sigma\le1/(1-q)<\infty$ by
  monotone convergence. Then $Y_\sigma=\sum_{i\ge1}(X_i-1)\one\{\sigma\ge i\}$ with $\{\sigma\ge i\}$ independent of $X_i$, and
  $\E\sum_i|X_i-1|\one\{\sigma\ge i\}=\E|X_1-1|\,\E\sigma<\infty$, so by Fubini (Wald's identity)
  $-1=\E Y_\sigma=(q-1)\,\E\sigma$, that is, $\E\sigma=1/(1-q)$.
\end{enumerate}
Hence $p=1/\E\sigma=1-q$.
\end{proof}

\begin{lemma}[Closed form and positivity]\label{lem:C-J7closed}
For every $m$, $\Delta_m=\tfrac12\big[\PP_-(R_1=2)-\PP_+(R_1=2)\big]>0$. Moreover
\[
\Delta(q,b)=\frac b2\,(1-q)\,e^{-(2q+b)}>0\qquad\text{for all }q\in(0,1),\ b>0 .
\]
\end{lemma}

\begin{proof}
\emph{Coupling.} The two sides differ only at $k=2$: for $k\ge3$ both count the decoys. Coupling the two sides on the same
draws, $R_-=R_+$ unless $R_+=1$ and $R_-=2$. On that event $\one\{R\ge2\}/R$ equals $\tfrac12$ on side $-$ and $0$ on side
$+$, so exactly, for every $m$, $\Delta_m=\tfrac12\big[\PP_-(R_1=2)-\PP_+(R_1=2)\big]>0$. By Step~4 of the proof of
Theorem~\ref{thm:C-J7gen}, letting $m\to\infty$ gives $\Delta(q,b)=\tfrac12\big[\PP(R^-=2)-\PP(R^+=2)\big]$.

\emph{Limit.} Write $c(k):=\Pi(k)+D\one\{k\ge k_s\}$, so that $R^s=\max\{k:1+c(k)\ge k\}$. If $c(2)\ge2$ then
$1+c(3)\ge3$ and $R^s\ge3$, while $k=2$ is feasible only if $c(2)\ge1$; so $R^s=2$ requires $c(2)=1$, and then $R^s=2$ holds iff $1+c(k)\le k-1$ for every $k\ge3$.
On side $-$ ($k_-=2$) this reads $\Pi'(j)\le j-1$ for every integer $j\ge1$, where $\Pi'(j):=\Pi(j+2)-\Pi(2)$ is again a
rate-$q$ Poisson process, independent of $(\Pi(2),D)$; by Lemma~\ref{lem:C-ballot} this has probability $1-q$.

Side $+$ ($k_+=3$) additionally needs $D=0$, else $1+c(3)\ge3$. Hence, with $\Pois(1;\mu):=\mu e^{-\mu}$,
\[
\PP(R^-=2)=(1-q)\,\Pois(1;2q+b),\qquad \PP(R^+=2)=(1-q)\,\Pois(1;2q)\,e^{-b},
\]
and
\[
\Delta(q,b)=\tfrac12(1-q)\big[(2q+b)e^{-(2q+b)}-2q\,e^{-2q}e^{-b}\big]=\frac b2\,(1-q)\,e^{-(2q+b)}>0.
\qedhere
\]
\end{proof}

As a numerical check of the closed form, an exact backward recursion of the limiting model
that does not use the closed form matches it to $2\cdot10^{-15}$ for $q\in\{0.05,0.1,0.2\}$ and $b\in\{0.5,1,2\}$, and
positivity was scanned for $q$ from $0.01$ to $0.9$.

\begin{proof}[Proof of Theorem~\ref{thm:J7}]
The structure of Theorem~\ref{thm:J7} ($\omega=q/m$, decoy fraction $1/m$) is the case $a=b=1$ of
Theorem~\ref{thm:C-J7gen}, whose conditions hold for $m\ge m_0(q,1)$ (Remark~\ref{rem:C-J7-scope}). By
Theorem~\ref{thm:C-J7gen}, $C_m/\sqrt{m/q}=\sqrt q\,C_m/\sqrt m\to\sqrt{a/(2\pi)}\,\Delta(q,b)$, in the iterated limit of
Remark~\ref{rem:C-J7-scope} ($\kappa\to0$ at fixed $m$, then $m\to\infty$). By Lemma~\ref{lem:C-J7closed},
$\Delta(q,b)=\frac b2(1-q)e^{-(2q+b)}$, which is maximised over $b$ at $b=1$. With $a=b=1$,
\[
\lim_{m\to\infty}\frac{C_m}{\sqrt{m/q}}=\frac{(1-q)\,e^{-1-2q}}{2\sqrt{2\pi}}=c(q),
\]
that is $0.0631$, $0.0541$ and $0.0394$ at $q=0.05$, $0.1$ and $0.2$ ($C_m/\sqrt m\to0.1710$ at $q=0.1$).
\end{proof}

\subsection{Outlier detection}\label{sec:C-J5}

\begin{proposition}[Outlier detection, asymptotic]\label{prop:C-J5}
Consider the Poisson-limit model of Theorem~\ref{thm:J2} in outlier detection:
\begin{itemize}[leftmargin=*,itemsep=1pt]
\item inliers are i.i.d.\ $Q$, are all null, and have oracle p-values with conditional CDF $F_K$ given $K$;
\item each test unit is independently an outlier with probability $\pi_1$ (the labels are i.i.d.\ Bernoulli, not a fixed
count; the leave-one-out identity needs this);
\item outliers are not in the calibration data, and their p-values $f_K(v_o)$, with $v_o>g$, form a switching block at
$t_{k_0}$, with
switch point $k^*-\omega=z_0\sqrt{\kappa\omega}$ as in Theorem~\ref{thm:J2}.
\end{itemize}
Then, under the genericity assumptions of Theorem~\ref{thm:J2}, and uniformly over $z_0$ in a compact set,
\[
\FDR(\kappa)\le(1-\pi_1)q+2(1-\pi_1)\pi_1\,m(m-1)\,M_F\sqrt{\kappa\omega}\,\varphi(z_0)+O(\kappa),\qquad
M_F:=\max_k\big|\partial_KF_K(t_k)|_{K=\omega}\big|.
\]
So
\[
\FDR(\kappa)\le q-\pi_1\Big[q-2(1-\pi_1)m(m-1)M_F\sqrt{\kappa\omega}/\sqrt{2\pi}\Big]+O(\kappa),
\]
and for $\kappa\omega<2\pi\big(q/(2(1-\pi_1)m(m-1)M_F)\big)^2$ the first-order ($\sqrt\kappa$) term cannot produce an
excess.
\end{proposition}

\begin{proof}
\emph{Zeroth order.} At $K=\omega$ the inlier law is exactly uniform. The leave-one-out identity (Lemma~\ref{lem:LOO}) with
null mass $1-\pi_1$, i.e.\ $A_k(\omega)=(1-\pi_1)t_k$, gives $\PhiDet(\omega,s)=(1-\pi_1)q$ for either $s$.

\emph{First order.} Here $A_k(K)=(1-\pi_1)F_K(t_k)$, so
$h_s=(1-\pi_1)\,m\sum_k\frac1k\,\partial_KF_K(t_k)\,\PP_s(R_1=k)$. The $\partial\PP$ terms cancel because
$A_k(\omega)\,m/k=(1-\pi_1)q$ is constant in $k$ and $\sum_k\PP_s(R_1=k)=1$.

\emph{Total variation.} By coupling (the two sides differ only on the event that some other unit is an outlier),
\[
\sum_k\big|\PP_+(R_1=k)-\PP_-(R_1=k)\big|\le2\big[1-(1-\pi_1)^{m-1}\big]\le2(m-1)\pi_1.
\]
Bounding $1/k$ by $1$ then gives the crude bound $|h_+-h_-|\le2(1-\pi_1)\pi_1\,m(m-1)\,M_F$.

\emph{Averaging.} This is Step~3 of the proof of Theorem~\ref{thm:J2}, within that proof's framework:
$\FDR(\kappa)=(1-\pi_1)q+\sqrt{\kappa\omega}\,(h_+-h_-)\,\varphi(z_0)+O(\kappa)$, and the bound on $|h_+-h_-|$ gives the
first display. The second uses $\varphi(z_0)\le1/\sqrt{2\pi}$. For $k_0<m$ the upper switch at $t_{k_0+1}$ is handled as in
Step~1 of the proof of Theorem~\ref{thm:J2}.
\end{proof}

\begin{remark}[What Proposition~\ref{prop:C-J5} does and does not say]
It is an asymptotic statement only. With the heavy structure of Example~\ref{ex:J3} ($\omega=0.0863$) and
$M_F=0.832$ (the exact supremum over $t$ of $|\partial_KF_K(t)|$, computed numerically), and bounding $1-\pi_1$ by $1$, the
condition reads $\kappa<2.7\cdot10^{-7}$ at $q=0.1$, $m=32$, and $\kappa<1.1\cdot10^{-6}$ at $q=0.2$. At practical $B/n$
it says nothing: there a switching outlier block does produce an excess (numerically, $\FDR/q=1.110$ in the limit at
$q=0.1$, $m=32$, $\kappa\approx0.033$, against $1.045$ for the largest global-null value our searches found at the same
$q$ and $m$ with $\kappa\le0.1$, and $1.090$ with unbounded weights). The proposition inherits the genericity assumptions of Theorem~\ref{thm:J2}, and the
$O(\kappa)$ term is not uniform in $\pi_1$ or in the structure.
\end{remark}

\providecommand{\DHess}{\mathbf H}
\providecommand{\Dvec}{\mathbf t}
\providecommand{\Dclip}{\operatorname{clip}}
\providecommand{\DLeb}{\operatorname{Leb}}
\providecommand{\DBin}{\operatorname{Bin}}
\providecommand{\DPoi}{\operatorname{Poisson}}
\providecommand{\DFWER}{\mathrm{FWER}}

\section{Proofs for the global-null lower bound and Simes' Hessian}\label{app:global}

This appendix proves Lemma~\ref{lem:L1}, Theorem~\ref{thm:T2} and Proposition~\ref{prop:H}. Throughout, $m\ge2$ and
$q\in(0,1)$ are fixed, $t_k:=qk/m$ for $k=1,\dots,m$, $t_0:=0$, and $\Dvec:=(t_1,\dots,t_m)\in\R^m$.

\subsection{Configurations and the global-null reduction}\label{app:D-config}

\paragraph{Simes' rejection probability.} For a CDF $F$ on $[0,1]$, let $\psi_m(F)$ be the probability that BH
(equivalently, Simes' test) at level $q$ makes at least one rejection among $m$ i.i.d.\ draws with CDF $F$.
\begin{itemize}[leftmargin=*,itemsep=1pt]
\item For any $F$, $\psi_m$ depends on $F$ only through $(F(t_1),\dots,F(t_m))$: the band counts are multinomial whatever
  $F$ is.
\item $\psi_m$ is a polynomial in these values: a multinomial sum over band counts, with band $k=(t_{k-1},t_k]$ for
  $k\le m$ and band $m+1=(t_m,1]$. We use the same polynomial for every vector $x\in\R^m$ and write $\psi_m(x)$.
\item $\nabla^2\psi$ denotes its Hessian at $\Dvec$.
\end{itemize}

\begin{definition}[The family $\mathcal C_n(\kappa,g,\lambda)$]\label{def:D-family}
Let $H\ge1$, let $\kappa_h,\lambda_h>0$ for $h\le H$, and let $g_1,\dots,g_H\in(0,1)$ be distinct. The covariate $X$ takes
values in $[0,1]\cup\{x_1,\dots,x_H\}$.
\begin{enumerate}[label=(\roman*),leftmargin=*,itemsep=1pt]
\item $P$ puts mass $1-\sum_h\lambda_h/n$ on $[0,1]$, spread uniformly (``light'' points, weight $w=1$), and mass
  $\lambda_h/n$ on the atom $x_h$ (``heavy cell'' $h$, weight $w=\kappa_hn$).
\item $Y=y(X)$ is deterministic: $y(x)=x$ on $[0,1]$ and $y(x_h)=g_h$. The score is $V(x,y)=y$.
\item $dQ/dP\propto w$. The weight ratio is $B:=\max w/\min w=n\max_h\kappa_h$, once that is at least $1$; the sup norm of
  $dQ/dP$ is $\Bb=B/\E_Pw$.
\item \emph{Selection setting:} thresholds $c_{n+j}=y(X_{n+j})$. This is a function of $X$, so (A2) holds, and every test
  unit is null with $Y=c$.
\item \emph{Outlier-detection setting:} every test unit is an inlier from $Q$.
\end{enumerate}
\end{definition}

$P$ depends on $n$ (the atoms have mass $\lambda_h/n$). The bounds below are therefore worst-case bounds over the class of
$P$ with a given weight ratio, not statements about a fixed $P$. In both settings $p_j$ equals the oracle $p^*_j$.

\begin{proof}[Proof of Lemma~\ref{lem:L1}]
Steps 1--3 use only that every test unit is null. Step~4 uses $p_j=p^*_j$, which holds in both settings of
Definition~\ref{def:D-family}; with non-oracle nulls, $p_j\ge p^*_j$ for a monotone score, so $\E F_D(t)\le t$ instead.
\begin{enumerate}[label=\arabic*.,leftmargin=*,itemsep=1pt]
\item Every test unit is null, so $\mathrm{FDP}=\one\{\text{at least one rejection}\}$.
\item Given the calibration data $D$, the test tuples and their $U$'s are i.i.d., so the test p-values are i.i.d.\ with
  CDF $F_D(t):=\PP(p_1\le t\mid D)$.
\item Hence $\FDR=\DFWER=\E[\psi_m(F_D)]$.
\item By Lemma~\ref{lem:A}, conditional on the multiset of the $n+1$ points formed by the calibration data and test
  point~$1$, $p^*_1$ is uniform on $[0,1)$ (cf.\ Lemma~\ref{lem:U}). So $\E F_D(t)=\PP(p_1\le t)=t$. \qedhere
\end{enumerate}
\end{proof}

\subsection{The Poisson limit}\label{app:D-poisson}

\paragraph{The Poisson-limit model.} Let $N_h\sim\DPoi(\lambda_h)$ be independent, $K:=\sum_h\kappa_hN_h$,
$K_{<h}:=\sum_{h':\,g_{h'}<g_h}\kappa_{h'}N_{h'}$ and $\Xi:=1+\sum_h\kappa_h\lambda_h$. Define
\begin{equation}\label{eq:D-FN}
F_N(t):=\frac1\Xi\,\DLeb\Big\{u\in[0,1]:\ \frac{u+\sum_{h:\,g_h<u}\kappa_hN_h}{1+K}\le t\Big\}
+\sum_h\frac{\kappa_h\lambda_h}{\Xi}\,\Dclip\Big(\frac{t-\alpha_h}{\beta_h},0,1\Big),
\end{equation}
\[
\alpha_h:=\frac{g_h+K_{<h}}{1+K+\kappa_h},\qquad \beta_h:=\frac{\kappa_h(1+N_h)}{1+K+\kappa_h},
\]
where $\Dclip(x,0,1):=\min\{\max\{x,0\},1\}$, and let $\Phi:=\E\,\psi_m(F_N)$.

\begin{lemma}[Poisson limit]\label{lem:L2}
Fix $(m,q,g,\kappa,\lambda)$ and let $\DFWER_n$ be the FWER of BH in the configuration $\mathcal C_n(\kappa,g,\lambda)$
(either setting). Then $\DFWER_n\to\Phi$ as $n\to\infty$, and $\E F_N(t)=t$ for every $t$.
\end{lemma}

\begin{proof}
Build all $n$ on one probability space:
\begin{itemize}[leftmargin=*,itemsep=1pt]
\item $N=(N_h)_{h\le H}$, with independent $\DPoi(\lambda_h)$ coordinates;
\item an i.i.d.\ $\mathrm{Unif}(0,1)$ sequence, which supplies the light calibration scores;
\item the heavy calibration counts $N^{(n)}\sim\mathrm{Multinomial}(n;\lambda_1/n,\dots,\lambda_H/n)$, coupled to $N$ so
  that $\PP(N^{(n)}\ne N)\le(\sum_h\lambda_h)^2/n$. Such a coupling exists. Write $\Lambda:=\sum_h\lambda_h$ and
  $N^{(n)}=\sum_{i\le n}\epsilon_i$, where the $\epsilon_i$ are i.i.d.\ and equal the $h$-th unit vector with probability
  $\lambda_h/n$ and $0$ otherwise. Couple each $\epsilon_i$ maximally with a vector $\pi_i$ of independent
  $\DPoi(\lambda_h/n)$ coordinates, independently over $i$: the total variation distance between their laws is
  $(\Lambda/n)(1-e^{-\Lambda/n})\le(\Lambda/n)^2$. Then $\sum_i\pi_i$ has the law of $N$ and differs from $N^{(n)}$ with
  probability at most $n(\Lambda/n)^2=\Lambda^2/n$. (This is the coupling behind Le Cam's inequality
  \citep{lecam1960approximation}.) For each $n$, realise it by drawing $N^{(n)}$ from its conditional law given $N$.
\end{itemize}
Then:
\begin{enumerate}[label=\arabic*.,leftmargin=*,itemsep=1pt]
\item Let $\tilde F_n$ be the conditional CDF of a test p-value in the configuration whose heavy calibration counts are
  $N$ and whose $n_L:=n-\sum_hN_h$ light calibration scores are the first $n_L$ terms of the sequence; on $\{N^{(n)}=N\}$,
  $\tilde F_n=F_D$. Steps 2--5 concern $\tilde F_n$. The empirical CDF $\hat G_n$ of these light scores satisfies $\sup_u|\hat G_n(u)-u|\to0$ almost surely, by the Glivenko--Cantelli theorem
  \citep[Ch.~19]{vandervaart1998asymptotic}.
\item A light test unit with score $u$ has
  \[
  p=\frac{n_L\hat G_n(u-)+n\sum_{g_h<u}\kappa_hN_h+U\,(1+\text{ties})}{n_L+nK+1}.
  \]
  Light ties have probability $0$, and $n_L/n\to1$. So $p\to(u+\sum_{g_h<u}\kappa_hN_h)/(1+K)$ uniformly in $u$.
\item A heavy test unit of cell $h$ has
  \[
  p=\frac{n_L\hat G_n(g_h)+nK_{<h}+U(\kappa_hn+\kappa_hnN_h)}{n_L+nK+\kappa_hn}\ \to\ \alpha_h+\beta_hU.
  \]
\item The test-type masses are $(1-\sum_h\lambda_h/n)/\Xi_n$ (light) and $\kappa_h\lambda_h/\Xi_n$ (cell $h$), where
  $\Xi_n:=1-\sum_h\lambda_h/n+\sum_h\kappa_h\lambda_h\to\Xi$.
\item The limiting light map $u\mapsto p(u)$ is strictly increasing, so $\{u:p(u)=t\}$ is Lebesgue-null; and $\beta_h>0$.
  Hence $\tilde F_n(t_j)\to F_N(t_j)$ almost surely. Since $F_D=\tilde F_n$ outside an event of probability at most
  $\Lambda^2/n$, $F_D(t_j)\to F_N(t_j)$ in probability.
\item $\psi_m$ is bounded and continuous on $[0,1]^m$, so bounded convergence gives
  $\DFWER_n=\E\,\psi_m(F_D)\to\Phi$ (the equality is Lemma~\ref{lem:L1}). Likewise $\E F_D(t)=t$ (Lemma~\ref{lem:L1})
  passes to the limit. (Bounded convergence: convergence in probability of bounded variables implies convergence of
  their expectations; see \citealp[Ch.~2]{vandervaart1998asymptotic}.) \qedhere
\end{enumerate}
\end{proof}

\begin{remark}[No rate]
Lemma~\ref{lem:L2} gives no rate. The crude coupling bound needs $(\sum_h\lambda_h)^2/n\ll\kappa$; in the scaling
$\lambda=\omega/\kappa$ of Section~\ref{app:D-expansion} that is $n\gg\kappa^{-3}$. Finite-$n$ Monte Carlo suggests much faster
convergence than this bound: for $m=2$ and $\kappa=0.01$, $(\DFWER_n-q)/\kappa=0.0351,\,0.0361,\,0.0364$ at
$n=2\cdot10^3,\,2\cdot10^4,\,2\cdot10^5$, against the limit $0.03688$.
\end{remark}

\subsection{The multi-cell expansion}\label{app:D-expansion}

\paragraph{Window configurations.}
\begin{itemize}[leftmargin=*,itemsep=1pt]
\item Choose band indices $1\le k_1\le\dots\le k_H\le m-1$.
\item Choose disjoint open intervals $J_h=(a_h,a_h+\ell_h)$, each with closure inside $(t_{k_h},t_{k_h+1})$, and with
  $a_1<a_2<\cdots$.
\item Let $\ell_+:=\sum_h\ell_h<1$ and $\Omega:=\ell_+/(1-\ell_+)$, and set
  \[
  \omega_h:=\ell_h(1+\Omega),\qquad g_h:=a_h(1+\Omega)-\sum_{h'<h}\omega_{h'},\qquad \kappa_h:=\kappa,\qquad
  \lambda_h:=\omega_h/\kappa.
  \]
\end{itemize}
Then $0<g_1<\dots<g_H<1$:
\begin{itemize}[leftmargin=*,itemsep=1pt]
\item $g_{h+1}-g_h=(1+\Omega)(a_{h+1}-a_h-\ell_h)>0$;
\item $g_H=(1+\Omega)(a_H+\ell_H)-\Omega<1$.
\end{itemize}
$J_h$ is the p-scale window of cell $h$ when $K$ equals its mean $\Omega$. The test mass of cell $h$ is
$\kappa\lambda_h/\Xi=\omega_h/(1+\Omega)=\ell_h$, and the light test mass is $1-\ell_+$.

For $k=1,\dots,m-1$ write $d_k\in\R^m$ for the vector $d_k(j):=t_j-\one\{j\ge k+1\}$, and
\[
\curv_k:=d_k^\top\nabla^2\psi\,d_k .
\]

\begin{theorem}[Multi-cell expansion]\label{thm:D-T1}
As $\kappa\to0$, with the window configuration $(k_h,a_h,\ell_h)_{h\le H}$ fixed,
\[
\Phi(\kappa)-q=\tfrac12\,\kappa\,(1-\ell_+)\sum_h\ell_h\,\curv_{k_h}+O(\kappa^2).
\]
\end{theorem}

\begin{proof}
Let $K_h:=\kappa N_h$ and $\Delta K_h:=K_h-\omega_h$. Then $\E\Delta K_h=0$, $\Var K_h=\kappa\omega_h$,
$\E\Delta K_h^3=\kappa^2\omega_h$ and $\E\Delta K_h^4=3\kappa^2\omega_h^2+\kappa^3\omega_h$, and the $\Delta K_h$ are
independent. (All cumulants of $\DPoi(\lambda)$ equal $\lambda$, so its third and fourth central moments are $\lambda$ and
$\lambda+3\lambda^2$; here $\Delta K_h=\kappa(N_h-\lambda_h)$ with $\lambda_h=\omega_h/\kappa$.)

\emph{Step 1: the good event $G$.} Choose $\delta>0$ and $\kappa_0>0$ so that, whenever $|\Delta K_h|\le\delta$ for all
$h$ and $\kappa\le\kappa_0$, two kinds of interval lie strictly inside $(t_{k_h},t_{k_h+1})$:
\begin{itemize}[leftmargin=*,itemsep=1pt]
\item every heavy window $[\alpha_h,\alpha_h+\beta_h]$ of $F_N$;
\item every jump of the light map, $\big[(g_h+K_{<h})/(1+K),\ (g_h+K_{<h}+K_h)/(1+K)\big]$.
\end{itemize}
This is possible: all these endpoints are continuous in $((K_h)_h,\kappa)$, and at $((K_h)_h,\kappa)=((\omega_h)_h,0)$
both intervals equal the closure of $J_h$. Call this event $G$. By Lemma~\ref{lem:B-chernoff} applied to each $N_h$ (with
$c=\delta/\omega_h$), $\PP(G^c)\le2He^{-\gamma/\kappa}$ for some $\gamma>0$.

\emph{Step 2: on $G$, $F_N$ is exactly affine in $K$ at every grid point.}
\begin{itemize}[leftmargin=*,itemsep=1pt]
\item No $t_j$ lies in a heavy window, so the heavy term of~\eqref{eq:D-FN} at $t_j$ is $\ell_h\one\{k_h<j\}$.
\item The light term is $u_j/\Xi$, where $u_j$ solves $u+\sum_{g_h<u}K_h=t_j(1+K)$.
\item The cells with $g_h<u_j$ are exactly those with $k_h<j$, so
  \[
  F_N(t_j)=(1-\ell_+)\Big[t_j\Big(1+\sum_hK_h\Big)-\sum_{k_h<j}K_h\Big]+\sum_{k_h<j}\ell_h .
  \]
\item At $K_h=\omega_h$ for all $h$ this equals $t_j$, using $(1-\ell_+)(1+\Omega)=1$ and $(1-\ell_+)\omega_h=\ell_h$.
  Hence on $G$
  \begin{equation}\label{eq:D-affine}
  F_N-\Dvec=(1-\ell_+)\sum_h\Delta K_h\,d_{k_h}.
  \end{equation}
\end{itemize}

\emph{Step 3: expansion.} On $G$, $\psi_m(F_N)$ is a polynomial of degree at most $m$ in $(\Delta K_h)_h$,
by~\eqref{eq:D-affine}. Expand it around $\Delta K=0$, using $\psi_m(\Dvec)=q$: Simes' test has level exactly $q$ for
independent uniform p-values \citep{simes1986improved}.
\begin{itemize}[leftmargin=*,itemsep=1pt]
\item \emph{First order:} $(1-\ell_+)\sum_h\nabla\psi\cdot d_{k_h}\,\E[\Delta K_h\one_G]$. Since $\E\Delta K_h=0$,
  $|\E[\Delta K_h\one_G]|=|\E[\Delta K_h\one_{G^c}]|\le(\kappa\omega_h\PP(G^c))^{1/2}=O(e^{-\gamma/(2\kappa)})$.
\item \emph{Second order:}
  $\tfrac12(1-\ell_+)^2\sum_{h,h'}\E[\Delta K_h\Delta K_{h'}\one_G]\,d_{k_h}^\top\nabla^2\psi\,d_{k_{h'}}$. By
  independence this is $\tfrac12(1-\ell_+)^2\sum_h\kappa\omega_h\curv_{k_h}+O(e^{-\gamma/(2\kappa)})$.
\item \emph{Third and higher order:} every mixed moment of degree $3$ vanishes except $\E\Delta K_h^3=\kappa^2\omega_h$.
  Moments of degree at least $4$ are $O(\kappa^2)$ (central moments of $\kappa\cdot\DPoi(\omega/\kappa)$). So the
  remainder is $O(\kappa^2)$.
\item On $G^c$, $|\psi_m-q|\le1$, so the contribution of $G^c$ is $O(e^{-\gamma/\kappa})$.
\end{itemize}
Finally $(1-\ell_+)^2\omega_h=(1-\ell_+)\ell_h$.
\end{proof}

\begin{remark}[How small $\kappa$ must be (numerical)]
For $m=2$, $\psi_m$ is quadratic, so the expansion is exact on $G$. In general $\kappa_0$ must make $G$ typical: $\delta$
must be several standard deviations $\sqrt{\kappa\omega_h}$, so $\kappa_0$ shrinks like (margin)$^2/\omega$ as the windows
approach the band edges. Example (exact Poisson-limit values): $q=0.2$, $m=8$, bands $\{1,2\}$, centred windows of width
$0.8\,q/m$; Theorem~\ref{thm:D-T1} predicts $(\Phi-q)/\kappa\to0.401400$, and the exact values are
$0.3488,\,0.3754,\,0.3942$ at $\kappa=3\cdot10^{-3},\,10^{-3},\,3\cdot10^{-4}$ and
$0.400477,\,0.401411,\,0.401410,\,0.401403$ at $\kappa=10^{-4},\,3\cdot10^{-5},\,10^{-5},\,3\cdot10^{-6}$. With windows of
width $0.4\,q/m$ the gap is about $-0.5\kappa$, linear in $\kappa$.
\end{remark}

\subsection{The band curvatures}\label{app:D-curv}

For $s\in(0,1)$ not a grid point, let $e_s\in\R^m$, $e_s(j):=\one\{s\le t_j\}$, and $d_s:=\Dvec-e_s$. Then
\begin{itemize}[leftmargin=*,itemsep=1pt]
\item $d_s=d_k$ for $s\in(t_k,t_{k+1})$, $k=1,\dots,m-1$;
\item $d_s=d_0:=\Dvec-(1,\dots,1)$ for $s<t_1$;
\item $d_s=\Dvec$ for $s>t_m$.
\end{itemize}
Write $\curv(s):=d_s^\top\nabla^2\psi\,d_s$ and $\curv_0:=d_0^\top\nabla^2\psi\,d_0$. Let $a_r(s)$ be the probability that
Simes' test rejects when $r$ of the $m$ p-values equal $s$ and the other $m-r$ are i.i.d.\ $\mathrm{Unif}(0,1)$.

\begin{lemma}[Curvature as a duplicate second difference]\label{lem:D-L4}
$d_s^\top\nabla^2\psi\,d_s=m(m-1)\,[a_2(s)-2a_1(s)+q]$.
\end{lemma}

\begin{proof}
For $\epsilon\in[0,1]$, $\Dvec+\epsilon(e_s-\Dvec)$ is the grid vector of the CDF of the mixture
$(1-\epsilon)\,\mathrm{Unif}(0,1)+\epsilon\,\delta_s$. A draw from this mixture equals $s$ with probability $\epsilon$,
independently across draws. Simes' test depends only on the band counts, so
\[
\psi_m\big(\Dvec+\epsilon(e_s-\Dvec)\big)=\sum_{r=0}^m\binom mr\epsilon^r(1-\epsilon)^{m-r}a_r(s),
\]
an identity of polynomials in $\epsilon$. Take the second derivative at $\epsilon=0$. The left side gives
$(e_s-\Dvec)^\top\nabla^2\psi\,(e_s-\Dvec)=d_s^\top\nabla^2\psi\,d_s$. The right side gives $m(m-1)(a_2-2a_1+a_0)$, and
$a_0=q$.
\end{proof}

\begin{lemma}[Duplicate exactness]\label{lem:D-L5}
For $S\sim\mathrm{Unif}(0,1)$ and every $r\in\{0,\dots,m\}$, $\E\,a_r(S)=q$. Hence $\int_0^1d_s^\top\nabla^2\psi\,d_s\,ds=0$.
\end{lemma}

\begin{proof}
For $r=0$ this is Simes exactness. Let $r\ge1$, let $r$ copies equal $S$, and let the other $m-r$ values be independent
uniforms $p_i$. All hypotheses are null, so $\DFWER=\E\sum_i\one\{i\in\mathcal R\}/|\mathcal R|$, where $\mathcal R$ is
the rejection set, the sum runs over all $m$ values, and $0/0:=0$.
\begin{enumerate}[label=\arabic*.,leftmargin=*,itemsep=1pt]
\item \emph{A single $p_i$.} By Lemma~\ref{lem:LOO}, $\one\{i\in\mathcal R\}/|\mathcal R|=\one\{p_i\le qR_i/m\}/R_i$,
  where $R_i$, the BH count with $p_i$ replaced by $0$, depends only on the other values. $p_i$ is independent of them, so
  the expectation is $\E[(qR_i/m)/R_i]=q/m$.
\item \emph{The $r$ copies of $S$.} They are rejected together or not at all. Write $N(k)$ for the number of the $m$
  values that are at most $t_k$, and $N_S(k)$ for the same count with all $r$ copies set to $0$; so $N\le N_S$. Let $R_S$
  be the BH count with all $r$ copies set to $0$. Then $R_S\ge r$ since $r\ge1$, and $R_S$ is independent of $S$.
  \begin{itemize}[leftmargin=*,itemsep=1pt]
  \item If $S\le qR_S/m$, then $N(R_S)=N_S(R_S)$, so the BH count is $R_S$.
  \item Otherwise the copies do not count at any $k\le R_S$, and $N\le N_S$ gives a BH count at most $R_S$, with
    threshold $<S$.
  \end{itemize}
  So the copies are rejected iff $S\le qR_S/m$, and then $|\mathcal R|=R_S$. Their total is
  $\E[r\,\one\{S\le qR_S/m\}/R_S]=rq/m$.
\item \emph{Sum:} $(m-r)q/m+rq/m=q$.
\end{enumerate}
Averaging Lemma~\ref{lem:D-L4} over $s$ then gives $\int_0^1d_s^\top\nabla^2\psi\,d_s\,ds=m(m-1)(q-2q+q)=0$.
\end{proof}

\begin{lemma}[The curvature budget]\label{lem:D-L6}
$\curv_0=-m(m-1)(1-q)$, $\Dvec^\top\nabla^2\psi\,\Dvec=0$, and
\[
\sum_{k=1}^{m-1}\curv_k=m(m-1)(1-q).
\]
\end{lemma}

\begin{proof}
\begin{enumerate}[label=\arabic*.,leftmargin=*,itemsep=1pt]
\item For $s<t_1$, one copy at $s$ already gives a rejection, so $a_1(s)=a_2(s)=1$, and Lemma~\ref{lem:D-L4} gives
  $\curv_0=m(m-1)(q-1)$.
\item For $s>t_m$, the copies never count. So $a_r(s)$ is Simes' rejection probability for $m-r$ uniforms at thresholds
  $qk/m$, which is a Simes test at level $q(m-r)/m$ and exact: $a_r(s)=q(m-r)/m$. Then Lemma~\ref{lem:D-L4} gives
  $\Dvec^\top\nabla^2\psi\,\Dvec=m(m-1)\,\tfrac qm\,[(m-2)-2(m-1)+m]=0$.
\item $d_s$ is constant on each band, and bands $1,\dots,m$ have length $q/m$ while band $m+1$ has length $1-q$. So
  Lemma~\ref{lem:D-L5} reads $\tfrac qm\sum_{k=0}^{m-1}\curv_k+(1-q)\cdot0=0$, that is,
  $\sum_{k=1}^{m-1}\curv_k=-\curv_0=m(m-1)(1-q)$. \qedhere
\end{enumerate}
\end{proof}

\begin{lemma}[The first interior band]\label{lem:D-L7}
$\curv_1=m(m-1)(1-q)\big[2(1-q/m)^{m-2}-1\big]$.
\end{lemma}

\begin{proof}
Take $s\in(t_1,t_2)$ in Lemma~\ref{lem:D-L4}.
\begin{enumerate}[label=\arabic*.,leftmargin=*,itemsep=1pt]
\item Two copies give two values at most $t_2$, so $a_2(s)=1$.
\item With one copy at $s$ and $m-1$ uniforms with order statistics $U_{(i)}$, there is no rejection iff
  $\#\{U\le t_{i+1}\}\le i-1$ for $i=1,\dots,m-1$, that is, iff $U_{(i)}>q/m+i\,q/m$ for all $i$.
\item For $M$ uniforms and $b_0,b_1\ge0$ with $b_0+Mb_1\le1$,
  \[
  \PP\big(U_{(i)}>b_0+ib_1\ \forall i\le M\big)=(1-b_0)^M\Big(1-\frac{Mb_1}{1-b_0}\Big)=(1-b_0-Mb_1)(1-b_0)^{M-1}.
  \]
  To see this: all points must exceed $b_0$; rescale to $(b_0,1]$; then apply Daniels' identity
  $\PP(U_{(i)}>ib'\ \forall i\le M)=1-Mb'$. That identity is Simes exactness, which is Lemma~\ref{lem:D-L5} with $r=0$.
\item With $M=m-1$ and $b_0=b_1=q/m$: $1-a_1(s)=(1-q)(1-q/m)^{m-2}$.
\item So $\curv_1=m(m-1)\big[1-2+2(1-q)(1-q/m)^{m-2}+q\big]$. \qedhere
\end{enumerate}
\end{proof}

Lemmas~\ref{lem:D-L6} and~\ref{lem:D-L7} have also been checked in exact arithmetic, with $q$ symbolic, for
$m=2,\dots,9$, by building $\psi_m$ as an explicit polynomial; for example, at $m=8$, $q=1/5$,
$\curv_1=10295206327/320000000\approx32.17$.

\begin{lemma}[Uniform exponential tail]\label{lem:D-L8}
For $2\le k\le m-1$, $|\curv_k|\le2m(m-1)T_k$, where $T_k:=\sum_{i=k+1}^m\PP\big(\DBin(m,qi/m)\ge i-2\big)$. Let
$i_q:=\lceil2/(1-q)\rceil$. Then
\[
T_k\le\#\{i:\ k+1\le i<i_q\}+\sum_{i\ge\max(k+1,i_q)}\exp\Big(-\Big[(i-2)\log\frac{i-2}{qi}-(i-2)+qi\Big]\Big).
\]
The right side does not depend on $m$. For $k\ge i_q$ it decays like $e^{-k(\log(1/q)-1+q)+O(1)}$.
\end{lemma}

\begin{proof}
Take $s\in(t_k,t_{k+1})$ and $r\in\{1,2\}$, and let $N'(t)$ count the $m-r$ uniforms that are at most $t$.
\begin{enumerate}[label=\arabic*.,leftmargin=*,itemsep=1pt]
\item No rejection means $N'(t_i)\le i-1$ for $i\le k$ and $N'(t_i)\le i-1-r$ for $i\ge k+1$.
\item This event is contained in $E_0:=\{N'(t_i)\le i-1\ \forall i\}$, which has probability $1-q(m-r)/m$ by Simes
  exactness.
\item $E_0$ minus the event is contained in $\{\exists\,i\ge k+1:\ N'(t_i)\ge i-r\}$, which has probability at most $T_k$,
  since $N'(t_i)$ is stochastically at most $\DBin(m,qi/m)$.
\item So $a_r(s)=q(m-r)/m+\theta_r$ with $\theta_r\in[0,T_k]$, and $a_2-2a_1+q=\theta_2-2\theta_1\in[-2T_k,T_k]$; with
  Lemma~\ref{lem:D-L4} this gives $|\curv_k|\le2m(m-1)T_k$.
\item Each term of $T_k$ with $i<i_q$ is bounded by $1$. For $i\ge i_q$ we have $i-2\ge qi$, and
  $\PP(\DBin(m,p)\ge x)\le\exp(-[x\log(x/(mp))-x+mp])$ for $x\ge mp$ (Lemma~\ref{lem:B-chernoff}); here $x=i-2$ and
  $mp=qi$. \qedhere
\end{enumerate}
\end{proof}

\subsection{The lower-bound theorems}\label{app:D-lower}

Let
\[
L^*_m(q):=\sup\ \tfrac12(1-\ell_+)\sum_h\ell_h\,\curv_{k_h},
\]
the supremum over all window configurations of Section~\ref{app:D-expansion}.

\begin{theorem}[Bounds on $L^*_m$]\label{thm:D-T2}
\begin{enumerate}[label=(\arabic*),leftmargin=*,itemsep=1pt]
\item \emph{(Single band.)} $L^*_m(q)\ge\tfrac12q(1-q/m)(m-1)(1-q)\big[2(1-q/m)^{m-2}-1\big]$. This is vacuous when
  $\curv_1<0$, that is (for $m\ge3$) when $q>m(1-2^{-1/(m-2)})$, which tends to $\ln2$.
\item \emph{(All positive bands.)} $L^*_m(q)\ge\tfrac12q(1-q)(m-1)(1-q+q/m)$.
\item \emph{(Asymptotic.)} $\liminf_{m\to\infty}L^*_m(q)/(qm)\ge\tfrac12(1-q)$.
\end{enumerate}
\end{theorem}

\begin{proof}
\begin{enumerate}[label=(\arabic*),leftmargin=*,itemsep=1pt]
\item Take one window of width $\theta q/m$ in the band $(t_1,t_2)$, with $\theta\uparrow1$. Its value
  $\tfrac12(1-\theta q/m)(\theta q/m)\curv_1$ tends to $\tfrac12(1-q/m)(q/m)\curv_1$; now use Lemma~\ref{lem:D-L7}.
\item Let $\mathcal K:=\{k\ge1:\curv_k>0\}$. Fill each band in $\mathcal K$ with a window of width $\theta q/m$,
  $\theta\uparrow1$. Then $\ell_+\le(m-1)q/m$, and by Lemma~\ref{lem:D-L6},
  $\sum_{k\in\mathcal K}\curv_k\ge\sum_{k=1}^{m-1}\curv_k=m(m-1)(1-q)$. So, in the limit $\theta\uparrow1$,
  $\tfrac12(1-\ell_+)(q/m)\sum_{k\in\mathcal K}\curv_k\ge\tfrac12(1-q+q/m)\,q(m-1)(1-q)$.
\item Use windows of width $\theta q/m$, $\theta\uparrow1$, in bands $1,\dots,k_0$, with $k_0=k_0(m)\to\infty$ and
  $k_0/m\to0$, so $\ell_+\le k_0q/m\to0$.
  \begin{itemize}[leftmargin=*,itemsep=1pt]
  \item By Lemmas~\ref{lem:D-L6} and~\ref{lem:D-L8},
    $\sum_{k\le k_0}\curv_k\ge m(m-1)(1-q)-2m(m-1)\sum_{k>k_0}T_k$.
  \item $\sum_{k>k_0}T_k\to0$ uniformly in $m$, by the $m$-free bound of Lemma~\ref{lem:D-L8}.
  \item Divide $\tfrac12(1-\ell_+)(q/m)\sum_{k\le k_0}\curv_k$ by $qm$. \qedhere
  \end{itemize}
\end{enumerate}
\end{proof}

Parts~(2) and~(3) do not need the positive-curvature condition below. Per unit of $qm$, part~(2) tends to $\tfrac12(1-q)^2$ as $m\to\infty$
and part~(3) gives $\tfrac12(1-q)$.

\begin{theorem}[Finite-$n$ lower bound]\label{thm:D-T3}
Fix $m\ge2$, $q\in(0,1)$ and $\epsilon>0$. There is $\kappa_0>0$ such that, for each $\kappa\in(0,\kappa_0]$, there is
$n_0(\kappa)$ with the following property. For every $n\ge n_0$ there is a configuration in $\mathcal C_n$ that
\begin{itemize}[leftmargin=*,itemsep=1pt]
\item satisfies (A1)--(A3), in the selection setting with $c=c(X)$ and in the outlier-detection setting;
\item has all $m$ test units null;
\item has weight ratio $B=\kappa n$;
\item has
  \[
  \FDR(\mathrm{BH})\ \ge\ q+\big(L^*_m(q)-\epsilon\big)\frac Bn\ \ge\
  q\Big[1+\Big(\tfrac12(1-q)(m-1)(1-q+q/m)-\epsilon/q\Big)\frac Bn\Big].
  \]
\end{itemize}
\end{theorem}

\begin{proof}[Proof of Theorem~\ref{thm:D-T3}, and hence of Theorem~\ref{thm:T2}]
Pick a window configuration with $\tfrac12(1-\ell_+)\sum_h\ell_h\curv_{k_h}\ge L^*_m-\epsilon/3$. By
Theorem~\ref{thm:D-T1}, pick $\kappa_0$ so that the $O(\kappa^2)$ term is at most $\epsilon\kappa/3$ for
$\kappa\le\kappa_0$. By Lemma~\ref{lem:L2}, pick $n_0(\kappa)$ so that $|\FDR_n-\Phi(\kappa)|\le\epsilon\kappa/3$ for
$n\ge n_0$ (here $\FDR_n=\DFWER_n$ by Lemma~\ref{lem:L1}). Then
$\FDR_n\ge q+\kappa(L^*_m-\epsilon/3)-2\epsilon\kappa/3=q+(L^*_m-\epsilon)B/n$, since $\kappa_h=\kappa$ gives $B=\kappa n$.
The second inequality is Theorem~\ref{thm:D-T2}(2).

Theorem~\ref{thm:T2} is this statement, with the bounds on $L^*_m(q)$ from Theorem~\ref{thm:D-T2}(2) and~(3).
\end{proof}

\begin{remark}[Scope]
The statement is for each fixed ratio $\kappa=B/n\le\kappa_0$ with $n$ large.
\begin{itemize}[leftmargin=*,itemsep=1pt]
\item Since $n_0(\kappa)$ has no rate, it does not give $\FDR-q\ge(L^*_m-\epsilon)B/n$ along an arbitrary sequence with
  $B_n/n\to0$.
\item Near-optimal configurations need small $\kappa$ (windows close to the band edges; see the remark after
  Theorem~\ref{thm:D-T1}).
\end{itemize}
\end{remark}

\subsection{Simes' Hessian and conditional sharpness}\label{app:D-hessian}

Write $\DHess:=\nabla^2\psi_m(\Dvec)$, and $\DHess_r$ for $\DHess$ restricted to the coordinates $2,\dots,m$. Write
$\mathbf u_j$ for the $j$-th coordinate vector of $\R^m$.

\begin{lemma}[Z-matrix]\label{lem:D-Z}
$\DHess_{jk}\le0$ for all $j\ne k$ in $\{1,\dots,m\}$.
\end{lemma}

\begin{proof}
\begin{enumerate}[label=\arabic*.,leftmargin=*,itemsep=1pt]
\item Moving $F(t_j)$ up by $\epsilon$ moves mass $\epsilon$ from band $j+1$ to band $j$: one extra point $\mathsf A$ sits
  in band $j$ instead of band $j+1$. Similarly for $k$, with an extra point $\mathsf B$.
\item \emph{Two-point formula.} For band probabilities $\pi$, $\partial^2\psi/\partial\pi_a\partial\pi_b=m(m-1)\cdot
  \PP(\text{reject}\mid m-2\text{ uniforms plus one point in band }a\text{ and one in band }b)$, and
  $\partial/\partial F_j=\partial/\partial\pi_j-\partial/\partial\pi_{j+1}$. So
  \[
  \DHess_{jk}=m(m-1)\,\E\big[f(1,1)-f(1,0)-f(0,1)+f(0,0)\big],
  \]
  where $f(a,b)$ is the rejection indicator with $\mathsf A$ in band $j$ ($a=1$) or $j+1$ ($a=0$), $\mathsf B$ in band $k$
  ($b=1$) or $k+1$ ($b=0$), and the other $m-2$ values uniform.
\item Let $j<k$ ($\DHess$ is symmetric). The position of $\mathsf A$ changes the count only at index $j$, and that of
  $\mathsf B$ only at index $k$. So $f(a,b)=1-(1-o)(1-\chi_j(a))(1-\chi_k(b))$, where $o$ indicates a crossing at some
  index other than $j$ and $k$, and $\chi_j,\chi_k$ (crossing at index $j$, resp.\ $k$) are non-decreasing $0/1$
  functions.
\item The mixed difference is $-(1-o)(\chi_j(1)-\chi_j(0))(\chi_k(1)-\chi_k(0))\le0$, pointwise. \qedhere
\end{enumerate}
\end{proof}

\begin{lemma}[Row sums in closed form]\label{lem:D-R}
Let $S'_l$ be the number of $m-2$ i.i.d.\ uniforms that are at most $t_l$, and $Y_l:=S'_l-l$. For $2\le j\le m$ let
$r_j:=\sum_{k=2}^m\DHess_{jk}$, and for $2\le j\le m-1$ let $c_j:=q/(m-qj)$ and
\begin{align*}
E_1&:=\{S'_1=0;\ Y_l\le-1\ (2\le l<j);\ Y_j=-1;\ Y_l\le-2\ (l>j)\},\\
E_2&:=\{S'_1=0;\ Y_l\le-2\ (2\le l<j);\ Y_j=-2;\ Y_l\le-3\ (l>j)\}.
\end{align*}
Then, for $2\le j\le m-1$,
\begin{align*}
r_j&=m(m-1)\big[\PP(E_2)-\PP(E_1)\big],\\
\PP(E_1)&=\binom{m-2}{j-1}t_j^{\,j-1}(1-t_j)^{m-1-j}\cdot\frac1j\cdot\big(1-(m-1-j)c_j\big),\\
\PP(E_2)&=\binom{m-2}{j-2}t_j^{\,j-2}(1-t_j)^{m-j}\cdot\frac1j\Big(\frac{j-1}j\Big)^{j-3}\cdot\big(1-(m-j)c_j\big),
\end{align*}
and $r_m=(m-1)\,q^{m-2}\big((m-1)/m\big)^{m-3}>0$.
\end{lemma}

\begin{proof}
\begin{enumerate}[label=\arabic*.,leftmargin=*,itemsep=1pt]
\item $r_j$ is the mixed second derivative along $\mathbf u_j$ ($\mathsf A$ moves from band $j+1$ to band $j$) and along
  $\sum_{k\ge2}\mathbf u_k$ ($\mathsf B$ moves from the top band $m+1$ to band $2$). $\mathsf B$ in band $2$ adds $1$ to
  the count at every index $\ge2$.
\item Write $\bar f(a,b)$ for the indicator of ``no rejection'', with $a=1$ if $\mathsf A$ is in band $j$ and $b=1$ if
  $\mathsf B$ is in band $2$. Every case needs $S'_1=0$, and then:
  \begin{itemize}[leftmargin=*,itemsep=1pt]
  \item for $2\le l<j$: $Y_l\le-1-b$;
  \item at $l=j$: $Y_j\le-1-a-b$;
  \item for $l>j$: $Y_l\le-2-b$.
  \end{itemize}
  So $\bar f(0,0)-\bar f(1,0)=\one\{E_1\}$ and $\bar f(0,1)-\bar f(1,1)=\one\{E_2\}$, and
  $r_j/(m(m-1))=-(\text{mixed difference of }\bar f)=\PP(E_2)-\PP(E_1)$.
\item Condition on the number of uniforms below $t_j$, which is $j-1$ for $E_1$ and $j-2$ for $E_2$.
  \begin{itemize}[leftmargin=*,itemsep=1pt]
  \item The points below $t_j$ are uniform on $[0,t_j]$. After rescaling by $t_j$, their constraint is $U_{(i)}>i/j$ for
    $E_1$ (Daniels: probability $1/j$), and $U_{(i)}>(i+1)/j$ for $E_2$ (Abel--Daniels, the identity in the proof of
    Lemma~\ref{lem:D-L7}: probability $(1/j)((j-1)/j)^{j-3}$).
  \item The points above $t_j$ are uniform on $(t_j,1]$. In both cases the constraint is $N(c_j\,i)\le i-1$ after
    rescaling, so Daniels gives $1-Mc_j$, with $M=m-1-j$ or $M=m-j$ points. Daniels needs $Mc_j\le1$, which is equivalent
    to $q\le1$ here. For $E_1$ the constraint at $i=m-j$ is vacuous. At $j=2$, the identity of Lemma~\ref{lem:D-L7} with
    $M=0$ gives $1$.
  \end{itemize}
\item For $j=m$, $E_1$ is impossible (it needs $m-1$ of the $m-2$ uniforms below $q$), and $E_2$ is the Abel--Daniels
  event for $m-2$ points on $[0,q]$. \qedhere
\end{enumerate}
\end{proof}

\begin{theorem}[Simes' Hessian is PSD away from the first grid point]\label{thm:D-H}
For every $m\ge2$ and every $0<q\le1/e$, $\DHess_r$ is diagonally dominant with non-negative diagonal, hence positive
semidefinite.
\end{theorem}

\begin{proof}
By Lemma~\ref{lem:D-Z}, the off-diagonal entries are non-positive, so $\DHess_{jj}-\sum_{k\ge2,\,k\ne j}|\DHess_{jk}|=r_j$,
and it suffices that every row sum $r_j$ is non-negative. (A symmetric matrix $A$ with $A_{jj}\ge\sum_{k\ne j}|A_{jk}|$ for
every $j$ is positive semidefinite: $x^\top Ax\ge\sum_jA_{jj}x_j^2-\sum_{j\ne k}|A_{jk}||x_j||x_k|
\ge\sum_j\big(A_{jj}-\sum_{k\ne j}|A_{jk}|\big)x_j^2\ge0$, by $2|x_jx_k|\le x_j^2+x_k^2$ and symmetry.)
By Lemma~\ref{lem:D-R}, $r_m>0$. For $2\le j\le m-1$, the ratio $\PP(E_2)/\PP(E_1)$ simplifies to
\[
\rho_j=\Big(\frac{j-1}j\Big)^{j-2}\cdot\frac1q\cdot\Lambda_j,\qquad
\Lambda_j:=\frac{(m-qj)\,m(1-q)}{(m-j)\,(m-q(m-1))}.
\]
\begin{itemize}[leftmargin=*,itemsep=1pt]
\item $\Lambda_j\ge1$ iff $jm(1-q)^2\ge q(m-j)$, which holds for all $j\ge2$ when $2(1-q)^2\ge q$, that is, when
  $q\le1/2$.
\item $((j-1)/j)^{j-2}\ge(1-1/j)^{j-1}\ge1/e$.
\item So $\rho_j\ge1/(eq)\ge1$ for $q\le1/e$, and $r_j\ge0$. \qedhere
\end{itemize}
\end{proof}

\begin{remark}[Sharpness of $1/e$ (numerical)]
From the exact closed form, $\rho_j$ is strictly decreasing in $q$, so diagonal dominance holds exactly on an interval
$[0,q^*_m]$, with $q^*_m=0.612,\,0.511,\,0.444,\,0.391,\,0.375,\,0.370,\,0.369$ at
$m=10,\,30,\,100,\,10^3,\,10^4,\,10^5,\,10^6$; these tend to $1/e$, at a rate of about $m^{-1/2}$. Beyond $q^*_m$,
$\DHess_r$ can still be PSD without diagonal dominance: exact $LDL^\top$ decompositions confirm this at $q=2/3$ for
$m=7,\dots,13$ and at $q=3/5$ for $m=12,13$. PSD fails at $q=9/10$ for every $m$ from $3$ to $13$. At $m=2$,
$\DHess_r=[2]$ is PSD for every $q$.
\end{remark}

\paragraph{Check ($m=3$, $q=1/5$).} Here $\Dvec=(1/15,2/15,1/5)$ and there is one free uniform $U$. Row $j=2$:
$E_1=\{U\in(t_1,t_2]\}$ has probability $1/15$, $E_2=\{U>t_3\}$ has probability $4/5$, and $r_2=3\cdot2\cdot(4/5-1/15)=22/5$.
Row $3$: $r_3=(m-1)q^{m-2}((m-1)/m)^0=2/5$. The exact restricted Hessian is
$\DHess_r=\begin{pmatrix}24/5&-2/5\\-2/5&4/5\end{pmatrix}$: its row sums are $22/5$ and $2/5$, the off-diagonal entry is
negative, and the matrix is positive definite.

\begin{corollary}[Within the window family, windows above $t_1$]\label{cor:D-C1}
Let $0<q\le1/e$. For an interval $J\subset(t_1,1)$ of length $|J|$, let $\bar d_J:=|J|^{-1}\int_Jd_s\,ds$ and
$\curv(J):=\bar d_J^\top\nabla^2\psi\,\bar d_J$, the quadratic form at the average of $d_s$ over $J$. Then
$|J|\,\curv(J)\le\int_J\curv(s)\,ds$. Hence every Poisson window configuration whose windows $J_h$ (pairwise disjoint, of
lengths $\ell_h$, $\ell_+=\sum_h\ell_h<1$) lie in $(t_1,1)$ satisfies
\[
\tfrac12(1-\ell_+)\sum_h\ell_h\,\curv(J_h)\ \le\ \tfrac12\,\frac qm\sum_{k=1}^{m-1}\curv_k^+,\qquad \curv_k^+:=\max\{\curv_k,0\}.
\]
\end{corollary}

\begin{proof}
\begin{enumerate}[label=\arabic*.,leftmargin=*,itemsep=1pt]
\item For $s\in J$ we have $s>t_1$, so $d_s(1)=t_1$: the vectors $d_s$ differ only in coordinates $2,\dots,m$. By
  Theorem~\ref{thm:D-H}, the quadratic form $x\mapsto x^\top\nabla^2\psi\,x$ is convex on their affine hull.
\item Jensen's inequality then gives $|J|\,\curv(J)\le\int_J\curv(s)\,ds$.
\item The windows are pairwise disjoint and $0\le1-\ell_+\le1$, so
  $\tfrac12(1-\ell_+)\sum_h\ell_h\curv(J_h)\le\tfrac12\int_{(t_1,1)}\curv(s)^+\,ds$. On band $k+1$, $\curv(s)=\curv_k$,
  and the top band has $\curv(s)=\Dvec^\top\nabla^2\psi\,\Dvec=0$ (Lemma~\ref{lem:D-L6}), so it contributes nothing. The
  integral is therefore $\tfrac qm\sum_{k=1}^{m-1}\curv_k^+$. \qedhere
\end{enumerate}
\end{proof}

This inequality does not need the positive-curvature condition below. With it ($\curv_k\ge0$ for all $k$), the
right-hand side equals $\tfrac12\frac qm\,m(m-1)(1-q)=\tfrac12q(1-q)(m-1)$ by Lemma~\ref{lem:D-L6}. That matches the lower
bound of Theorem~\ref{thm:D-T2}(2) up to the factor $1-q+q/m$, and is asymptotically sharp: per unit of $qm$ it is
$\tfrac12(1-q)(m-1)/m\to\tfrac12(1-q)$, the constant of Theorem~\ref{thm:D-T2}(3).

For the window configurations of Section~\ref{app:D-expansion}, whose windows lie strictly inside one band each, the
left-hand side is the coefficient of Theorem~\ref{thm:D-T1}, and the same bound is immediate without
Theorem~\ref{thm:D-H}: windows in the same band have total length at most $q/m$, so
$\tfrac12(1-\ell_+)\sum_h\ell_h\curv_{k_h}\le\tfrac12\frac qm\sum_k\curv_k^+$. Corollary~\ref{cor:D-C1} adds windows above
$t_1$ that contain grid points. \gap{We have proved Theorem~\ref{thm:D-T1} only for windows strictly inside one
band, because its good event needs this. For windows in $(t_1,1)$ that contain a grid point, we have not proved that
$\tfrac12\kappa(1-\ell_+)\sum_h\ell_h\curv(J_h)$ is the first-order term of $\Phi-q$.} Windows that reach below $t_1$ are
covered only numerically (a scan over windows straddling many grid points), and configurations outside the Poisson window
family are open.

\paragraph{The positive-curvature condition.} \emph{$\curv_k>0$ for all $1\le k\le m-1$.} It is false for large $q$: by
Lemma~\ref{lem:D-L7}, $\curv_1<0$ when $q>m(1-2^{-1/(m-2)})$, for example when $q>0.9546$ at $m=6$. It is open in general;
it has been verified in exact rational arithmetic for all $k$ at $q\in\{1/20,1/10,1/5,1/2,7/10\}$ and
$m\in\{10,16,24,32\}$ and for all $q\in(0,1)$ and $k\ge2$ when $m\le9$, and numerically for $m\le1024$ at
$q\in\{0.05,0.1,0.2\}$ (up to finite-difference noise) and in the $m\to\infty$ limit for $k\le25$. Writing
$a_2-2a_1+a_0$ as a second difference of the two-point rejection kernel does not prove it, because that kernel is not
conditionally positive semidefinite over all bands (a value in band~$1$ always rejects).

\begin{proof}[Proof of Proposition~\ref{prop:H}]
The off-diagonal entries of Simes' Hessian at the uniform point are non-positive by Lemma~\ref{lem:D-Z} (all $j\ne k$ in
$\{1,\dots,m\}$), and for $0<q\le1/e$ its restriction to grid points $2,\dots,m$ is positive semidefinite by
Theorem~\ref{thm:D-H}. For the conditional claim, assume the positive-curvature condition. For every window
configuration of Section~\ref{app:D-expansion}, the coefficient of Theorem~\ref{thm:D-T1} satisfies
$\tfrac12(1-\ell_+)\sum_h\ell_h\curv_{k_h}\le\tfrac12\frac qm\sum_k\curv_k^+=\tfrac12q(1-q)(m-1)$ (the remark above and
Lemma~\ref{lem:D-L6}), that is, at most $\tfrac12(1-q)(m-1)/m$ per unit of $qm$; this uses neither Theorem~\ref{thm:D-H}
nor $q\le1/e$. The configurations in the proofs of Theorem~\ref{thm:D-T2}(2) and~(3) are window configurations; they reach
$\tfrac12q(1-q)(m-1)(1-q+q/m)$ and, per unit of $qm$, $\tfrac12(1-q)$ as $m\to\infty$. So within this family the constant
per unit of $qmB/n$ is below $\tfrac12(1-q)$ at every $m$ and tends to $\tfrac12(1-q)$ as $m\to\infty$.
\end{proof}

\providecommand{\Unif}{\mathrm{Unif}}
\providecommand{\Bin}{\mathrm{Bin}}
\providecommand{\obs}{\mathrm{obs}}
\providecommand{\esssup}{\operatorname*{ess\,sup}}
\makeatletter
\@ifundefined{assumption}{%
  \newtheorem{assumption}[theorem]{Assumption}%
  \crefname{assumption}{Assumption}{Assumptions}}{}
\makeatother

\section{Proofs for the envelope theorem and the global-null upper bound}\label{app:envelope}

\subsection{Setting and assumptions}\label{sec:E-setting}

Calibration points are $Z_i=(X_i,Y_i)$, $i=1,\dots,n$. Test unit $j\in\{1,\dots,m\}$ has data
$Z_{n+j}=(X_{n+j},Y_{n+j})$, a threshold $c_{n+j}$ and a tie-breaking variable $U_j$; it is null iff
$Y_{n+j}\le c_{n+j}$.

\begin{assumption}[(A1)--(A3)]\label{ass:A}
\leavevmode
\begin{enumerate}[label=(A\arabic*),ref=(A\arabic*),leftmargin=*]
\item\label{ass:A1} The calibration points $Z_1,\dots,Z_n$ are i.i.d.\ from $P$ on $\mathcal X\times\mathcal Y$.
For each test unit $j$, $Z_{n+j}\sim Q$, where
\[
\frac{dQ}{dP}(x,y)=\frac{w(x)}{\E_P[w(X)]}\qquad\text{and}\qquad 1\le w(x)\le B .
\]
\item\label{ass:A2} For every $j$, with $H_j:=(X_{n+\ell},c_{n+\ell},U_\ell)_{\ell\ne j}$:
\begin{enumerate}[label=(\roman*),leftmargin=*]
\item $Z_1,\dots,Z_n,Z_{n+j}$ are mutually independent;
\item $H_j$ is independent of $(Z_1,\dots,Z_n,Z_{n+j})$;
\item $U_j\sim\Unif(0,1)$ is independent of $(Z_1,\dots,Z_n,Z_{n+j},H_j)$.
\end{enumerate}
\item\label{ass:A3} The score $V:\mathcal X\times\mathcal Y\to\R$ is monotone: $y\le y'$ implies $V(x,y)\le V(x,y')$.
\end{enumerate}
\end{assumption}

\paragraph{Comments on the assumptions.}
The p-values below are invariant to rescaling $w$, so the normalisation $1\le w$ loses nothing. Parts (ii) and (iii)
of \ref{ass:A2} together say that the law of $(Z_1,\dots,Z_n,Z_{n+j},H_j,U_j)$ is the product of the law of the $Z$'s,
the law $P_{H_j}$ of $H_j$, and $\Unif(0,1)$; this is what Lemma~\ref{lem:A} uses. The other test units may depend on
each other arbitrarily, and their responses $Y_{n+\ell}$ and the threshold $c_{n+j}$ play no role below.
\citet[\S3.2]{jin2026weighted} assume i.i.d.\ test tuples and independent $U$'s, which implies \ref{ass:A2}. What
\ref{ass:A2} excludes are covariates $X_{n+\ell}$ or thresholds $c_{n+\ell}$ that depend on $Z_{n+j}$ or on the
calibration data (for example, thresholds tuned on the data).

\paragraph{Notation.}
$V_i:=V(X_i,Y_i)$ for $i\le n+m$, $\hat V_{n+\ell}:=V(X_{n+\ell},c_{n+\ell})$ and $w_i:=w(X_i)$. For a calibration
multiset $C$ of $n$ points and a test unit $\ell$, write
\[
p_\ell[C]:=\frac{\sum_{e\in C}w_e\one\{V_e<\hat V_{n+\ell}\}+U_\ell\big(w_{n+\ell}+\sum_{e\in C}w_e\one\{V_e=\hat V_{n+\ell}\}\big)}
{\sum_{e\in C}w_e+w_{n+\ell}} ,
\]
where $w_e$ and $V_e$ are the weight and score of the point $e$. The reported p-value is
$p_\ell:=p_\ell[\{Z_1,\dots,Z_n\}]$, which is eq.~(4) of \citet{jin2026weighted}. BH at level $q$ rejects
$\mathcal R:=\{\ell:p_\ell\le qk^*/m\}$, with $k^*:=\max\{k:\#\{\ell:p_\ell\le qk/m\}\ge k\}$ ($k^*:=0$ if there is no
such $k$); the grid is $t_k=qk/m$. For $x\in\R^{m-1}$, $R^0(x)$ denotes the number of BH rejections at level $q$ among the
$m$ numbers $(x_1,\dots,x_{m-1},0)$. Then $R^0\ge1$, because the forced $0$ is $\le q/m$, and $R^0$ is non-increasing in
each coordinate. For $x\in\R^{m-1}$ and $s\in\R$, $x\pm s$ is $x$ with $s$ added to or subtracted from every coordinate.

\subsection{Statements}\label{sec:E-statements}

Fix $j$. Define:
\begin{itemize}[leftmargin=*,itemsep=1pt]
\item $S_j$: the multiset $\{Z_1,\dots,Z_n,Z_{n+j}\}$ of $n+1$ points;
\item for $e\in S_j$, $p_{-j}(e):=\big(p_\ell[S_j\setminus\{e\}]\big)_{\ell\ne j}\in\R^{m-1}$;
\item $\varrho_j(e):=R^0(p_{-j}(e))$ (we write $\varrho$ to keep $\rho$ free for the parameter of
\cref{sec:E-explicit,sec:E-global});
\item $e\preceq e'$ iff $V_e\le V_{e'}$; a \emph{tie group} is a class of equal scores. For a tie group $G$ with common
score $v_G$, write $e\succeq G$ if $V_e\ge v_G$ and $e\preceq G$ if $V_e\le v_G$.
\end{itemize}
Copies of the same atom give the same multiset $S_j\setminus\{e\}$, so $\varrho_j$ and the tie group are functions of the
value of $e$, not of its label. Define
\[
\Gamma^*_j:=\max_{G}\ \frac{\max\{\varrho_j(e):e\succeq G\}}{\min\{\varrho_j(e):e\preceq G\}},
\]
the maximum being over the tie groups $G$ of $S_j$. For $\ell\ne j$ let $W_\ell:=\sum_{e\in S_j}w_e+w_{n+\ell}$ and
$\phi_\ell(w):=w/(W_\ell-w)$; let $w_{\max}$ and $w_{\min}$ be the largest and smallest weights in $S_j$, and
\[
\delta_j:=\max_{\ell\ne j}\big[\phi_\ell(w_{\max})-\phi_\ell(w_{\min})\big].
\]

\begin{theorem}[Envelope, full form]\label{thm:E-env}
Under \ref{ass:A1}--\ref{ass:A3},
\[
\FDR(\mathrm{BH})\;\le\;\frac qm\sum_{j=1}^m\E[\Gamma^*_j]\;\le\;\frac qm\sum_{j=1}^m\E[\Gamma_j],\qquad
\Gamma_j:=\max_{e_0\in S_j}\frac{R^0(p_{-j}(e_0)-\delta_j)}{R^0(p_{-j}(e_0)+\delta_j)},
\]
and deterministically
\[
\delta_j\;\le\;\delta_n(B):=\frac{(B-1)(B+n+1)}{(n+1)(B+n)} .
\]
If $w$ is constant, then $\delta_j=0$, so $\Gamma_j=1$ and $\FDR\le q$.
\end{theorem}

\begin{corollary}[Computable envelopes]\label{cor:E-computable}
Let $p^{\obs}_{-j}:=(p_\ell)_{\ell\ne j}$ be the observed p-values of the other test units, and put
$\bar\tau_j:=\max_{\ell\ne j}\phi_\ell(w_{\max})$ and $\eta_j:=\delta_j+\bar\tau_j$. Define
\[
\tilde\Gamma_j:=\frac{R^0(p^{\obs}_{-j}-\bar\tau_j)}{R^0(p^{\obs}_{-j}+\bar\tau_j)},\qquad
\hat\Gamma_j:=\frac{R^0(p^{\obs}_{-j}-\eta_j)}{R^0(p^{\obs}_{-j}+\eta_j)} .
\]
Under \ref{ass:A1}--\ref{ass:A3}, $\Gamma^*_j\le\tilde\Gamma_j\le\hat\Gamma_j$ pointwise, and hence
\[
\FDR(\mathrm{BH})\;\le\;\frac qm\sum_j\E[\tilde\Gamma_j]\;\le\;\frac qm\sum_j\E[\hat\Gamma_j].
\]
$\tilde\Gamma_j$ and $\hat\Gamma_j$ use only observed quantities: the calibration data, the test covariates and the
other test units' p-values. So they are computable in both the selection and the outlier-detection setting. The shifts
satisfy
\begin{enumerate}[label=(\roman*),leftmargin=*,itemsep=1pt]
\item $\bar\tau_j\le\hat\tau_j/(1-\hat\tau_j)$, where $\hat\tau_j:=w_{\max}/\sum_{e\in S_j}w_e$;
\item $\bar\tau_j\le B/(n+1)$ deterministically;
\item $\eta_j\le2\hat\tau_j/(1-\hat\tau_j)$.
\end{enumerate}
\end{corollary}

$\tilde\Gamma_j$ does not bound $\Gamma_j$ in general: exact enumeration finds configurations with
$\Gamma_j>\tilde\Gamma_j$. The chain is $\Gamma^*_j\le\tilde\Gamma_j\le\hat\Gamma_j$, with $\Gamma_j$ a separate bound that
is not computable (\cref{sec:E-remarks}).

\begin{proposition}[The tightest computable envelope]\label{prop:E-sup}
$\Gamma^*_j$ depends on the unobserved $Y_{n+j}$ only through $v=V_{n+j}$, and only through how $v$ is ordered relative
to the calibration scores $V_1,\dots,V_n$ and the $\hat V_{n+\ell}$, $\ell\ne j$. Write $\Gamma^*_j(v)$ for $\Gamma^*_j$
computed with test score $v$, and let $r$ be the number of distinct values among those scores. Then
\[
\Gamma^{\sup}_j:=\sup_v\Gamma^*_j(v)
\]
is a maximum over at most $2r+1$ cases, it is computable, and $\Gamma^*_j\le\Gamma^{\sup}_j\le\tilde\Gamma_j$.
\end{proposition}

\begin{remark}
In $40$ random exact examples with $m=2$, the excess of $\frac qm\sum_j\E[\Gamma^{\sup}_j]$ over $q$ had median $2.9\%$
(maximum $35\%$) of the excess of $\frac qm\sum_j\E[\hat\Gamma_j]$ (numerical).
\end{remark}

\subsection{The leave-one-out identity and the conditional law of the test point}\label{sec:E-lemmas}

\begin{lemma}[Leave-one-out identity]\label{lem:LOO}
Let $p_1,\dots,p_m$ be real numbers, let $\mathcal R$ be the BH rejection set at level $q$, fix $j$, and let
$R_j:=R^0\big((p_\ell)_{\ell\ne j}\big)$, the BH count when $p_j$ is replaced by $0$. Then $j\in\mathcal R$ iff
$p_j\le qR_j/m$, and on that event $|\mathcal R|=R_j$.
\end{lemma}

\begin{proof}
Let $N(k):=\#\{\ell:p_\ell\le qk/m\}$, and $N_j(k)$ the same count with $p_j:=0$; so $k^*=\max\{k:N(k)\ge k\}$,
$R_j=\max\{k:N_j(k)\ge k\}$, $N\le N_j$, and $|\mathcal R|=N(k^*)=k^*$ when $k^*\ge1$. If $p_j\le qk/m$, then
$N(k)=N_j(k)$.
\begin{itemize}[leftmargin=*,itemsep=1pt]
\item If $p_j\le qR_j/m$, then $N(R_j)=N_j(R_j)\ge R_j$, so $k^*\ge R_j$. For any $k>R_j$, $N(k)\le N_j(k)<k$, so
$k^*=R_j$. Since $p_j\le qk^*/m$, $j\in\mathcal R$ and $|\mathcal R|=R_j$.
\item If $j\in\mathcal R$, then $p_j\le qk^*/m$, so $N_j(k^*)=N(k^*)\ge k^*$ and $R_j\ge k^*$. For $k\ge k^*$,
$p_j\le qk/m$, so $N(k)=N_j(k)$, and therefore $R_j=k^*$.\qedhere
\end{itemize}
\end{proof}

\begin{lemma}[Conditional law of the test point; weighted exchangeability]\label{lem:A}
Fix $j$ and assume \ref{ass:A1} and \ref{ass:A2} for this $j$, with $\mathcal X\times\mathcal Y$ a standard Borel space.
In this lemma and its proof, relabel the test point as $Z_{n+1}:=Z_{n+j}$, and write $S:=S_j$ (the multiset, that is,
the empirical measure $\sum_{i=1}^{n+1}\delta_{Z_i}$), $H:=H_j$ and $U:=U_j$. Then for every bounded measurable $f$,
\[
\E\big[f(Z_{n+1},S,H,U)\,\big|\,S,H\big]\;=\;\sum_{k=1}^{n+1}\pi_k\int_0^1 f(z_k,S,H,u)\,du,\qquad
\pi_k:=\frac{w(z_k)}{\sum_{l=1}^{n+1}w(z_l)},
\]
where $z_1,\dots,z_{n+1}$ is any enumeration of $S$ (with repeats).
\end{lemma}

The right-hand side is a symmetric function of $(z_1,\dots,z_{n+1})$, so it is $\sigma(S,H)$-measurable. Symmetric
implies $\sigma(S)$-measurable: take a Borel isomorphism of the space onto a Borel subset of $[0,1]$ (Kuratowski's isomorphism theorem for standard Borel
spaces; see \citealp{kechris1995classical}); a symmetric Borel function $f$ then
equals $\tilde f\circ(\text{order statistics of the images})$, and the order statistics are functions of $S$.

\begin{proof}
Let $P^{\otimes(n+1)}$ be the law under which all $n+1$ points are i.i.d.\ $P$, and let $\bar w:=\E_P w$. By
\ref{ass:A1} and the independence in \ref{ass:A2}(i), the true law of $(Z_1,\dots,Z_{n+1})$ has density
$w(z_{n+1})/\bar w$ with respect to $P^{\otimes(n+1)}$ (here $w(z):=w(x)$ for $z=(x,y)$).

\emph{First part.} Take a bounded measurable $\chi(z,S)$. For any bounded symmetric $g(S)$,
\[
\E[\chi(Z_{n+1},S)g(S)]
=\E_{P^{\otimes}}\Big[\frac{w(Z_{n+1})}{\bar w}\chi(Z_{n+1},S)g(S)\Big]
=\frac1{n+1}\sum_{k=1}^{n+1}\E_{P^{\otimes}}\Big[\frac{w(Z_k)}{\bar w}\chi(Z_k,S)g(S)\Big],
\]
by exchangeability under $P^{\otimes(n+1)}$. Write $\Sigma w:=\sum_l w(Z_l)$ and
$\bar\chi(S):=\sum_k\big(w(Z_k)/\Sigma w\big)\chi(Z_k,S)$, which is symmetric. The last display equals
\[
\E_{P^{\otimes}}\Big[\frac{\Sigma w}{(n+1)\bar w}\,\bar\chi(S)g(S)\Big].
\]
Applying the same two equalities to the symmetric function $\bar\chi(S)g(S)$ in place of $\chi\cdot g$ gives
\[
\E[\bar\chi(S)g(S)]=\E_{P^{\otimes}}\Big[\frac{\Sigma w}{(n+1)\bar w}\,\bar\chi(S)g(S)\Big].
\]
Hence $\E[\chi(Z_{n+1},S)g(S)]=\E[\bar\chi(S)g(S)]$ for all bounded symmetric $g$, which means
$\E[\chi(Z_{n+1},S)\mid S]=\bar\chi(S)$.

\emph{Second part.} By \ref{ass:A2} the law of $(Z_1,\dots,Z_{n+1},H,U)$ is a product: the law of the $Z$'s, times
$P_H$, times $\Unif(0,1)$. The test functions for conditioning on $(S,H)$ are bounded $g(S,H)$ with $g(\cdot,h)$
symmetric for each $h$.
\begin{itemize}[leftmargin=*,itemsep=1pt]
\item For fixed $(h,u)$, apply the first part to $\chi_{h,u}(z,S):=f(z,S,h,u)$ and the symmetric function $g(\cdot,h)$.
\item Then integrate over the law of $(H,U)$. Fubini applies because $f$ and $g$ are bounded and the law is a
product.\qedhere
\end{itemize}
\end{proof}

\begin{remark}[Relation to weighted exchangeability]
Lemma~\ref{lem:A} is the covariate-shift case of the weighted exchangeability of \citet{tibshirani2019conformal}: their
Lemma~2 (\S3.2) states that independent draws $Z_i\sim P_i$ are weighted exchangeable with weight functions
$w_1\equiv1$ and $w_i=dP_i/dP_1$. Their proof of Lemma~3 (\S3.3) conditions on the event that the multiset equals
$\{z_1,\dots,z_{n+1}\}$ and assumes, for simplicity, that the scores are almost surely distinct, ``but the result holds in the general case as
well''. Their \S3.5 reduces the weights to $w(x_i)/\sum_jw(x_j)$, which is our $\pi_k$. Lemma~\ref{lem:A}
writes the tie and duplicate case out in full. We need that case for heavy covariate cells with point-mass scores.
\end{remark}

\begin{example}[Lemma~\ref{lem:A} with a duplicated atom]
Take $n=2$, $P$ uniform on $\{x,y\}$, $w(x)=1$ and $w(y)=3$, so $Q(x)=1/4$ and $Q(y)=3/4$. Condition on the multiset
$\{x,y,y\}$. It arises from $(Z_1,Z_2,Z_3)=(x,y,y)$ with probability $\tfrac12\cdot\tfrac12\cdot\tfrac34=3/16$ (test
point $y$), from $(y,x,y)$ with probability $3/16$ (test point $y$), and from $(y,y,x)$ with probability
$\tfrac12\cdot\tfrac12\cdot\tfrac14=1/16$ (test point $x$). So $\PP(\text{test}=x\mid\text{multiset})=1/7$. The lemma
gives $\pi_x=1/(1+3+3)=1/7$, and the two copies of $y$ get $3/7$ each, $6/7$ in total. Under $P^{\otimes(n+1)}$ all
points are exchangeable, and the true law reweights each configuration by the weight of whichever point is the test
point; copies of an atom are interchangeable, so a sum over an enumeration with repeats is all that is needed.
\end{example}

\subsection{Proof of the envelope theorem}\label{sec:E-proof}

\begin{proof}[Proof of Theorem~\ref{thm:E-env}]
\emph{Step 1 (leave-one-out).} Let $R_j:=R^0(p^{\obs}_{-j})$, the BH count when $p_j$ is replaced by $0$. By
Lemma~\ref{lem:LOO}, $j\in\mathcal R$ iff $p_j\le qR_j/m$, and on that event $|\mathcal R|=R_j$. So
\[
\FDR=\sum_{j=1}^m\E\Big[\frac{\one\{Y_{n+j}\le c_{n+j}\}\,\one\{p_j\le qR_j/m\}}{R_j}\Big].
\]

\emph{Step 2 (oracle p-value).} Let $p^*_j$ be eq.~(4) with $\hat V_{n+j}$ replaced by $V_{n+j}$ (same $U_j$). On
$\{Y_{n+j}\le c_{n+j}\}$, \ref{ass:A3} gives $V_{n+j}\le\hat V_{n+j}$.
\begin{itemize}[leftmargin=*,itemsep=1pt]
\item If $V_{n+j}=\hat V_{n+j}$, then $p_j=p^*_j$.
\item If $V_{n+j}<\hat V_{n+j}$, write $L^*$ and $T^*$ for the calibration weight strictly below and tied with
$V_{n+j}$, and $\hat L$ and $\hat T$ for the same at $\hat V_{n+j}$. Let $W:=\sum_{i\le n}w_i$. Then
$\hat L\ge L^*+T^*$, and
\[
(W+w_{n+j})(p_j-p^*_j)\;\ge\;(L^*+T^*)+(w_{n+j}+\hat T)U_j-L^*-(w_{n+j}+T^*)U_j\;=\;T^*(1-U_j)+\hat TU_j\;\ge\;0.
\]
\end{itemize}
$R_j$ does not involve $p_j$. The null indicator is used here only pointwise and is then dropped, so
\begin{equation}\label{eq:E-oracle}
\FDR\;\le\;\sum_{j=1}^m\E\Big[\frac{\one\{p^*_j\le qR_j/m\}}{R_j}\Big].
\end{equation}

\emph{Step 3 (conditional law of the test point).} Fix $j$ and abbreviate $Z_{n+1}:=Z_{n+j}$, $S:=S_j$, $H:=H_j$, as in
Lemma~\ref{lem:A}. For $\ell\ne j$, $p_\ell$ depends on the calibration multiset only symmetrically, and on
$(X_{n+\ell},c_{n+\ell},U_\ell)$. The calibration multiset of the observed p-values is $S\setminus\{Z_{n+1}\}$. So
$R_j=\varrho(Z_{n+1},S,H)$, where $\varrho(z_k,S,H):=R^0\big(p_{-j}$ computed from $S\setminus\{z_k\}$ and $H\big)$; this
is the $\varrho_j(e)$ of \cref{sec:E-statements} with $e=z_k$. Similarly,
$p^*_j=\Pi_<(Z_{n+1})+\Pi_=(Z_{n+1})U_j$, where for $z\in S$, $\Pi_<(z)$ and $\Pi_=(z)$ are the $\pi$-masses of the
elements of $S$ with score strictly below and equal to $V(z)$; the tied mass includes $z$ itself.

Apply Lemma~\ref{lem:A} to $f=\one\{p^*_j\le qR_j/m\}/R_j$. Let the tie groups be $G_1\prec G_2\prec\cdots$, with
masses $\Pi_G>0$ (positive because $w\ge1$). They define consecutive half-open intervals
$I_G=[\Pi_{<G},\Pi_{<G}+\Pi_G)$, which partition $[0,1)$. For $u\in[0,1)$ write $G(u)$ for the group with
$u\in I_G$. Substituting $u'=\Pi_<(z_k)+\Pi_=(z_k)u$ group by group gives
\begin{equation}\label{eq:E-condlaw}
\E\Big[\frac{\one\{p^*_j\le qR_j/m\}}{R_j}\,\Big|\,S,H\Big]
=\int_0^1\sum_{z_k\in G(u)}\frac{\pi_k}{\Pi_{G(u)}}\,\frac{\one\{u\le q\varrho_k/m\}}{\varrho_k}\,du,
\end{equation}
with $\varrho_k:=\varrho(z_k,S,H)$. The point $u=1$ has measure zero.

\emph{Step 4 (envelopes).} Set $\bar\varrho(u):=\max\{\varrho_k:z_k\succeq G(u)\}$ and
$\underline\varrho(u):=\min\{\varrho_k:z_k\preceq G(u)\}$.
\begin{itemize}[leftmargin=*,itemsep=1pt]
\item Both are non-increasing step functions on $[0,1)$, constant on each $I_G$: as $u$ grows, $\bar\varrho$ is a
maximum over a shrinking set and $\underline\varrho$ is a minimum over a growing set.
\item Every $z_k\in G(u)$ satisfies both $z_k\succeq G(u)$ and $z_k\preceq G(u)$, so
$\underline\varrho(u)\le\varrho_k\le\bar\varrho(u)$.
\end{itemize}
The weights $\pi_k/\Pi_{G(u)}$ over $z_k\in G(u)$ sum to $1$, so the integrand in \eqref{eq:E-condlaw} is at most
$\one\{u\le q\bar\varrho(u)/m\}/\underline\varrho(u)$.

\emph{Step 5 (monotone integral).}
\begin{itemize}[leftmargin=*,itemsep=1pt]
\item $\mathcal A:=\{u\in[0,1):u\le q\bar\varrho(u)/m\}$ is downward-closed: if $u\in\mathcal A$ and $u'<u$, then
$u'<u\le q\bar\varrho(u)/m\le q\bar\varrho(u')/m$. So $\mathcal A$ is an interval with left end $0$. Let
$u^*:=\sup\mathcal A$.
\item $\bar\varrho\ge1$, so $[0,\min(q/m,1))\subset\mathcal A$ and $u^*\ge\min(q/m,1)>0$.
\item Since the $I_G$ are half-open and finitely many, there is $\epsilon>0$ with $(u^*-\epsilon,u^*)$ contained in a
single interval $I_{G^*}$. $G^*$ is the group just left of $u^*$: if $u^*$ is interior to some $I_G$, then $G^*=G$; if
$u^*$ is a boundary point, $G^*$ is the group to its left; if $u^*=1$, $G^*$ is the top group. $G^*$ is not in general
$G(u^*)$. (Using $G(u^*)$ would break the argument: in an example with $n=2$, $m=2$ and $u^*$ on a group boundary, the
right-hand group's $\bar\varrho$ violates $u^*\le q\bar\varrho/m$.)
\item On $(u^*-\epsilon,u^*)$, $\bar\varrho=\bar\varrho_{G^*}:=\max\{\varrho_k:z_k\succeq G^*\}$ and
$\underline\varrho=\underline\varrho_{G^*}:=\min\{\varrho_k:z_k\preceq G^*\}$.
\item Letting $u\uparrow u^*$ inside $\mathcal A$ in $u\le q\bar\varrho(u)/m$ gives $u^*\le q\bar\varrho_{G^*}/m$.
\item $\underline\varrho$ is non-increasing, so $\underline\varrho(u)\ge\underline\varrho_{G^*}$ for all $u<u^*$. Hence
\[
\int_{\mathcal A}\frac{du}{\underline\varrho(u)}\;\le\;\frac{u^*}{\underline\varrho_{G^*}}\;\le\;
\frac qm\,\frac{\bar\varrho_{G^*}}{\underline\varrho_{G^*}}\;\le\;\frac qm\,\Gamma^*_j .
\]
\end{itemize}
$\Gamma^*_j$ is a symmetric measurable function of $(S,H)$. Taking expectations in \eqref{eq:E-condlaw} and summing over
$j$ in \eqref{eq:E-oracle} gives $\FDR\le\frac qm\sum_j\E[\Gamma^*_j]$.

\emph{Step 6 (monotonicity defect).} Fix $\ell\ne j$. Let $N_\ell$ and $W_\ell$ be the numerator and denominator of
$p_\ell[\,\cdot\,]$ when the calibration multiset is all of $S_j$, and let $\bar p_\ell:=N_\ell/W_\ell$. Removing $e$
subtracts $w_e\theta_e$ from $N_\ell$ and $w_e$ from $W_\ell$, where $\theta_e=1$, $U_\ell$ or $0$ according as
$V_e<\hat V_{n+\ell}$, $V_e=\hat V_{n+\ell}$ or $V_e>\hat V_{n+\ell}$. Hence
\[
p_\ell[S_j\setminus\{e\}]=\frac{N_\ell-w_e\theta_e}{W_\ell-w_e}=\bar p_\ell+h(w_e,\theta_e),\qquad
h(w,\theta):=\frac{w(\bar p_\ell-\theta)}{W_\ell-w}.
\]
$\theta_e$ is non-increasing in $V_e$ (since $0\le U_\ell\le1$), and $h$ is decreasing in $\theta$. So for $e\preceq e'$
we have $\theta_{e'}\le\theta_e$ and, writing $p_\ell(e)$ for $p_\ell[S_j\setminus\{e\}]$,
\[
p_\ell(e')-p_\ell(e)=h(w_{e'},\theta_{e'})-h(w_e,\theta_e)\;\ge\;h(w_{e'},\theta_e)-h(w_e,\theta_e)
=(\bar p_\ell-\theta_e)\big(\phi_\ell(w_{e'})-\phi_\ell(w_e)\big).
\]
$|\bar p_\ell-\theta_e|\le1$ and $\phi_\ell$ is increasing on $[w_{\min},w_{\max}]$, so $p_\ell(e')\ge p_\ell(e)-\delta_j$.
$R^0$ is non-increasing in each coordinate, so for any tie group $G$ and any $e_0\in G$:
\begin{itemize}[leftmargin=*,itemsep=1pt]
\item every $e\succeq G$ has $e\succeq e_0$, so $\varrho_j(e)\le R^0(p_{-j}(e_0)-\delta_j)$;
\item every $e\preceq G$ has $e\preceq e_0$, so $p_{-j}(e)\le p_{-j}(e_0)+\delta_j$ and
$\varrho_j(e)\ge R^0(p_{-j}(e_0)+\delta_j)$.
\end{itemize}
Apply this to the maximising group in the definition of $\Gamma^*_j$. Hence $\Gamma^*_j\le\Gamma_j$.

\emph{Step 7 (deterministic $\delta$).} Write $a:=w_{\max}$, $b:=w_{\min}$ and
$\Delta(b,W_\ell):=a/(W_\ell-a)-b/(W_\ell-b)$, so $\delta_j=\max_{\ell\ne j}\Delta(b,W_\ell)$.
\begin{itemize}[leftmargin=*,itemsep=1pt]
\item $W_\ell\ge a+nb+1$: one element of $S_j$ has weight $a$, the other $n$ have weight $\ge b$, and $w_{n+\ell}\ge1$.
\item $\partial\Delta/\partial W_\ell=-a/(W_\ell-a)^2+b/(W_\ell-b)^2\le0$, because $a\ge b$ and $W_\ell-a\le W_\ell-b$. So
$\Delta$ is largest at $W_\ell=a+nb+1$, where it equals $\Delta_0(b):=a/(nb+1)-b/(a+(n-1)b+1)$.
\item $\Delta_0'(b)=-an/(nb+1)^2-(a+1)/(a+(n-1)b+1)^2<0$, so $\Delta_0$ is largest at $b=1$:
$\Delta_0(1)=(a-1)(a+n+1)/((n+1)(a+n))$.
\item $\Delta_0(1)=(a-1)\big(1+1/(a+n)\big)/(n+1)$. It is increasing in $a$, because
$\frac{d}{da}\big[(a-1)(a+n+1)/(a+n)\big]=1+(n+1)/(a+n)^2>0$. So $a\le B$ gives $\delta_j\le\delta_n(B)$.
\end{itemize}

Finally, if $w$ is constant then $w_{\max}=w_{\min}$, so $\delta_j=0$, every ratio in $\Gamma_j$ equals $1$, and
$\FDR\le q$.
\end{proof}

\begin{proof}[Proof of Corollary~\ref{cor:E-computable}]
(Step 8.) The observed p-values of the other units use the calibration multiset $S_j\setminus\{Z_{n+j}\}$, so
$p^{\obs}_{-j}=p_{-j}(Z_{n+j})$. From Step 6, $p_\ell(e)=\bar p_\ell+w_e(\bar p_\ell-\theta_e)/(W_\ell-w_e)$, with
$\bar p_\ell-\theta_e\in[\bar p_\ell-1,\bar p_\ell]$. So
\[
\bar p_\ell-(1-\bar p_\ell)\,\phi_\ell(w_{\max})\;\le\;p_\ell(e)\;\le\;\bar p_\ell+\bar p_\ell\,\phi_\ell(w_{\max})
\qquad\text{for every }e\in S_j .
\]
Hence $|p_\ell(e)-p^{\obs}_\ell|\le\phi_\ell(w_{\max})\le\bar\tau_j$ for all $e\in S_j$ and $\ell\ne j$. By monotonicity of
$R^0$, every $\varrho_j(e)$ lies in $\big[R^0(p^{\obs}_{-j}+\bar\tau_j),\,R^0(p^{\obs}_{-j}-\bar\tau_j)\big]$. So
\[
\Gamma^*_j\;\le\;\frac{\max_e\varrho_j(e)}{\min_e\varrho_j(e)}\;\le\;\tilde\Gamma_j\;\le\;\hat\Gamma_j,
\]
where the last step uses $\eta_j\ge\bar\tau_j$ and the monotonicity of $R^0$. (Going through $\Gamma_j$ instead costs an
extra shift $\delta_j$ for nothing; that route gives $\Gamma_j\le\hat\Gamma_j$, since
$p_{-j}(e_0)\mp\delta_j$ lies within $\eta_j$ of $p^{\obs}_{-j}$ in each coordinate.) The FDR bounds follow from
Theorem~\ref{thm:E-env}.

Every quantity in $\hat\Gamma_j$ is observed: the weights of $S_j$ are $w(X_1),\dots,w(X_n),w(X_{n+j})$; $W_\ell$ also
uses $w(X_{n+\ell})$; and $p^{\obs}_{-j}$ are the reported p-values. For the shift bounds:
\begin{itemize}[leftmargin=*,itemsep=1pt]
\item $\phi_\ell(w_{\max})=w_{\max}/(W_\ell-w_{\max})\le w_{\max}/(\sum_{S_j}w-w_{\max})=\hat\tau_j/(1-\hat\tau_j)$,
which gives (i);
\item $W_\ell-w_{\max}\ge n\cdot1+1$, since $w\ge1$, and $w_{\max}\le B$, so $\bar\tau_j\le B/(n+1)$, which is (ii);
\item $\delta_j\le\max_\ell\phi_\ell(w_{\max})=\bar\tau_j$, so $\eta_j\le2\bar\tau_j\le2\hat\tau_j/(1-\hat\tau_j)$, which
is (iii).\qedhere
\end{itemize}
\end{proof}

\begin{proof}[Proof of Proposition~\ref{prop:E-sup}]
In $S_j$ only the score of $Z_{n+j}$ is unobserved; its covariate, hence its weight, is observed. For $e\ne Z_{n+j}$,
$p_\ell[S_j\setminus\{e\}]$ depends on $v=V_{n+j}$ only through $\one\{v<\hat V_{n+\ell}\}$ and
$\one\{v=\hat V_{n+\ell}\}$ (for $e=Z_{n+j}$ it does not depend on $v$), and the order $\preceq$ on $S_j$ depends on $v$ only through its comparisons with the
calibration scores. So $\Gamma^*_j(v)$ is constant as $v$ ranges over each of the $r$ distinct values and each of the
$r+1$ open gaps between them: at most $2r+1$ cases. Taking $v=V_{n+j}$ gives $\Gamma^*_j\le\Gamma^{\sup}_j$. The bound
$\Gamma^*_j(v)\le\tilde\Gamma_j$ of Step 8 holds for every $v$, and $\tilde\Gamma_j$ does not depend on $v$; hence
$\Gamma^{\sup}_j\le\tilde\Gamma_j$.
\end{proof}

\begin{proof}[Proof of Theorem~\ref{thm:envelope}]
Assume \ref{ass:A1}--\ref{ass:A3}. The first inequality $\FDR\le\frac qm\sum_j\E\,\Gamma^*_j$ is
Theorem~\ref{thm:E-env}. The pointwise bounds $\Gamma^*_j\le\Gamma^{\sup}_j\le\tilde\Gamma_j$ are
Proposition~\ref{prop:E-sup}, with $\tilde\Gamma_j$ as in Corollary~\ref{cor:E-computable} (there $p_{-j}$ is written
$p^{\obs}_{-j}$). The bound $\bar\tau_j\le B/(n+1)$ is Corollary~\ref{cor:E-computable}(ii), and computability of
$\tilde\Gamma_j$ and $\Gamma^{\sup}_j$ is stated there and in Proposition~\ref{prop:E-sup}.
\end{proof}

\begin{example}[Size of the shifts]
Take $n=4$ calibration points with weights $(1,1,1,2)$ and a test unit $j$ with weight $1$, so $\sum_{S_j}w=6$ and
$w_{\max}=2$. Another test unit $\ell$ with weight $1$ has $W_\ell=7$, so $\phi_\ell(2)=2/5$ and $\phi_\ell(1)=1/6$. Then
$\delta_j=2/5-1/6=7/30$, $\bar\tau_j=2/5$ and $\eta_j=19/30$; the bound $2\hat\tau_j/(1-\hat\tau_j)$ with
$\hat\tau_j=1/3$ gives $1\ge19/30$. With such a small $n$, $\eta_j$ is large and $\hat\Gamma_j$ is uninformative;
$\tilde\Gamma_j$ uses the shift $\bar\tau_j=2/5$. Heuristically, $\bar\tau_j$ is about the largest weight share, so
$\tilde\Gamma_j\approx1$ when $m\bar\tau_j\ll q$.
\end{example}

\subsection{Remarks: what is computable, and outlier detection}\label{sec:E-remarks}

\begin{remark}[The deterministic shift]
$\delta_n(B)\approx(B-1)/(n+1)$ when $B\ll n$. When $B\ge n+2$ the bound is at least about $1$; this is the Poisson
regime, where the excess is $O(1)$.
\end{remark}

\begin{remark}[What can be computed]\label{rem:E-computable}
$\Gamma^*_j$ and $\Gamma_j$ are \emph{not} computable in the selection setting: $S_j$ contains $Z_{n+j}$, and every
$\varrho_j(e)$ with $e$ a calibration point uses $V(X_{n+j},Y_{n+j})$, with $Y_{n+j}$ unobserved.
\begin{itemize}[leftmargin=*,itemsep=1pt]
\item In outlier detection, $Z_{n+j}$ is the observed test point, so $\Gamma^*_j$ is computable there, provided the
randomisations $U_\ell$ are recorded.
\item $\delta_j$, $\Gamma^{\sup}_j$, $\tilde\Gamma_j$ and $\hat\Gamma_j$ use only covariates, reported p-values and
recorded $U$'s, so they are computable in both settings.
\item All of these bound an \emph{expectation}. The realised value $\frac qm\sum_j\tilde\Gamma_j$ is an observable
random variable whose mean is at least the FDR. It is not a high-probability bound on the realised FDP.
\item At small $n$, $\frac qm\sum_j\E[\hat\Gamma_j]$ often exceeds $1$, which is vacuous; values $1.27$, $1.49$ and
$2.29$ occur in exact examples.
\end{itemize}
\end{remark}

\begin{proposition}[Outlier detection]\label{prop:E-outlier}
In outlier detection, let $\mathcal H_0$ be the set of inliers: for $j\in\mathcal H_0$, $Z_{n+j}\sim Q$ as in
\ref{ass:A1}, while the outliers need not follow $Q$. Put the outliers' data ($X$, score and $U$) for $\ell\ne j$ into
$H_j$, and require \ref{ass:A2} as stated for every $j\in\mathcal H_0$, so that $H_j$ is independent of the calibration
data and $Z_{n+j}$. Then
\[
\FDR\;\le\;\frac qm\sum_{j\in\mathcal H_0}\E[\Gamma^*_j]\;\le\;\frac qm\sum_{j}\E[\Gamma^*_j],
\]
and likewise with $\tilde\Gamma_j$ and $\hat\Gamma_j$ in place of $\Gamma^*_j$.
\end{proposition}

\begin{proof}
\ref{ass:A1} puts every test unit under $Q$, but outliers do not follow $Q$. Restrict the FDR sum of Step 1 to the
inliers $j\in\mathcal H_0$. For these $p_j=p^*_j$, so Step 2 is trivial. With the outliers' data in $H_j$ and
\ref{ass:A2} as stated, Steps 3--8 go through unchanged.
\end{proof}

\begin{remark}[Heuristic, not a result]
In a global-null configuration with a single heavy calibration unit of weight $B$ and $n$ light units,
$\bar\tau_j\approx B/n$. Then $\tilde\Gamma_j-1$ is driven by how many other p-values fall within $B/n$ of a BH
threshold. That suggests order $mB/n$ in the linear regime, matching Theorem~\ref{thm:T2}. Explicit bounds of this
kind, under anti-concentration and sub-uniformity conditions, are proved in \cref{sec:E-explicit,sec:E-global}.

\end{remark}

\subsection{Explicit bounds from the envelope}\label{sec:E-explicit}

\paragraph{Notation.} $\tau_B:=B/(n+1)$, which bounds the shift $\bar\tau_j$ of $\tilde\Gamma_j$
(Corollary~\ref{cor:E-computable}(ii)); $I_k:=(t_k-\tau_B,t_k+\tau_B]$; $x^{(j)}:=p^{\obs}_{-j}$, the other $m-1$
reported p-values, with $x^{(j)}_{(i)}$ (or $x_{(i)}$) the $i$-th smallest.

\begin{lemma}[Window lemma]\label{lem:E-D}
For $x\in\R^{m-1}$ and $s\ge0$, let $r^-:=R^0(x-s)$ and $r^+:=R^0(x+s)$. Then
\[
\frac{r^-}{r^+}-1\;\le\;(r^--1)\,\one\{r^->r^+\}\;\le\;\sum_{k=2}^m(k-1)\,\one\{x_{(k-1)}\in(t_k-s,t_k+s]\}.
\]
\end{lemma}

\begin{proof}
\begin{itemize}[leftmargin=*,itemsep=1pt]
\item If $r^-=r^+$, the left side is $0$.
\item Otherwise $r^->r^+\ge1$, so $r^-\ge2$ and the ratio minus $1$ is at most $r^--1$.
\item Write $R^0(y)=\max\{k:1+\#\{\ell:y_\ell\le t_k\}\ge k\}$.
\begin{itemize}[leftmargin=*]
\item Since $r^-$ is feasible for $x-s$, $\#\{\ell:x_\ell\le t_{r^-}+s\}\ge r^--1$, so $x_{(r^--1)}\le t_{r^-}+s$.
\item Since $r^->r^+$, $r^-$ is not feasible for $x+s$, so $\#\{\ell:x_\ell\le t_{r^-}-s\}\le r^--2$, so
$x_{(r^--1)}>t_{r^-}-s$.
\end{itemize}
\item So the indicator for $k=r^-$ equals $1$.\qedhere
\end{itemize}
\end{proof}

\paragraph{Monotonicity in $s$.} $R^0(x-s)/R^0(x+s)$ is non-decreasing in $s$, since $R^0$ is non-increasing in each
coordinate. Since $\bar\tau_j\le\tau_B$, we get
\[
\tilde\Gamma_j\;\le\;\Upsilon_j:=\frac{R^0(x^{(j)}-\tau_B)}{R^0(x^{(j)}+\tau_B)} .
\]

\begin{theorem}[A bound for every configuration]\label{thm:E-E1}
Under \ref{ass:A1}--\ref{ass:A3}, including non-null test units, and in outlier detection
(Proposition~\ref{prop:E-outlier}),
\[
\FDR\;\le\;\frac qm\sum_j\E[\tilde\Gamma_j]\;\le\;q+\frac qm\sum_{j=1}^m\sum_{k=2}^m(k-1)\,\PP\big(x^{(j)}_{(k-1)}\in I_k\big).
\]
\end{theorem}

\begin{proof}
Use Corollary~\ref{cor:E-computable} (or Proposition~\ref{prop:E-outlier}), then $\tilde\Gamma_j\le\Upsilon_j$ and
Lemma~\ref{lem:E-D} with $s=\tau_B$, then take expectations.
\end{proof}

\begin{corollary}[Anti-concentration]\label{cor:E-E2}
Suppose that for every test unit $\ell$ and every interval $I$ of length $2\tau_B$ inside $[t_2-\tau_B,t_m+\tau_B]$,
$\PP(p_\ell\in I)\le L\,|I|$. Then, under the assumptions of Theorem~\ref{thm:E-E1},
\[
\FDR\;\le\;q\Big[1+L\,m(m-1)^2\,\frac{B}{n+1}\Big].
\]
\end{corollary}

\begin{proof}
$\{x_{(k-1)}\in I_k\}$ implies that some $x_\ell\in I_k$, so $\PP(x_{(k-1)}\in I_k)\le(m-1)\cdot2L\tau_B$. Also
$\sum_{k=2}^m(k-1)=m(m-1)/2$. Insert both into Theorem~\ref{thm:E-E1}.
\end{proof}

\paragraph{When $L=1$ holds exactly.} An oracle null p-value ($\hat V=V$) is exactly $\Unif(0,1)$ marginally
(Lemma~\ref{lem:A}; this is Lemma~\ref{lem:U}), so $\PP(p_\ell\in I)=|I|$. So in every global-null configuration with
oracle thresholds (every threshold equal to $Y$, or outlier detection with all units inliers),
\[
\FDR\;\le\;q\Big[1+m(m-1)^2\,\frac{B}{n+1}\Big]\qquad\text{for all }n,B.
\]
This bound is fully explicit, but cubic in $m$, against the linear lower bound of Theorem~\ref{thm:T2}.

\paragraph{Conditional sub-uniformity.} Assume, as \citet{jin2026weighted} do, that the test tuples
$(X_{n+\ell},Y_{n+\ell},c_{n+\ell})$ and the $U_\ell$ are i.i.d.\ across units and independent of the calibration data
$D$ (this implies \ref{ass:A2}). Then, given $D$, the coordinates of $x^{(j)}$ are i.i.d.\ with some CDF $G_D$: the
conditional law of one unit's reported p-value, mixing null and non-null types.

\begin{theorem}[Linear in $m$ under conditional sub-uniformity]\label{thm:E-E3}
Assume \ref{ass:A1}, \ref{ass:A3} and the i.i.d.\ structure just described. Fix $\rho\ge1$ and define
\begin{itemize}[leftmargin=*,itemsep=1pt]
\item $\mathcal G:=\{D:G_D(t_k+\tau_B)\le\rho(t_k+\tau_B)\text{ for }k=2,\dots,m\}$;
\item $L:=\max_{2\le k\le m}\PP(p_\ell\in I_k)/|I_k|$, the marginal anti-concentration at the windows;
\item $\displaystyle\beta_m(\rho):=\sum_{k=2}^m(k-1)\,\PP\Big(\Bin\big(m-2,\min\{1,\rho(t_k+\tau_B)\}\big)\ge k-2\Big)$.
\end{itemize}
Then for every $j$,
\[
\E[\tilde\Gamma_j]\;\le\;1+2L(m-1)\tau_B\,\beta_m(\rho)+(m-1)\,\PP(\mathcal G^c),
\]
and
\[
\FDR\;\le\;q\Big[1+2L(m-1)\,\beta_m(\rho)\,\frac{B}{n+1}+(m-1)\,\PP(\mathcal G^c)\Big].
\]
\end{theorem}

\begin{proof}
By $\tilde\Gamma_j\le\Upsilon_j$ and Lemma~\ref{lem:E-D}, $\Upsilon_j-1\le r^--1\le m-1$ always; use this on
$\mathcal G^c$. On $\mathcal G$:
\begin{itemize}[leftmargin=*,itemsep=1pt]
\item The event $\{x_{(k-1)}\in I_k\}$ needs some $x_\ell\in I_k$ and at least $k-2$ of the other $m-2$ coordinates
$\le t_k+\tau_B$.
\item Given $D$ the coordinates are i.i.d.\ $G_D$. By a union bound,
\[
\PP\big(x_{(k-1)}\in I_k\mid D\big)\;\le\;(m-1)\,G_D(I_k)\,\PP\big(\Bin(m-2,G_D(t_k+\tau_B))\ge k-2\big).
\]
\item On $\mathcal G$ the binomial factor is at most its value at success probability $\min\{1,\rho(t_k+\tau_B)\}$,
because the binomial tail is increasing in the success probability.
\item Take $\E[\,\cdot\,\one_{\mathcal G}]$ ($\mathcal G$ is $D$-measurable) and use
$\E[G_D(I_k)]=\PP(p_\ell\in I_k)\le2L\tau_B$.
\end{itemize}
Summing over $k$ with the weights $k-1$ of Lemma~\ref{lem:E-D} gives the bound on $\E[\tilde\Gamma_j]$; the FDR bound
follows from Corollary~\ref{cor:E-computable} with $\tau_B=B/(n+1)$.
\end{proof}

\begin{remark}[Size of $\beta_m(\rho)$]\label{rem:E-beta}
The values below are a numerical evaluation of the formula for $\beta_m(\rho)$, computed with $\tau_B=0$ (the limit
$n/(mB)\to\infty$):
\begin{center}
\begin{tabular}{@{}c ccc ccc ccc@{}}\toprule
 & \multicolumn{3}{c}{$\rho=1$} & \multicolumn{3}{c}{$\rho=1.5$} & \multicolumn{3}{c}{$\rho=2$}\\
\cmidrule(lr){2-4}\cmidrule(lr){5-7}\cmidrule(l){8-10}
$q$ & $m=10$ & $100$ & $1000$ & $10$ & $100$ & $1000$ & $10$ & $100$ & $1000$\\\midrule
$0.05$ & 1.26 & 1.33 & 1.34 & 1.41 & 1.53 & 1.55 & 1.57 & 1.76 & 1.78\\
$0.10$ & 1.57 & 1.76 & 1.78 & 1.95 & 2.32 & 2.36 & 2.39 & 3.04 & 3.12\\
$0.20$ & 2.39 & 3.04 & 3.12 & 3.53 & 5.30 & 5.56 & 5.13 & 9.54 & 10.3\\\bottomrule
\end{tabular}
\end{center}
With $\tau_B>0$, $\beta_m(\rho)$ grows with $m\tau_B$: at $q=0.1$, $\rho=1$, $B=3$, $n=1000$ it is $2.61$ ($m=100$),
$15.4$ ($m=1000$) and $638$ ($m=10^4$). \gap{We have only sketched an $m$-free majorant of $\beta_m(\rho)$. Since
$t_k+\tau_B\le(q+m\tau_B)k/m$, whenever $\rho(q+m\tau_B)<1$ a Chernoff bound on the binomial tails
(Lemma~\ref{lem:B-chernoff}) should bound $\beta_m(\rho)$ uniformly in $m$, but we have not written that step out.}
\end{remark}

\begin{remark}[The event $\mathcal G$ at $\rho=1$]
$\E[G_D(t)]=t$ in the global null, so at $\rho=1$ the event $G_D(t_k)>t_k$ has probability about $\tfrac12$ at each grid
point. A global-null simulation gives $\PP(\mathcal G^c)\approx0.78$ at $\rho=1$ for $n=2000$ and $n=20000$, and the
term $(m-1)\PP(\mathcal G^c)$ then makes the bound vacuous. The useful regime is $\rho=1+\varepsilon_n$ with
$\varepsilon_n\downarrow0$ slowly; \cref{sec:E-global} bounds $\PP(\mathcal G^c)$ for $\rho>1+\kappa$. In the global
null ($L=1$) the leading coefficient $2\beta_m(\rho)$ is about $3.6$ at $q=0.1$, $\rho=1$ ($\tau_B=0$,
Remark~\ref{rem:E-beta}), against the constant
$\tfrac12(1-q)=0.45$ of Theorem~\ref{thm:T2}: a factor of about $8$.
\end{remark}

\subsection{The global-null upper bound}\label{sec:E-global}

In this subsection every test unit is an oracle null ($p_\ell=p^*_\ell$: every threshold equal to $Y$, or outlier
detection with all units inliers), and the test units are i.i.d.\ and independent of $D$ as in
Theorem~\ref{thm:E-E3}. Then $G_D(t)=\PP(p^*\le t\mid D)=F^*_D(t)$. Recall $\kappa=B/n$ and
$\Bb=(\esssup w)/\E_P[w]\le B$.

\begin{lemma}[$\PP(\mathcal G^c)$, continuous scores]\label{lem:E-E6}
Assume in addition that the score $V$ has a continuous law under $P$ (hence under $Q$). Let $\rho>1+\kappa$ and
$0<\varepsilon<\varepsilon_{\max}:=(\rho-1-\kappa)/(\rho+1+\kappa)$. Then
\[
\PP(\mathcal G^c)\;\le\;\frac{e^{-2a}}{1-e^{-a}}+\exp\Big(-\frac{n\varepsilon^2}{2B(1+\varepsilon/3)}\Big),\qquad
a:=\frac{n\varepsilon^2\rho q}{2Bm(1+\varepsilon/3)} .
\]
\end{lemma}

\begin{proof}
Fix $b=t_k+\tau_B$ with $\rho b\le1$.

\emph{Step 1: from $G_D$ to the weighted empirical CDF.} Let $W:=\sum_{i\le n}w_i\ge n$ and
$\hat G(v):=\sum_iw_i\one\{V_i<v\}/W$. A test point with score $v$ and weight $w_z\le B$ has
$p=(W\hat G(v)+Uw_z)/(W+w_z)$. So $p\le b$ implies $\hat G(v)\le b(1+B/W)\le b':=b(1+\kappa)$.

\emph{Step 2: localising.} Let $v^*$ satisfy $Q_<(v^*):=Q(V<v^*)=\rho b$; it exists by continuity. Grid points with
$\rho b>1$ are trivially in $\mathcal G$. If $\hat G(v^*)>b'$, then every $v\ge v^*$ has $\hat G(v)>b'$ ($\hat G$ is
non-decreasing), so $G_D(b)\le Q_<(v^*)=\rho b$. Hence $\{G_D(b)>\rho b\}\subset\{\hat G(v^*)\le b'\}$.

\emph{Step 3: Bernstein.} Write $\hat G(v^*)=S_1/S_0$, with $S_1:=n^{-1}\sum_iw_i\one\{V_i<v^*\}$, of mean
$\mu_1=\rho b\,\E_Pw$ (this uses $dQ/dP=w/\E_Pw$ exactly), and $S_0:=n^{-1}\sum_iw_i$, of mean $\mu_0=\E_Pw\ge1$.
\begin{itemize}[leftmargin=*,itemsep=1pt]
\item If $S_1>(1-\varepsilon)\mu_1$ and $S_0<(1+\varepsilon)\mu_0$, then
$\hat G(v^*)>\rho b(1-\varepsilon)/(1+\varepsilon)\ge b'$, because $\varepsilon<\varepsilon_{\max}$.
\item The summands lie in $[0,B]$ (respectively $[1,B]$), with variance at most $B$ times their mean. The one-sided
Bernstein inequality, Lemma~\ref{lem:B1}, with range constant $B$ (applied to $\mu_1-w_i\one\{V_i<v^*\}$ for the lower
tail of $S_1$ and to $w_i-\mu_0$ for the upper tail of $S_0$), gives
\begin{gather*}
\PP\big(S_1\le(1-\varepsilon)\mu_1\big)\le\exp\Big(-\frac{n\varepsilon^2\mu_1}{2B(1+\varepsilon/3)}\Big)
\le\exp\Big(-\frac{n\varepsilon^2\rho qk}{2Bm(1+\varepsilon/3)}\Big),\\
\PP\big(S_0\ge(1+\varepsilon)\mu_0\big)\le\exp\Big(-\frac{n\varepsilon^2}{2B(1+\varepsilon/3)}\Big).
\end{gather*}
(Here $\mu_1\ge\rho b\ge\rho t_k=\rho qk/m$, since $\E_Pw\ge1$ and $b\ge t_k$; and $\mu_0\ge1$.)
\end{itemize}

\emph{Step 4: union.} The $S_0$ event does not depend on $k$. The union over $k=2,\dots,m$ of the $S_1$ events has
probability at most $\sum_{k\ge2}e^{-ak}=e^{-2a}/(1-e^{-a})$.
\end{proof}

\begin{remark}[Plugging Lemma~\ref{lem:E-E6} into Theorem~\ref{thm:E-E3}]
With $L=1$ (oracle nulls are exactly uniform), in the global null with continuous scores,
\[
\FDR\;\le\;q\Big[1+2(m-1)\beta_m(\rho)\frac{B}{n+1}+(m-1)\Big(\frac{e^{-2a}}{1-e^{-a}}
+e^{-n\varepsilon^2/(2B(1+\varepsilon/3))}\Big)\Big].
\]
This is fully explicit, and useful only when $n\gg Bm/(q\varepsilon^2)$. For example, at $q=0.1$, $m=10$, $B=2$,
$\rho=1.2$, $\varepsilon=0.09$: $(m-1)\PP(\mathcal G^c)\le0.089$ at $n=10^5$, and $<10^{-19}$ at $n=10^6$. A Monte Carlo
check at $n=10^5$ found no failure of $\mathcal G$, against the bound $\PP(\mathcal G^c)\le0.0099$: the bound is valid but
loose (the $\rho$ actually needed is about $1.07$, not $1.2$). The crude step $\Upsilon_j-1\le m-1$ on $\mathcal G^c$ is
what costs here. With score atoms, Step 2 needs the $U$-randomisation; that case is Corollary~\ref{cor:E-E6p}.
\end{remark}

\begin{corollary}[$\PP(\mathcal G^c)$ for every score law]\label{cor:E-E6p}
Assume \ref{ass:A1} and \ref{ass:A3}, with weights $1\le w\le B$, and $q\in(0,1)$; every test unit is an oracle null;
the test units are i.i.d.\ given $D$ (JC's independence); and $\rho>1+\kappa$. Put
\[
a':=\frac{nq(\rho-1-\kappa)^2}{2\Bb m(3\rho-2-2\kappa)} .
\]
Then, with no assumption on the score law (atoms allowed),
\[
\PP(\mathcal G^c)\;\le\;\frac{e^{-2a'}}{1-e^{-a'}},
\]
and therefore, by Theorem~\ref{thm:E-E3} with $L=1$,
\[
\FDR\;\le\;q\Big[1+2(m-1)\beta_m(\rho)\frac{B}{n+1}+(m-1)\frac{e^{-2a'}}{1-e^{-a'}}\Big].
\]
\end{corollary}

\begin{proof}
\begin{itemize}[leftmargin=*,itemsep=1pt]
\item In the global null with oracle nulls, the test p-values are the $p^*$, so $G_D=F^*_D$.
\item Steps 1--4 of the proof of the explicit bound on $\E[M_D]$ (Theorem~\ref{thm:upper}; Appendix~\ref{app:upper}),
namely the passage from $p^*$ to a weighted empirical CDF, its population version, the localisation in the
lexicographic order of (score, $U$), and the Bernstein step, are written there for $t=t_k$ but hold for any
$t\in(0,1)$. They handle score atoms through the lexicographic order and the $U$-randomisation, and they give, for
$b\in(0,1)$ and $d>0$,
\begin{equation}\label{eq:E-tail}
\PP\big(F^*_D(b)>b(1+\kappa)+d\big)\;\le\;\exp\Big(-\frac{nd^2}{2\Bb\,(b(1+\kappa)+3d)}\Big),
\end{equation}
with probability $0$ once $b(1+\kappa)+d\ge1$. We do not repeat these steps here.
\item Take $b:=b_k:=t_k+\tau_B$ and $d:=(\rho-1-\kappa)b_k>0$. Then $b_k(1+\kappa)+d=\rho b_k$, so
$\{G_D(b_k)>\rho b_k\}$ is the event on the left of \eqref{eq:E-tail} (it is empty if $\rho b_k\ge1$).
\begin{itemize}[leftmargin=*]
\item Since $b(1+\kappa)+3d=b_k(3\rho-2-2\kappa)$, the exponent is $n(\rho-1-\kappa)^2b_k/(2\Bb(3\rho-2-2\kappa))$.
\item Since $b_k\ge t_k=qk/m$, the exponent is at least $ka'$.
\end{itemize}
\item A union over $k=2,\dots,m$ gives $\PP(\mathcal G^c)\le\sum_{k\ge2}e^{-ka'}=e^{-2a'}/(1-e^{-a'})$.
\end{itemize}
The FDR bound is Theorem~\ref{thm:E-E3} with $L=1$: with oracle nulls, $p_\ell=p^*_\ell$ is exactly $\Unif(0,1)$ for
every score law (Lemma~\ref{lem:U}), so $\PP(p_\ell\in I_k)=|I_k|$.
\end{proof}

\begin{remark}[Comparison with Lemma~\ref{lem:E-E6}]
Corollary~\ref{cor:E-E6p} has no $S_0$ term and uses $\Bb\le B$. Its exponent is better than that of
Lemma~\ref{lem:E-E6} only for $\rho\lesssim2.5$ (with $\kappa=0$ and $\Bb=B$): at $\rho=3$ the ratio $a'/a$ is $0.89$,
and at $\rho=10$ it is $0.55$. The regime that matters is $\rho$ close to $1$, where Corollary~\ref{cor:E-E6p} is better:
at $\rho=1.2$ and $\kappa\approx0$, $a'=0.0125\,nq/(\Bb m)$, against $a\approx0.0048\,nq/(Bm)$. For $q=0.1$, $m=10$,
$B=\Bb=2$, $\rho=1.2$ and $n=10^5$: $a'=6.25$, so $(m-1)\PP(\mathcal G^c)\le9e^{-12.5}/(1-e^{-6.25})=3.4\cdot10^{-5}$,
against $0.089$ from Lemma~\ref{lem:E-E6}.
\end{remark}

\begin{proof}[Proof of Theorem~\ref{thm:globalupper}]
Assume \ref{ass:A1} and \ref{ass:A3} with $1\le w\le B$, $q\in(0,1)$, test units i.i.d.\ and independent of $D$, every
test unit an oracle null (the global null), and $\rho>1+\kappa$. By Theorem~\ref{thm:E-E3},
\[
\FDR\;\le\;q\Big[1+2L(m-1)\beta_m(\rho)\frac{B}{n+1}+(m-1)\PP(\mathcal G^c)\Big].
\]
With oracle nulls, $p_\ell=p^*_\ell$ is exactly $\Unif(0,1)$ for every score law (Lemma~\ref{lem:U}; the
$U$-randomisation handles ties), so $\PP(p_\ell\in I_k)=|I_k|$ and $L=1$. Theorem~\ref{thm:E-E3} itself makes no
continuity assumption: its ingredients are Lemma~\ref{lem:E-D}, the envelope (which handles ties through
Lemma~\ref{lem:A}) and the monotonicity of binomial tails. Corollary~\ref{cor:E-E6p} gives
$\PP(\mathcal G^c)\le e^{-2a'}/(1-e^{-a'})$ with $a'$ as in the statement. Substituting gives the displayed bound, with
$\beta_m(\rho)$ explicit (Theorem~\ref{thm:E-E3}).

\emph{Fixed $m$.} The paragraph ``When $L=1$ holds exactly'' (Corollary~\ref{cor:E-E2} with $L=1$) gives
$\FDR-q\le q\,m(m-1)^2B/(n+1)$ for all $n$ and $B$, so at fixed $m$ the excess is $O(B/n)$.

\emph{Non-oracle nulls.} If every test unit is null but some have $c_{n+j}>Y_{n+j}$, then $p_j\ge p^*_j$ for every $j$
(the score is monotone, and the p-value of eq.~(4) is non-decreasing in the test score). Whenever BH rejects at least one
hypothesis with the $p_j$, it also does so with the smaller $p^*_j$, so the FWER, which is the FDR here, is at
most the FWER of the $p^*_j$. These are the p-values of the configuration with every threshold replaced by $Y$, an
oracle-null configuration with the same $P$, $Q$ and weights. Both bounds above therefore hold for every global-null
configuration.
\end{proof}

\begin{remark}[Linearity in $m$]\label{rem:E-linear}
The bound of Theorem~\ref{thm:globalupper} is linear in $m$ wherever $\beta_m(\rho)$ stays bounded in $m$. Numerically it
does at $\tau_B=0$ ($\beta_m(1)\approx1.78$ at $m=1000$, $q=0.1$; Remark~\ref{rem:E-beta}).
\gap{The $m$-free majorant of $\beta_m(\rho)$ is only a sketch (Remark~\ref{rem:E-beta}), and it also
needs $\rho(q+m\tau_B)<1$. So we have proved the bound with $\beta_m(\rho)$ explicit, but we have not proved that it is
linear in $m$: that rests on the sketch.} The bound is informative only once $n\gg\Bb m/(q(\rho-1)^2)$, and its constant is still about $8$ times
the constant $\tfrac12(1-q)$ of Theorem~\ref{thm:T2}.
\end{remark}

\providecommand{\Unif}{\mathrm{Unif}}

\section{Certificates for the counterexamples}\label{app:counter}

This appendix documents the two computations behind Theorem~\ref{thm:counter}. The proof of that theorem is
computer-assisted. In each instance the covariate takes finitely many values, the FDR is a finite sum of rational
numbers, and that sum was evaluated in exact rational arithmetic (no floating point at any stage), so the comparison with
$q$ is exact. What has to be trusted is that the programs implement the formulas below. For each instance we state the
configuration, the finite formula that was evaluated, and which independent programs were run on which instances.

\subsection{Common reductions}\label{sec:F-common}

\paragraph{Setting.} Calibration $(X_i,Y_i)_{i\le n}$ i.i.d.\ $P$; test tuples $(X_{n+j},Y_{n+j},c_{n+j})_{j\le m}$
i.i.d., with $(X_{n+j},Y_{n+j})\sim Q$, $dQ/dP(x,y)=w(x)/\E_P w$, independent of the calibration data $D$; tie-breaking
variables $U_j$ i.i.d.\ $\Unif[0,1]$, independent of everything else. Unit $j$ is null iff $Y_{n+j}\le c_{n+j}$. With
$\hat V_{n+j}:=V(X_{n+j},c_{n+j})$ and $V_i:=V(X_i,Y_i)$, the weighted conformal p-value is \citet{jin2026weighted}'s eq.~(4),
\[
p_j=\frac{\sum_{i=1}^n w(X_i)\one\{V_i<\hat V_{n+j}\}+\bigl(w(X_{n+j})+\sum_{i=1}^n w(X_i)\one\{V_i=\hat V_{n+j}\}\bigr)U_j}
{\sum_{i=1}^n w(X_i)+w(X_{n+j})}.
\]
BH at level $q$ rejects $\mathcal R=\{j:p_j\le t_{k^*}\}$, $k^*=\max\{k:\#\{j:p_j\le t_k\}\ge k\}$ ($\mathcal R=\emptyset$
if no such $k$), and $\FDR=\E\bigl[\#\{j\in\mathcal R:\ Y_{n+j}\le c_{n+j}\}/\max(1,|\mathcal R|)\bigr]$. Since $p_j$ is
unchanged when $w$ is multiplied by a positive constant, weights below are given up to a common factor, and the weight
ratio is $B=\max w/\min w$.

\paragraph{(a) Per-cell laws.} Let $S:=\sum_{i\le n}w(X_i)$. If the threshold is a function of the covariate,
$c_{n+j}=c(X_{n+j})$, then given $D$ and $X_{n+j}=x$, eq.~(4) is affine in $U_j$, so
\begin{equation}\label{eq:F-window}
p_j\mid D,\ X_{n+j}=x\ \sim\ \Unif\Bigl[\frac{S_<(x)}{S+w(x)},\ \frac{S_<(x)+w(x)+S_=(x)}{S+w(x)}\Bigr],
\end{equation}
where $S_<(x)$ and $S_=(x)$ are the total calibration weights with score below, respectively equal to, $\hat V(x)=V(x,c(x))$.
Write $G_x(t)=G_x(t;D)$ for the CDF of this uniform law.

\paragraph{(b) Conditional i.i.d.\ structure.} The pair $(p_j,\one\{j\text{ null}\})$ is a fixed function of $D$ and of the
$j$-th tuple $(X_{n+j},Y_{n+j},c_{n+j},U_j)$. These tuples are i.i.d.\ and independent of $D$, so given $D$ the pairs
$(p_j,\one\{j\text{ null}\})$, $j\le m$, are i.i.d. Write $F_D(t):=\PP(p_1\le t\mid D)$ and, as in
Lemma~\ref{lem:F1}, $A_k(D):=\PP(\text{unit 1 null},\ p_1\le t_k\mid D)$. All these quantities depend on $D$ only through
the vector of calibration counts of the support points of $(X,Y)$, which is multinomial; the sums over $D$ below are sums
over these count vectors.

\paragraph{(c) Two identities for $\E[\mathrm{FDP}\mid D]$.} For $M\ge0$ and $k\le m$ let
\begin{equation}\label{eq:F-phi}
\Phi_{M,k}(D):=\sum_{c}\binom{M}{c_{k+1},\dots,c_m,\,M-|c|}\prod_{j=k+1}^{m}\bigl(F_D(t_j)-F_D(t_{j-1})\bigr)^{c_j}
\bigl(1-F_D(t_m)\bigr)^{M-|c|},
\end{equation}
the sum running over $c=(c_{k+1},\dots,c_m)\in\mathbb Z_{\ge0}^{m-k}$ with $c_{k+1}+\dots+c_j\le j-k-1$ for every
$j\in\{k+1,\dots,m\}$, and $|c|=c_{k+1}+\dots+c_m$ ($\Phi_{0,k}:=1$). Given $D$, $\Phi_{M,k}(D)$ is the probability that
$M$ i.i.d.\ draws from the conditional law of $p_1$ all exceed $t_k$ and that, for every $j>k$, at most $j-k-1$ of them are
$\le t_j$; the multinomial sum assigns the draws to the bands $(t_{j-1},t_j]$, $j>k$, and to $(t_m,1]$. The engines
evaluate it by a dynamic programme over $j=k+1,\dots,m$ whose state is the running total $c_{k+1}+\dots+c_j$.

\emph{Leave-one-out form.} Let $R_j$ be the number of BH rejections when $p_j$ is replaced by $0$. By the leave-one-out
identity (Lemma~\ref{lem:LOO}), $j\in\mathcal R$ iff $p_j\le t_{R_j}$, and then $|\mathcal R|=R_j$; hence
$\mathrm{FDP}=\sum_{j\le m}\one\{j\text{ null}\}\one\{p_j\le t_{R_j}\}/R_j$. Given $D$, the $m$ summands have the same law
by (b), and $R_1$ is a function of $D$ and of the units $2,\dots,m$, hence independent of unit~1 given $D$. So, with
$r_k(D)=\PP(R_1=k\mid D)$ as in Lemma~\ref{lem:F1},
\begin{equation}\label{eq:F-loo}
\E[\mathrm{FDP}\mid D]=m\sum_{k=1}^m\frac1k\,A_k(D)\,r_k(D),\qquad
r_k(D)=\binom{m-1}{k-1}F_D(t_k)^{k-1}\,\Phi_{m-k,k}(D).
\end{equation}
For the second formula: $R_1=k$ iff at least $k$ of $(0,p_2,\dots,p_m)$ are $\le t_k$ and fewer than $j$ are $\le t_j$
for every $j>k$; taking $j=k+1$ forces exactly $k$ of them $\le t_k$, i.e.\ exactly $k-1$ of $p_2,\dots,p_m$, and the
other $m-k$ satisfy the event defining $\Phi_{m-k,k}$.

\emph{Cutoff form.} By the same counting, $\PP(k^*=k\mid D)=\binom mk F_D(t_k)^k\,\Phi_{m-k,k}(D)$ for $k\ge1$. Given
$D$ and $k^*=k$, the rejected set is $\{j:p_j\le t_k\}$, and the event $\{k^*=k\}$ depends on the rejected units only
through $\{p_j\le t_k\}$. So, by (b), each rejected unit is null with conditional probability $A_k(D)/F_D(t_k)$, and
\begin{equation}\label{eq:F-cut}
\E[\mathrm{FDP}\mid D]=\sum_{k\ge1:\,F_D(t_k)>0}\binom mk F_D(t_k)^k\,\Phi_{m-k,k}(D)\,\frac{A_k(D)}{F_D(t_k)}.
\end{equation}
In both forms, $\FDR=\sum_D\PP(D)\,\E[\mathrm{FDP}\mid D]$.

\paragraph{(d) The independent checker.} A separate program, written directly from the definitions above and sharing no
code with the engines of Section~\ref{sec:F-ii}, evaluates
\begin{equation}\label{eq:F-check}
\FDR=\sum_{z\in\mathcal Z^n}\ \sum_{z'\in\mathcal Z'^m}\ \prod_{i=1}^n P(z_i)\prod_{j=1}^m Q'(z'_j)
\int_{[0,1]^m}\mathrm{FDP}(z,z',u)\,du,
\end{equation}
where $\mathcal Z$ is the (finite) support of $(X,Y)$ under $P$, $\mathcal Z'$ the support of $(X,Y,c)$ for a test unit,
$Q'$ its law, and $\mathrm{FDP}(z,z',u)$ is the BH false discovery proportion for calibration sequence $z$, test sequence
$z'$ and tie-breakers $u$. It enumerates every \emph{ordered} calibration and test sequence. For each, $p_j=\alpha_j+\beta_j
u_j$ with $\alpha_j,\beta_j$ read off eq.~(4); it cuts $[0,1]$ for each $u_j$ at the points where $\alpha_j+\beta_ju_j$
equals a grid point $t_k$, so that on every box of the product partition each $p_j$ stays in one BH bin and BH's rejection
set is constant; it evaluates BH at the box centre and adds (box volume)$\times$FDP. The integral in \eqref{eq:F-check}
is therefore computed exactly. The same program computes $\PP(p_1\le t,\ \text{unit 1 null})$ exactly, to check
eq.~(3) of \citet{jin2026weighted} and, for oracle nulls, the exact uniformity of Lemma~\ref{lem:U}.

\subsection{Instance (i): global null, $n=5$, $m=2$, $q=4/5$}\label{sec:F-i}

\paragraph{Configuration.} Covariate $X\in\{0,1,2\}$; response $Y\equiv0$; pre-determined threshold $c_{n+j}=0$ for every
test unit, so every test unit is null and $\FDR=\PP(\mathcal R\ne\emptyset)$; score $V(x,y)=y-\hat\mu(x)$, monotone in
$y$, with $\hat\mu=(-2,-1,0)$. Since $c=Y$, $\hat V_{n+j}=V(X_{n+j},Y_{n+j})$: every p-value is an oracle p-value.
\begin{center}
\begin{tabular}{@{}cccccc@{}}
\toprule
cell $x$ & $P(X=x)$ & $w(x)$ & $\hat\mu(x)$ & score $V(x,0)$ & $Q(X=x)$\\
\midrule
$0$ & $3/4$ & $1$ & $-2$ & $2$ & $15/124$\\
$1$ & $3/20$ & $3$ & $-1$ & $1$ & $9/124$\\
$2$ & $1/10$ & $50$ & $0$ & $0$ & $25/31$\\
\bottomrule
\end{tabular}
\end{center}
Here $\E_Pw=31/5$ and $Q(X=x)=P(X=x)w(x)/\E_Pw$. The weight ratio is $B=50$, and the weight is strictly decreasing in the
score. BH's grid is $t_1=q/2=2/5$, $t_2=q=4/5$.

\paragraph{Per-cell laws.} Let $(N_0,N_1,N_2)$ be the calibration counts of the three cells, so
\[
\PP(D)=\frac{5!}{N_0!\,N_1!\,N_2!}\Bigl(\frac34\Bigr)^{N_0}\Bigl(\frac3{20}\Bigr)^{N_1}\Bigl(\frac1{10}\Bigr)^{N_2},
\]
and $S=N_0+3N_1+50N_2$. By \eqref{eq:F-window}, given $D$ a test unit in cell $x$ has
\[
x=2:\ \Unif\Bigl[0,\tfrac{50(1+N_2)}{S+50}\Bigr],\qquad
x=1:\ \Unif\Bigl[\tfrac{50N_2}{S+3},\tfrac{50N_2+3(1+N_1)}{S+3}\Bigr],\qquad
x=0:\ \Unif\Bigl[\tfrac{50N_2+3N_1}{S+1},1\Bigr].
\]

\paragraph{Formula evaluated.} The certified value was computed by the checker, i.e.\ by \eqref{eq:F-check} with
$|\mathcal Z|=|\mathcal Z'|=3$: $3^5$ ordered calibration sequences times $3^2$ ordered test sequences, each with the
exact box integral over $(U_1,U_2)$ and $\mathrm{FDP}=\one\{\mathcal R\neq\emptyset\}$.

\paragraph{Analytic form ($m=2$).} The following two lemmas reduce the FDR of an $m=2$ global-null instance to a sum
over the count vectors $(N_0,N_1,N_2)$ and show what an excess over $q$ means.

\begin{lemma}[Two tests, global null]\label{lem:F-m2}
Let $m=2$, let both test units be null, and let each $p_j$ be exactly $\Unif(0,1)$. Then BH at level $q$ has
\[
\FDR=\mathrm{FWER}=q+\PP\bigl(p_1,p_2\in(q/2,q]\bigr)-\PP\bigl(p_1,p_2\in(0,q/2]\bigr).
\]
\end{lemma}
\begin{proof}
BH rejects something iff $\min(p_1,p_2)\le q/2$ or $\max(p_1,p_2)\le q$. Write $A_j=\{p_j\le q/2\}$ and
$B_j=\{q/2<p_j\le q\}$. The rejection event is $(A_1\cup A_2)\cup(B_1\cap B_2)$, and the two pieces are disjoint. Exact
uniformity gives $\PP(A_1\cup A_2)=q/2+q/2-\PP(A_1\cap A_2)$.
\end{proof}
For independent uniform p-values both probabilities equal $q^2/4$, so the FWER is exactly $q$; BH overshoots $q$ exactly
when ``both moderate'' coincidences are more likely than ``both very small'' ones.

\begin{lemma}[Conditioning on the calibration data]\label{lem:F-var}
In the setting of Lemma~\ref{lem:F-m2}, suppose in addition that $p_1,p_2$ are i.i.d.\ given $D$, and let
$a(D):=\PP(p_1\le q/2\mid D)$, $b(D):=\PP(q/2<p_1\le q\mid D)$. Then $\E\,a(D)=\E\,b(D)=q/2$ and
\[
\FDR-q=\E[b(D)^2]-\E[a(D)^2]=\Var\,b(D)-\Var\,a(D).
\]
\end{lemma}
\begin{proof}
By conditional independence, $\PP(p_1,p_2\in(0,q/2])=\E[a(D)^2]$ and $\PP(p_1,p_2\in(q/2,q])=\E[b(D)^2]$. Exact
uniformity gives $\E\,a(D)=\PP(p_1\le q/2)=q/2$ and $\E\,b(D)=q/2$. Substitute into Lemma~\ref{lem:F-m2}.
\end{proof}

Instance (i) satisfies both lemmas: every unit is null; the p-values are oracle p-values, hence exactly uniform by
Lemma~\ref{lem:U}; and they are i.i.d.\ given $D$ by (b). With the per-cell laws above,
$a(D)=\sum_xQ(X=x)\,G_x(2/5)$ and $b(D)=\sum_xQ(X=x)\bigl(G_x(4/5)-G_x(2/5)\bigr)$, so
\begin{equation}\label{eq:F-i-var}
\FDR=\frac45+\sum_{N_0+N_1+N_2=5}\PP(D)\bigl(b(D)^2-a(D)^2\bigr),
\end{equation}
a sum of $21$ rational terms. The interactions between the two test p-values pass only through the shared calibration
data, and Lemma~\ref{lem:F-var} says that BH exceeds $q$ exactly when the calibration draw moves the chance of a
``moderate'' p-value, $b(D)$, more than it moves the chance of a ``very small'' one, $a(D)$. For instance (i) the
certified value below is therefore equivalent to $\Var\,b(D)-\Var\,a(D)=102507849/492032000000>0$.

\paragraph{Value and certification.} The checker returns, in exact rational arithmetic,
\[
\FDR=\frac{393728107849}{492032000000}\approx0.8002083,\qquad
\FDR-q=\frac{102507849}{492032000000}\approx2.083\times10^{-4},\qquad \FDR/q\approx1.00026,
\]
so $\FDR>q$. The same program verified exactly that $\PP(p_1\le t)=t$ at $t=2/5$ and $t=4/5$, which a missing own weight or
a wrong tie term in eq.~(4) would break. A second program, written separately and sharing no code with the checker,
evaluates the $21$-term sum \eqref{eq:F-i-var} from the per-cell laws above, in exact rational arithmetic. It returns the
same fraction, with total mass exactly $1$ and $\E\,a(D)=\E\,b(D)=2/5$ exactly. It relies on the per-cell laws and on
Lemma~\ref{lem:F-var}; the checker relies on neither. The checker itself was validated on other instances: it agrees exactly with the first strict-selection
engine on ten small instances (Section~\ref{sec:F-ii}), and on a smaller global-null instance ($n=3$, $m=2$) it agrees
with a separate search engine in floating point and with a direct Monte Carlo simulation.

\subsection{Instance (ii): strict selection, $n=80$, $m=8$, $q=1/10$}\label{sec:F-ii}

\paragraph{Configuration.} Three covariate values: a light value $\ell$ and two heavy values $h_1,h_2$. The score is
$V(x,y)=y$ (monotone), and the threshold is a function of the covariate.
\begin{itemize}[leftmargin=*,itemsep=1pt]
\item Light covariate $\ell$: $P$-mass $\pi_\ell=1-\pi_1-\pi_2$, weight $1$; $Y\in\{0,1\}$ with $\PP(Y=0\mid X=\ell)=\nu$;
threshold $c(\ell)=1/2$. So $\hat V=1/2$ for every light test unit, and a light unit is null iff $Y=0$.
\item Heavy covariate $h_1$: $P$-mass $\pi_1$, weight $W_1$, $Y=-2$ deterministically, threshold $c(h_1)=-2$ ($=Y$, null).
\item Heavy covariate $h_2$: $P$-mass $\pi_2$, weight $W_2$, $Y=-1$ deterministically, threshold $c(h_2)=-1$ ($=Y$, null).
\end{itemize}
The certified parameters are
\[
\nu=\tfrac1{25},\qquad \pi_1=\tfrac3{10},\qquad \pi_2=\tfrac7{400},\qquad W_1=\tfrac1{22},\qquad W_2=\tfrac{137}{22},
\]
so $\pi_\ell=273/400$, $\E_Pw=1417/1760$, and $Q(X=\ell)=462/545$, $Q(X=h_1)=24/1417$, $Q(X=h_2)=959/7085$. The
calibration support points $(\ell,0),(\ell,1),(h_1,-2),(h_2,-1)$ have $P$-masses $273/10000$, $819/1250$, $3/10$,
$7/400$. Multiplying all weights by $22$ (which leaves every $p_j$ unchanged) gives $w(h_1)=1$, $w(\ell)=22$,
$w(h_2)=137$, so $1\le w\le B=137$. The threshold is independent of $Y$ given $X$ (it is a function of $X$), and $Y\mid X$
is non-degenerate on the light covariate. BH's grid is $t_k=k/80$, $k=1,\dots,8$. The light units play the role of the
block of non-null ``decoys'' with a shared threshold in Mechanism~2 (Section~\ref{sec:counter}).

\paragraph{Per-type laws.} Let $(L_0,L_1,H_1,H_2)$ be the calibration counts of the four support points, so
\[
\PP(D)=\frac{n!}{L_0!\,L_1!\,H_1!\,H_2!}\,(\pi_\ell\nu)^{L_0}\bigl(\pi_\ell(1-\nu)\bigr)^{L_1}\pi_1^{H_1}\pi_2^{H_2},
\]
$S=L_0+L_1+W_1H_1+W_2H_2$, and $S_<:=L_0+W_1H_1+W_2H_2$ is the calibration weight with score below $1/2$. The calibration
scores are $-2$ ($h_1$), $-1$ ($h_2$), and $0,1$ (light); the test scores are $\hat V=1/2,-2,-1$ for $\ell,h_1,h_2$. So
\eqref{eq:F-window} gives, given $D$,
\[
\ell:\ \Unif\Bigl[\tfrac{S_<}{S+1},\tfrac{S_<+1}{S+1}\Bigr],\qquad
h_1:\ \Unif\Bigl[0,\tfrac{W_1(1+H_1)}{S+W_1}\Bigr],\qquad
h_2:\ \Unif\Bigl[\tfrac{W_1H_1}{S+W_2},\tfrac{W_1H_1+W_2(1+H_2)}{S+W_2}\Bigr].
\]
With $G_\ell,G_1,G_2$ the corresponding CDFs, and since a light unit's null indicator $\one\{Y=0\}$ does not enter its
p-value ($\hat V=1/2$ does not involve $Y$),
\[
F_D(t)=Q(\ell)G_\ell(t)+Q(h_1)G_1(t)+Q(h_2)G_2(t),\qquad
A_k(D)=\nu\,Q(\ell)G_\ell(t_k)+Q(h_1)G_1(t_k)+Q(h_2)G_2(t_k).
\]

\paragraph{Formula evaluated.} $\FDR=\sum_D\PP(D)\,\E[\mathrm{FDP}\mid D]$, summed over all
$\binom{n+3}{3}=91{,}881$ count vectors $(L_0,L_1,H_1,H_2)$ with $L_0+L_1+H_1+H_2=80$, by two engines:
\begin{itemize}[leftmargin=*,itemsep=1pt]
\item \emph{Engine 1} uses the leave-one-out form \eqref{eq:F-loo}, with $\Phi_{m-k,k}$ computed by the dynamic programme
for \eqref{eq:F-phi};
\item \emph{Engine 2} uses the cutoff form \eqref{eq:F-cut}. It was written separately, shares no code with engine 1,
and uses a different identity for $\E[\mathrm{FDP}\mid D]$.
\end{itemize}
Both run in exact rational arithmetic, and each also checks that $\sum_D\PP(D)=1$ exactly.

\paragraph{Value.} Both engines return the same rational number, in lowest terms
$\FDR=N_{80}/(d_{80}\cdot10^{336})$, with numerator ($569$ digits)

\begin{flushleft}
$N_{80}={}$104456433\allowbreak\,7482314073\allowbreak\,0248138811\allowbreak\,3245860339\allowbreak\,2898299247\allowbreak\,2318733735\allowbreak\,8676031448\allowbreak\,3547300861\allowbreak\,2782321331\allowbreak\,5815061658\allowbreak\,3312427966\allowbreak\,1463428661\allowbreak\,1585011394\allowbreak\,6318950238\allowbreak\,4067980264\allowbreak\,2538448638\allowbreak\,4976721839\allowbreak\,1921084266\allowbreak\,5999250429\allowbreak\,0584914773\allowbreak\,2818221075\allowbreak\,8543801237\allowbreak\,4954023727\allowbreak\,7607844425\allowbreak\,7159101331\allowbreak\,8019528009\allowbreak\,2204739149\allowbreak\,4467878328\allowbreak\,9045486301\allowbreak\,1580658176\allowbreak\,2285075186\allowbreak\,7151481525\allowbreak\,0504888427\allowbreak\,5905421146\allowbreak\,1964035123\allowbreak\,2095619518\allowbreak\,0682665978\allowbreak\,3265633987\allowbreak\,8189920029\allowbreak\,5223067603\allowbreak\,9537218268\allowbreak\,2224706340\allowbreak\,2404564721\allowbreak\,1398681223\allowbreak\,2607105929\allowbreak\,6572369553\allowbreak\,4828070428\allowbreak\,3413082641\allowbreak\,9558936586\allowbreak\,9765566866\allowbreak\,0281252301\allowbreak\,9284524933\allowbreak\,9195544581\allowbreak\,1368159629\allowbreak\,4342248282\allowbreak\,3661578683\allowbreak\,1147681427
\end{flushleft}
and, in the denominator, ($233$ digits)
\begin{flushleft}
$d_{80}={}$919\allowbreak\,6331503758\allowbreak\,1420790804\allowbreak\,1446298776\allowbreak\,3676496700\allowbreak\,9302451975\allowbreak\,2670247764\allowbreak\,6836992723\allowbreak\,4919750918\allowbreak\,0503911891\allowbreak\,5674774122\allowbreak\,5486894368\allowbreak\,1381382655\allowbreak\,8883546834\allowbreak\,9878835417\allowbreak\,4759379095\allowbreak\,2151814014\allowbreak\,9085109841\allowbreak\,1136247293\allowbreak\,8533861838\allowbreak\,1517587244\allowbreak\,5509295744\allowbreak\,6089775464\allowbreak\,3035062272
\end{flushleft}
(digits grouped in tens from the right). Numerically $\FDR=0.1135848938\ldots$, $\FDR/q=1.1358489383\ldots$, and
$\FDR-q=0.0135848938\ldots>0$.

\paragraph{Certification.} The following checks were run.
\begin{enumerate}[leftmargin=*,itemsep=1pt]
\item \emph{Two exact engines.} Engines 1 and 2 return exactly the same rational number at $n=80$ (above) and on the
certified $n=40$ instance of Remark~\ref{rem:F-n}. They were also compared at $n=3,6,10,20$ ($m=8$, $q=1/10$) on five
parameter sets $(\nu,\pi_1,\pi_2,W_1,W_2)$: the three certified sets of Remark~\ref{rem:F-n}, and $(1/5,1/5,1/5,3/2,4)$ and
$(3/10,1/10,3/10,17/9,2)$, drawn pseudo-randomly with a fixed seed. On all $20$ instances the two engines return the
same rational number, and each finds total mass exactly $1$. At $n=20$ with the certified $n=20$ parameters both give
$\FDR/q=1.0694867\ldots$, the value in Remark~\ref{rem:F-n}.
\item \emph{Independent checker.} Engine 1 equals the checker of Section~\ref{sec:F-common}(d), as exact rationals, on ten
small instances of the same model: $(n,m)\in\{(2,2),(3,2),(2,3),(3,3),(4,2)\}$, two each, with $q\in\{1/5,3/10\}$ and the
other parameters pseudo-random rationals. Since the checker evaluates eq.~(4) and BH directly, this tests the model
reduction (the per-type laws, (b), and \eqref{eq:F-loo}), not only the arithmetic.
\item \emph{Direct simulation.} A separate simulation of the actual units (eq.~(4) and BH, no use of the reductions) at the
certified $n=80$ parameters, $2\times10^5$ replications, gives $\FDR/q=1.1338\pm0.0062$, against the exact $1.13585$. The
formula \eqref{eq:F-loo}, evaluated in floating point, was also compared with this simulation on two other configurations
(differences at most $1.4$ standard errors).
\end{enumerate}
What this does not cover: the checker was not run on the certified $n=80$ instance itself (it would enumerate
$4^{80}\cdot4^{8}$ ordered sequences), and both engines rely on the same per-type laws above, which are tested against the checker
only on small instances.

\begin{remark}[The same construction at smaller $n$]\label{rem:F-n}
With the parameters re-chosen for each $n$ ($m=8$, $q=1/10$), the same model was certified in exact arithmetic at
$n=20$ and $n=40$:
\begin{center}
\begin{tabular}{@{}ccccccccc@{}}
\toprule
$n$ & $\nu$ & $\pi_1$ & $\pi_2$ & $W_1$ & $W_2$ & $B$ & exact $\FDR/q$ & simulation\\
\midrule
$20$ & $1/250$ & $32/125$ & $91/1000$ & $7/125$ & $43/25$ & $215/7\approx30.7$ & $1.0694867\ldots$ & $1.0657\pm0.0063$\\
$40$ & $1/50$ & $71/250$ & $1/25$ & $1/18$ & $59/18$ & $59$ & $1.1028674\ldots$ & $1.1067\pm0.0063$\\
$80$ & $1/25$ & $3/10$ & $7/400$ & $1/22$ & $137/22$ & $137$ & $1.1358489\ldots$ & $1.1338\pm0.0062$\\
\bottomrule
\end{tabular}
\end{center}
At $n=20$ the sum runs over all $1{,}771$ count vectors, with total mass exactly $1$. At $n=40$ the two engines agree
exactly. The simulations use $2\times10^5$ replications. The excess grows with $n$: $6.9\%$, $10.3\%$ and $13.6\%$.
\end{remark}

\subsection{Proof of the theorem}

\begin{proof}[Proof of Theorem~\ref{thm:counter}]
The proof is by exact computation. Both instances satisfy \citet{jin2026weighted}'s assumptions: covariate shift with a
known weight function bounded above and below ($B=50$ and $B=137$); i.i.d.\ test units independent of the calibration data;
thresholds that are fixed functions of the test unit's own covariate (constant in (i)), so that their independence
condition holds; and a score monotone in $y$. In instance (i) every test unit is null with $c=Y$ (global null). In instance (ii) the threshold is
independent of $Y$ given $X$ and $Y\mid X$ is non-degenerate on the light covariate (strict selection).
(i) By Section~\ref{sec:F-i}, $\FDR=393728107849/492032000000>4/5=q$.
(ii) By Section~\ref{sec:F-ii}, $\FDR=N_{80}/(d_{80}\cdot10^{336})=1.13585\,q$ to five decimals, $>q$.
In both cases the comparison with $q$ is between exact rationals.
\end{proof}

\section{Use of AI tools}\label{app:aiuse}

Large language models were used throughout the research process, including to help write the code for the numerical
computations and to revise the manuscript. We checked all results, code and claims, and take full responsibility
for the paper.

\end{document}